\documentclass{aa}
\usepackage{graphicx}
\usepackage{hyperref}
\usepackage{natbib}
\bibpunct{(}{)}{;}{a}{}{,} 
\usepackage{afterpage}
\usepackage{xspace}
\usepackage{textcomp}
\usepackage{xcolor}
\usepackage{amssymb}
\usepackage{multirow}
\usepackage{soul,xcolor}
\usepackage{xtab}
\usepackage{arydshln}
\usepackage{longtable}
\usepackage{txfonts}

\newcommand{\cc}{\ensuremath{{\rm cm}^{-3}}\xspace}
\newcommand{\amm}{\ensuremath{{\rm NH}_3}\xspace}
\newcommand{\tex}{\ensuremath{{T_{\rm ex}}}\xspace}
\newcommand{\tk}{\ensuremath{T_{\rm K}}\xspace}
\newcommand{\sig}{\ensuremath{\sigma_{\rm v}}\xspace}
\newcommand{\vel}{\ensuremath{\rm{v_{ LSR}}}\xspace}
\newcommand{\htw}{\ensuremath{\rm H_2}\xspace}
\newcommand{\msun}{\ensuremath{M_\odot}\xspace}
\newcommand{\td}{\ensuremath{T_{\rm dust}}\xspace}
\newcommand{\kms}{\ensuremath{{\rm km\,s}^{-1}}\xspace}
\newcommand{\ms}{\ensuremath{{\rm m s}^{-1}}\xspace}
\defcitealias{choudhury2020_letter}{Paper~I}

\begin{document}

   \title{Transition from Coherent Cores to Surrounding Cloud in L1688}

   \author{
    Spandan Choudhury \inst{1}
        \and
    Jaime E. Pineda \inst{1}
        \and
    Paola Caselli \inst{1}
        \and
    Stella S. R. Offner \inst{2}
        \and
    Erik Rosolowsky \inst{3}
        \and
    Rachel K. Friesen \inst{4}
        \and
    Elena Redaelli \inst{1}
        \and
    Ana Chac\'on-Tanarro \inst{5}
        \and
    Yancy Shirley \inst{6}
        \and
    Anna Punanova \inst{7}
        \and
    Helen Kirk \inst{8,9}
    }
    \institute{
        Max-Planck-Institut f\"ur extraterrestrische Physik, Giessenbachstrasse 1, D-85748 Garching, Germany\\
        \email{spandan@mpe.mpg.de}
        \and
        Department of Astronomy, The University of Texas at Austin, Austin, TX 78712, USA
        \and
        Department of Physics, 4-181 CCIS, University of Alberta, Edmonton, AB T6G 2E1, Canada
        \and
        Department of Astronomy \& Astrophysics, University of Toronto, 50 St. George St., Toronto, ON M5S 3H4, Canada
        \and
        Observatorio Astron\'omico Nacional (OAN-IGN), Alfonso XII 3, 28014, Madrid, Spain
        \and
        Steward Observatory, 933 North Cherry Ave., Tucson, AZ 85721, USA
        \and
        Ural Federal University, 620002 Mira st. 19, Yekaterinburg, Russia
        \and
        Department of Physics and Astronomy, University of Victoria, 3800 Finnerty Rd., Victoria, BC V8P 5C2, Canada
        \and
        Herzberg Astronomy and Astrophysics, National Research Council of Canada, 5071 West Saanich Rd., Victoria, BC V9E 2E7, Canada 
        }

   \date{}

 
  \abstract
   {Stars form in cold dense cores showing subsonic velocity dispersions. The parental molecular clouds display higher temperatures and supersonic velocity dispersions.
   The transition from core to cloud has been observed in velocity dispersion, but temperature and abundance variations are unknown.}
   {We aim to measure the temperature and velocity dispersion across cores and ambient cloud in a single tracer to study the transition between the two regions.}
   {We use \amm (1,1) and (2,2) maps in L1688 from the Green Bank Ammonia Survey, smoothed to 1\arcmin, and determine the physical properties by fitting the spectra.
   We identify the coherent cores and study the changes in temperature and velocity dispersion from cores to the surrounding cloud.}
   {We obtain a kinetic temperature map extending beyond dense cores and tracing the cloud, improving from previous maps tracing mostly the cores. 
   The cloud is 4--6 K warmer than the cores, and shows a larger velocity dispersion ($\Delta$\sig = 0.15--0.25 \kms).
   Comparing to $\emph{Herschel}$-based dust temperatures, we find that cores show kinetic temperature $\approx$1.8 K lower than the dust temperature; while the gas temperature is higher than the dust temperature in the cloud.
   We find an average p-\amm fractional abundance (with respect to \htw) of (4.2$\pm$0.2) $\times 10^{-9}$ towards the coherent cores, and (1.4$\pm$0.1) $\times 10^{-9}$ outside the core boundaries.
   Using stacked spectra, we detect two components, one narrow and one broad, towards cores and their neighbourhoods. We find the turbulence in the narrow component to be correlated to the size of the structure (Pearson-r=0.54).
   With these unresolved regional measurements, we obtain a turbulence-size relation of $\sigma_{\rm v,NT} \propto r^{0.5}$, similar to previous findings using multiple tracers.
}
   {
   We discover that the subsonic component extends up to 0.15 pc beyond the typical coherent boundaries, unveiling larger extents of the coherent cores and showing gradual transition to coherence over $\sim$ 0.2 pc. 
   }


   \maketitle
%

\section{Introduction}

Star formation takes place in dense cores in molecular clouds. Detailed studies of dense cores unveil their physical and chemical properties, which provide the initial conditions in the process of star formation. Across different molecular clouds, the star-forming cores are characterised by higher density and lower temperatures, compared to the ambient cloud. Many of the cores are also shown to exhibit subsonic turbulence. However, the transition from core to cloud is still not well-understood.

Hyperfine structure of \amm inversion transitions allows its individual components to remain optically thin at high column densities \citep{caselli_2017_amm_abun}. Unlike carbon-bearing species, such as CO and HCO$^+$, \amm appears to not deplete at high densities and cold temperatures characteristic of cores \citep{bergin_langer_97_depletion}. Therefore, \amm is an important and useful high-density tracer of cold gas. Using \amm (1,1) line emission, \citet{coh_core_barranco_goodman_1998} found that the linewidth inside the four cores studied was roughly constant, and slightly greater than the pure thermal value. They also reported that at the edge of the cores, the linewidths begin to increase. 
By analysing the cores and their environments, \citet{coh_core_goodman_barranco_1998} suggested that a transition to coherence might mark the boundaries of the dense cores.
Using \amm (1,1) observations with the Green Bank Telescope, \citet{pineda2010} studied the transition from inside a core to the surrounding gas, for the first time in the same tracer, and reported a similar, but sharper transition to coherence (an increase in the dispersion by a factor of 2 in a scale less than 0.04 pc) in the B5 region in Perseus. 
However, the exact nature of the transition from the subsonic cores to the surrounding molecular cloud is not well-known. It is important to study these transition regions, as this could give us clues on how dense cores form and accrete material from the surrounding cloud.

In the Green Bank Ammonia Survey \citep[GAS,][]{GASDR1}, star forming regions in the Gould Belt were observed using \amm hyperfine transitions. Their first data release included four regions in nearby molecular clouds: B18 in Taurus, NGC1333 in Perseus, L1688 in Ophiuchus, and Orion A North. Using the results from this survey and \htw column densities derived with $\emph{Herschel}$, \citet{chen2019} identified 18 coherent structures (termed ``droplet'') in L1688 and B18. They observed that these droplets show gas at high density ($\rm \langle n_{H} \rangle \approx 5 \times 10^4$ cm$^{-3}$; from masses and effective radii of the droplets, assuming a spherical geometry) and near-constant, almost-thermal velocity dispersion, with a sharp transition in dispersion around the boundary. 

The results from \citet{pineda2010} and \citet{chen2019} suggest that we can define the boundaries of coherent cores systematically as the regions with subsonic non-thermal linewidths. It is to be noted that this approach does not necessarily define cores identically to those defined using continuum emission. Therefore, not all of the ``coherent cores'' will have continuum counterparts.

The temperature profile inside cores has been studied, and the cores are found to be usually at a temperature $\approx$10 K \citep[e.g., ][]{tafalla_2002}, and more dynamically evolved cores show a temperature drop towards the centre \citep{crapsi_2007_temp_drop_core, pagani_2007_core_temp}. \citet{crapsi_2007_temp_drop_core} observed a temperature drop down to $\approx$ 6$\,$K towards the centre of L1544 \citep[see also][]{young_2004_tmp_drp_l1544}. \citet{pagani_2007_core_temp} and \citet{lht_2013_core_temp} also reported gradients in temperature outwards from the centres of cores with observations of $\rm{N_{2}H^{+}}-\rm{N_{2}D^{+}}$, and FIR-submillimetre continuum, respectively. However, the transition in gas temperature from cores to their immediate surrounding has not been studied. 
This is important, as dust and gas are not thermally coupled at volume densities below 10$^5$ cm$^{-3}$ \citep{goldsmith_2001_dust_gas_coup}, as expected in inter-core material (average density in L1688 is $\sim 4 \times 10^3\, \rm{cm}^{-3}$, see Section \ref{sec_amm_data}). Therefore, the gas temperature can give us important constraints on the cosmic-ray ionisation rate (with cosmic rays being the main heating agents of dark clouds), as discussed in \citet{alexei_2019_cr_ir}.
This transition in temperature, as well as in other physical properties, such as velocity dispersion, density and \amm abundance, from core to cloud, is the focus of the present paper.

In our previous paper \citep[][hereafter Paper I]{choudhury2020_letter}
we reported a faint supersonic component along with the narrow core component, towards all cores. We suggested that the broad component traces the cloud surrounding the cores, and therefore, presents an opportunity to study the gas in the neighbourhood of the cores with the same density tracer. Here, we extend that analysis to study the changes in physical properties of these two components from cores to their surrounding.

In this project, we use the data from GAS in L1688, smoothed to a larger beam, to study the transition in both velocity dispersion and temperature, from coherent cores, to the extended molecular cloud, using the same lines (\amm (1,1) and (2,2)). In Section \ref{sec_data}, the primary \amm data from GAS and complimentary $\emph{Herschel}$ continuum maps, are briefly described. Section \ref{sec_analy} explains the procedure used to determine the physical parameters in the cloud. The improved integrated intensity maps and the parameter maps are presented in Section \ref{sec_res}. In Section \ref{sec_diss}, the selection of coherent cores, considered in this paper is explained, followed by discussion of the observed transition of physical parameters and spectra, from cores to their surrounding.


\section{Data}
\label{sec_data}

\subsection{Ammonia maps}
\label{sec_amm_data}

\begin{figure*}[!ht]
    \centering
    \includegraphics[width=0.8\textwidth]{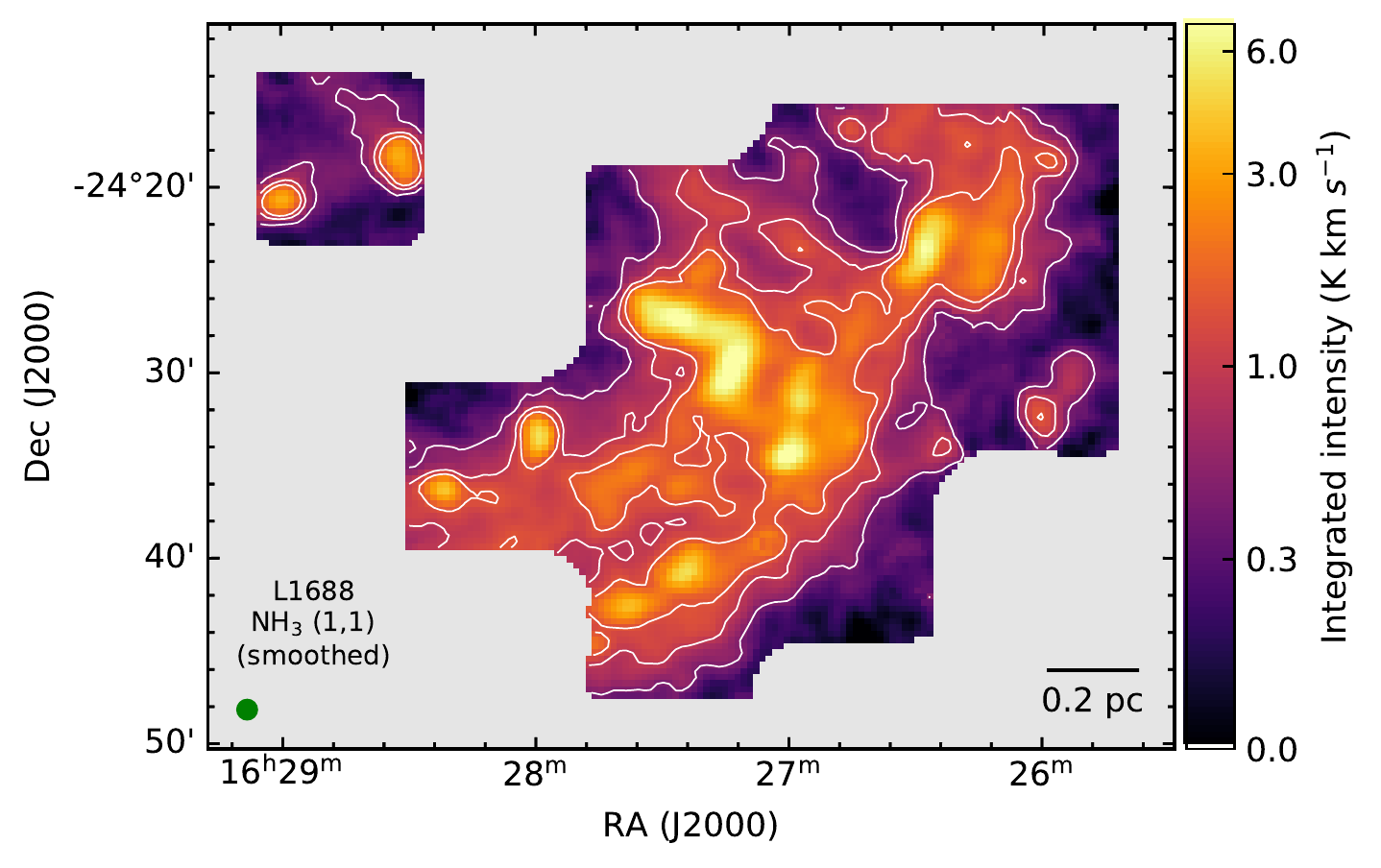}
    \includegraphics[width=0.8\textwidth]{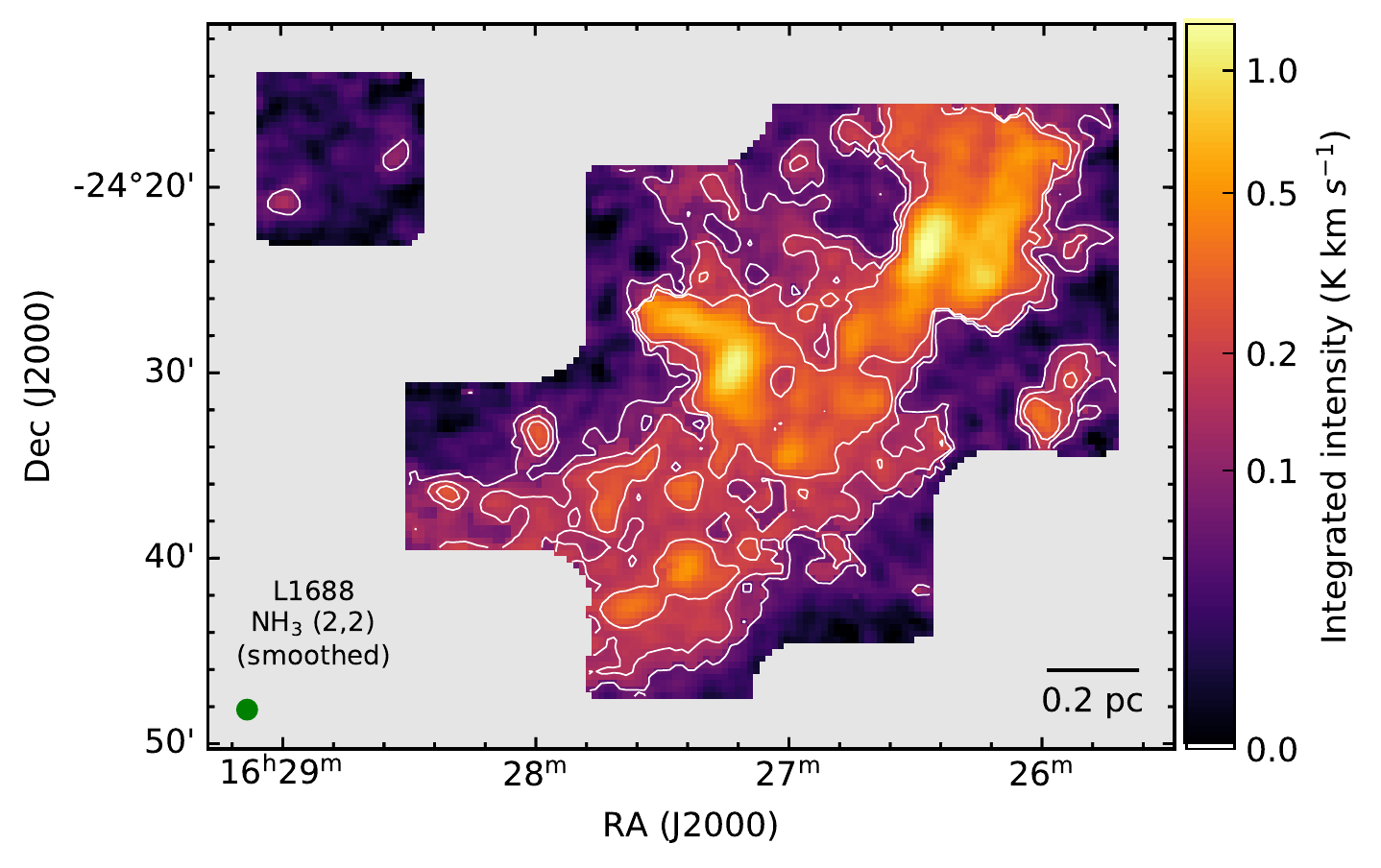}
    \caption{Integrated intensity maps of \amm (1,1) (top) and (2,2) (bottom) lines. For \amm (1,1), the contour levels indicate 15$\rm{\sigma}$,30$\rm{\sigma}$ and 45$\rm{\sigma}$, and for \amm (2,2), contours are shown for 6$\rm{\sigma}$,12$\rm{\sigma}$ and 18$\rm{\sigma}$; where $\rm \sigma$ is the median error in emitted intensity, calculated from signal-free spectral range in each pixel, and converted to the error in integrated intensity. The moment maps were improved by considering only the spectral range containing emission, as described in Section \ref{sec_mom_map}. The 1\arcmin\ beam, which the data was convolved to, is shown in green in the bottom left corner, and the scale bar is shown in the bottom right corner.\label{integ_inten}}
\end{figure*}

As our primary data-set, we use the \amm (1,1) and (2,2) maps, taken from the first data release of Green Bank Ammonia Survey \citep[GAS,][]{GASDR1}. The observations were carried out using the Green Bank Telescope (GBT) to map \amm in the star forming regions in the Gould Belt with $A_V > 7$ mag, using the seven-beam K-Band Focal Plane Array(KFPA) at the $\emph{GBT}$. Observations were done in frequency switching mode with a frequency throw of 4.11 MHz ($\approx$52 \kms at 23.7 GHz). The spectral resolution of the data is 5.7 kHz, which corresponds to $\approx$ 72.1 \ms at 23.7 GHz (approximate frequency of observations). The extents of the maps were selected using continuum data from $\emph{Herschel}$ or $\emph{JCMT}$, or extinction maps derived from 2MASS (Two Micron All Sky Survey). To convert the spectra from frequency to velocity space, central frequencies for \amm (1,1) and (2,2) lines were considered as 23.6944955 GHz and 23.7226333 GHz, respectively \citep{amm_db}.

Out of the four regions in the GAS DR1, we focus on L1688 in this paper, as we were able to obtain an extended kinetic temperature map of the cloud for this region. L1688 is part of the Ophiuchus molecular cloud, at a distance of $\sim 138.4 \pm 2.6 $  pc \citep{l1688_dist_ortiz-leon}. The cloud mapped in \amm is $\sim$1 pc in radius with a mass of $\approx$980 \msun \citep{ladje_2020_hgbs}. Assuming spherical geometry and a mean molecular weight of 2.37 amu \citep{jens_2008_mu}, the average gas density is then $\sim 4 \times 10^3\,$\cc.

The parameter maps of L1688, released in DR1, are not very extended, particularly for kinetic temperature, as a robust temperature measurement is restricted by the signal-to-noise  ratio (SNR) of the (2,2) line. Therefore, in order to have a good detection, especially of the (2,2) line, in the outer and less dense part of the cloud, the data was smoothed by convolving it to a beam of 1\arcmin\ ($\emph{GBT}$ native beam at 23 GHz is $\approx$31$\arcsec$). The data cube was then re-gridded to avoid oversampling. The relative pixel size was kept the same as the original GAS maps, at one-third the beam-width. Mean rms (root-mean-square) noise level achieved as a result is 0.041 K and 0.042 K in the (1,1) and  (2,2), respectively. For comparison, the mean noise level in the GAS DR1 maps was 0.17 K.
Figure \ref{integ_inten} shows the integrated intensity maps obtained for \amm (1,1) and (2,2).

\subsection{\htw column density and dust temperature maps}
\label{sec_h2_td_data}

\begin{figure*}[!ht]  
\centering
\includegraphics[width=0.76\textwidth]{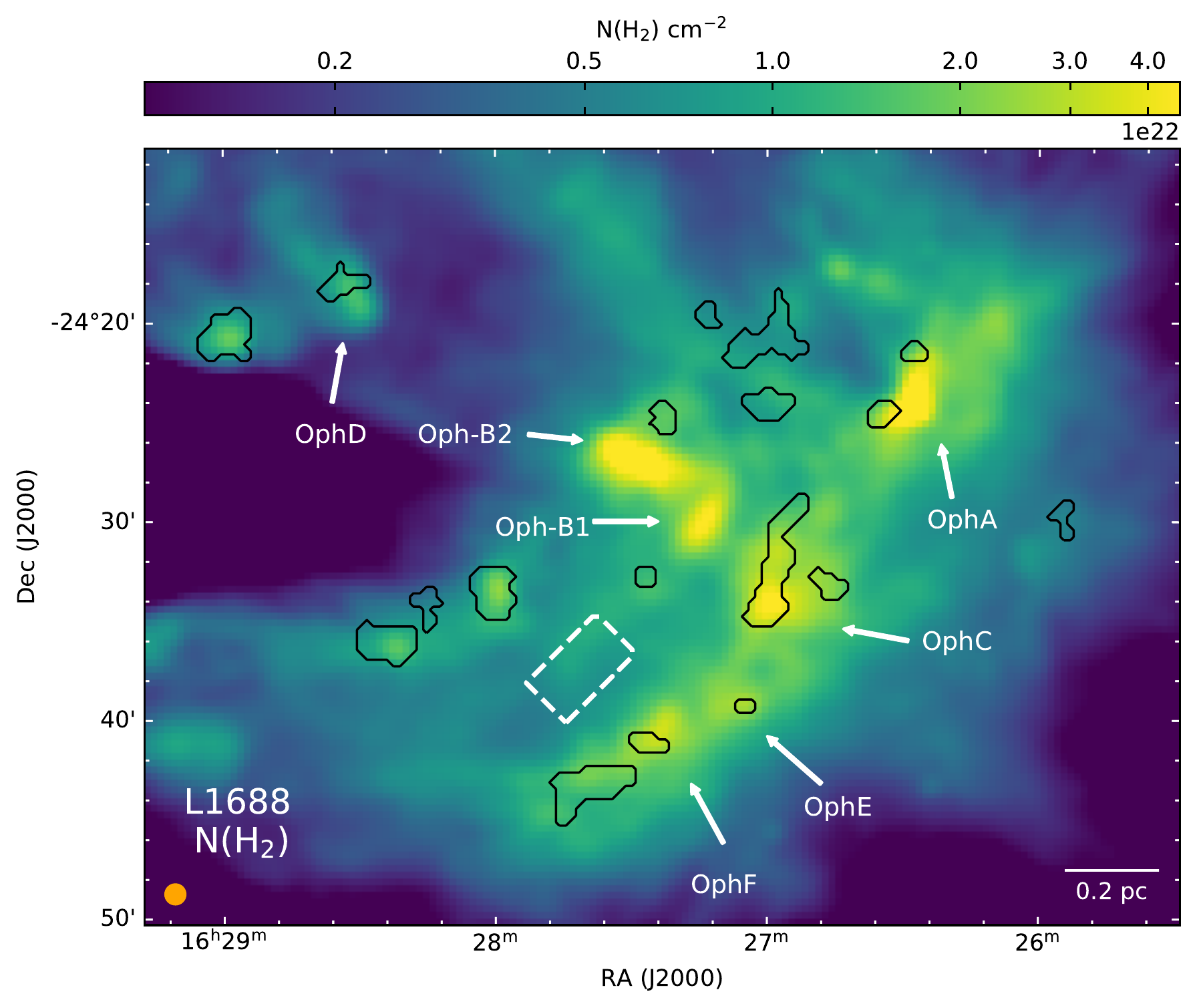}
  \caption{N(\htw) in L1688, taken from Herschel Gould Belt Survey archive. Positions of the continuum cores reported in \citep{motte1998} are indicated by arrows. The dashed rectangle shows the area used to calculate mean cloud properties (see Section \ref{sec_tk_sig}). The solid black contours show the coherent cores in the region (described later in Section \ref{sec_id_coh}). 
  The 1\arcmin\ beam, which the data was convolved to, is shown in orange in the bottom left corner, and the scale bar is shown in the bottom right corner.}
     \label{h2_col} 
     
\end{figure*}

\begin{figure*}[!ht]  
\centering
\includegraphics[width=0.76\textwidth]{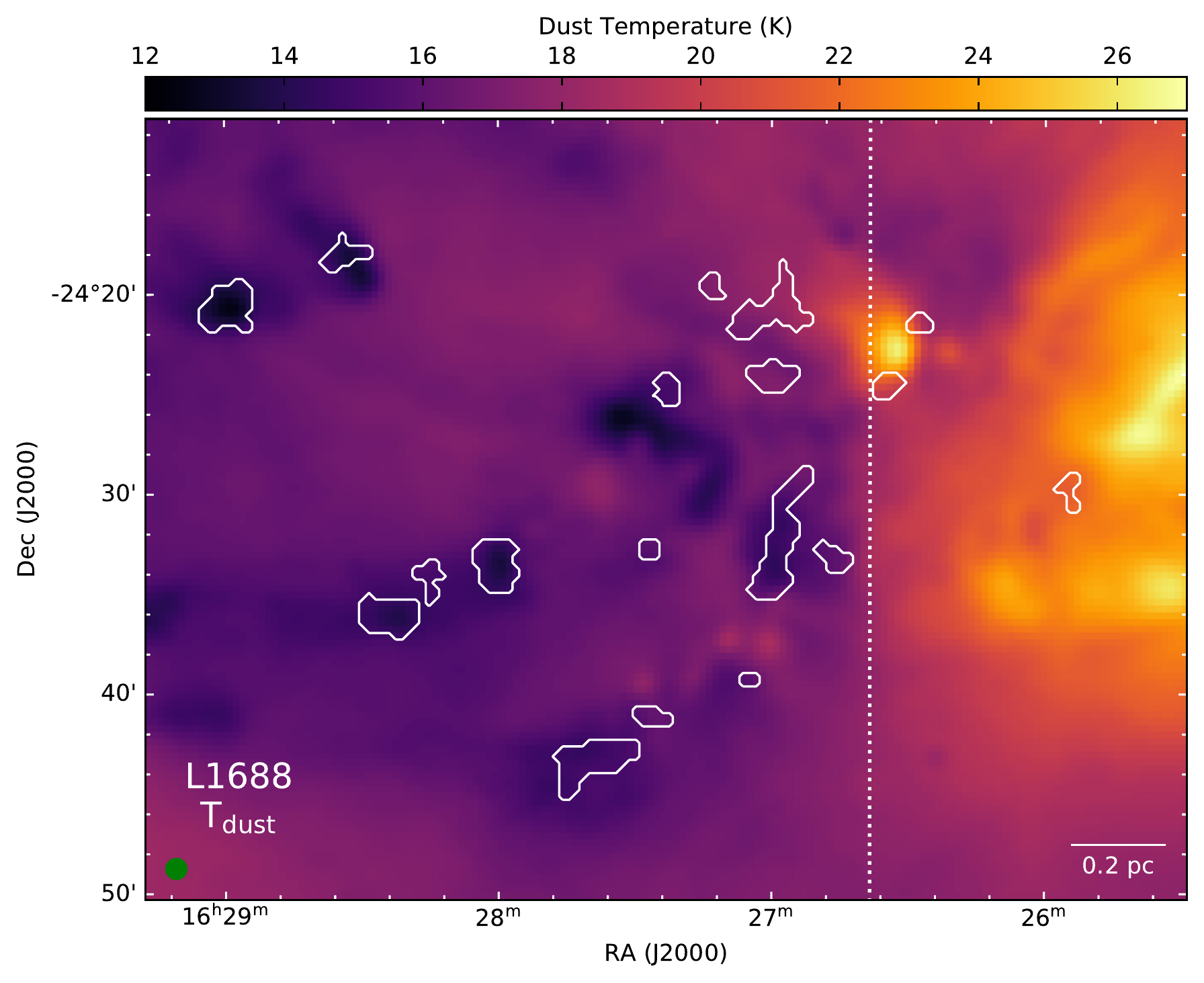}
  \caption{Dust temperature in L1688, taken from Herschel Gould Belt Survey archive. 
  The solid white contours show the coherent cores in the region (described later in Section \ref{sec_id_coh}).
  The vertical dotted line roughly separates the dark cloud (to the left of the line) from the molecular material affected by the external illumination, due to irradiation from HD147889 \citep{habart2003}.
  The 1\arcmin\ beam, which the data was convolved to, is shown in green in the bottom left corner, and the scale bar is shown in the bottom right corner.}
     \label{dust_temp} 
     
\end{figure*}

L1688 was observed in dust continuum using the $\emph{Herschel}$ space observatory as part of the Herschel Gould Belt Survey  \citep[HGBS,][]{hgbs_andre2010}. In order to compare the dust derived properties to our results with \amm, we use the dust temperature and \htw column density in Ophiuchus from the HGBS archive\footnote{\url{www.herschel.fr/cea/gouldbelt/en/archives}} \citep{hgbs_ladj2016}. These maps were also convolved to a 1\arcmin\ beam, and then regridded to the same grid as the \amm maps. Figures \ref{h2_col} and \ref{dust_temp} show the smoothed and re-gridded N(\htw) and dust temperature maps, respectively.

\section{Analysis}
\label{sec_analy}

\subsection{Line fitting}
\label{sec_line_fit}

We fit \amm line profiles to the data with the \verb+pyspeckit+ package \citep{pyspeckit}, which uses a forward modelling approach. We follow the fitting process described in \citet{GASDR1}. The range in velocity to fit is determined from the average spectra of the entire region. Since the ortho-to-para ratio in the region is not known, we only report the p-\amm column density here, and do not attempt to convert it to total \amm column density\footnote{To compare to the total \amm column densities reported in other works, an easy conversion is to multiply N(p-\amm) by 2, if the ortho-to-para ratio is assumed to be the LTE value of unity.}.

The model produces synthetic spectra based on the guesses provided for the input parameters : excitation temperature (\tex), kinetic temperature (\tk), para-ammonia column density (N(p-\amm)), velocity dispersion (\sig) and line-of-sight central velocity (\vel) of the gas \citep[see Section 3.1 in][]{GASDR1}. A non-linear gradient descent algorithm, MPFIT \citep{markwardt2009} is then used to determine the best-fit model, and corresponding values of the parameters. A good set of initial conditions is necessary to ensure that the non-linear least-squares fitting does not get stuck in a local minimum. The value of \vel is critical, therefore we use the first-order moments of the (1,1) line as initial guesses. The second order moments are used as guesses for velocity dispersion. For the other parameters, we use :
$\log_{10} (N_{p-\amm}/ {\rm cm^{-2}}) = 14$, $\tk = 20$\,K and $\tex = 5$\,K. These numbers are within the range of values reported in GAS DR1 maps, and therefore, are reasonable guesses. As a test, we checked the fits with varying guesses, and determined that the exact values of the initial parameters did not affect the final results, as long as they were within a reasonable range.

In this work, we use the \verb+cold_ammonia+ model in \verb+pyspeckit+ library, which makes the assumption that only the (1,1), (2,2) and (2,1) levels are occupied in p-\amm. It is also assumed that the radiative excitation temperature, \tex is the same for (1,1) and (2,2) lines, as well as their hyperfine components.

\subsection{Data masking}
\label{sec_data_mask}

To ensure that the parameters are well-determined from the fit, several flags are applied on each parameter, separately. For a good fit to the (1,1) and (2,2) lines, a clear detection is necessary. For LSR velocity, velocity dispersion and excitation temperature, we mask the pixels with a SNR<5 in the integrated intensity of the (1,1) line. Similarly, we remove pixels with a SNR<3 in the peak intensity of the central hyperfine of the (2,2), in the kinetic temperature and p-\amm column density maps. These lower limits for the SNR in the (1,1) and (2,2) lines were chosen after a visual inspection of the spectra and the model fits in randomly sampled pixels, in order to determine which flags most effectively filtered out the unreliable fits. Since we use the pixels with a SNR of 5 or more in the integrated intensity of the (1,1), we expect the excitation temperature, centroid velocity and velocity dispersion to be determined with reasonably high certainty. Similarly, since for kinetic temperature, we only use the pixels with detection of the (2,2) with SNR of 3 in the peak intensity, \tk should also be well-determined. Therefore, we mask the pixels with a fit determined error >20\% of the corresponding parameter from the \tk, \vel and \sig maps. Calculation of N(p-\amm) requires a good determination of excitation temperature, and 
\texttt{pyspeckit} reports log(N(p-\amm)), and the error in log(N(p-\amm)). We still expect N(p-\amm) to be determined with reasonable accuracy, and hence, we set the relative error cut in N(p-\amm) at 33\%, or 3$\sigma$.

\section{Results}
\label{sec_res}

\subsection{Integrated Intensity maps}
\label{sec_mom_map}

While calculating the integrated intensity of the region, it is better to use a narrow range of velocities around the velocity at each pixel, than using a broad, average velocity range for the entire cloud, since there is a gradient in velocity present among different parts in L1688. Therefore, we employ the procedure described in Section 3.2 of \citet{GASDR1} in deriving the moment maps. We use the best-fit model at each pixel, and select the spectral range for which the model shows an intensity $> 0.0125$ K \footnote{Although this threshold is below the noise level ($\approx$ 1/4th), it was chosen since it significantly recovered the flux in the model.}. In places where the fit was not satisfactory (based on fit-determined errors), we use the average value of \vel and \sig for the cloud. We then use the known relative positions of the hyperfines for the lines to define a range where any emission is expected. We then calculate the integrated emission (and the associated error) using only these channels. 

The integrated intensity maps thus obtained for the smoothed data are presented in Figure \ref{integ_inten}.

\subsection{Property Maps}
\label{sec_prop_map}

\begin{figure*}[!ht]
    \centering
    \includegraphics[width=0.76\textwidth]{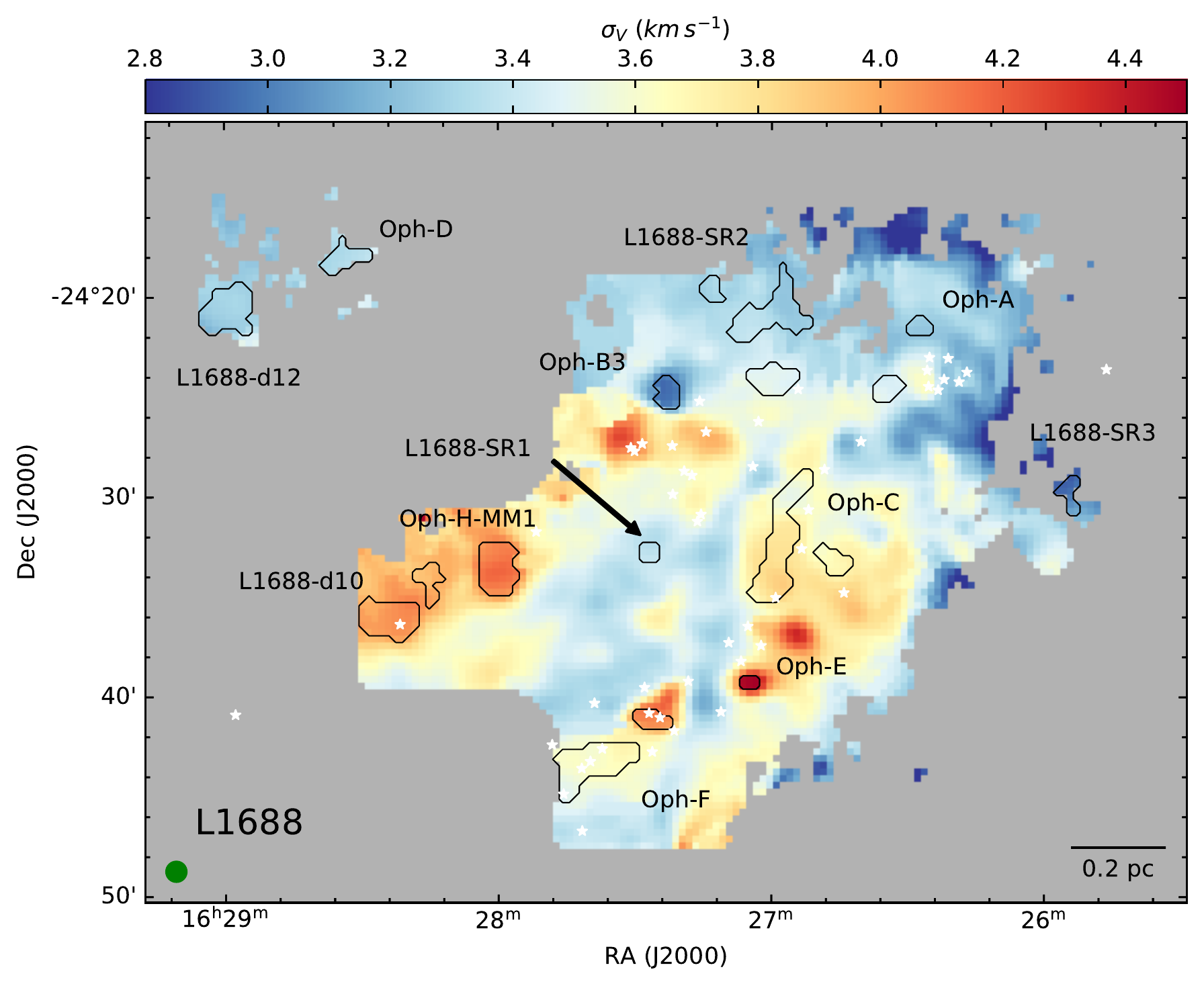}
    \includegraphics[width=0.76\textwidth]{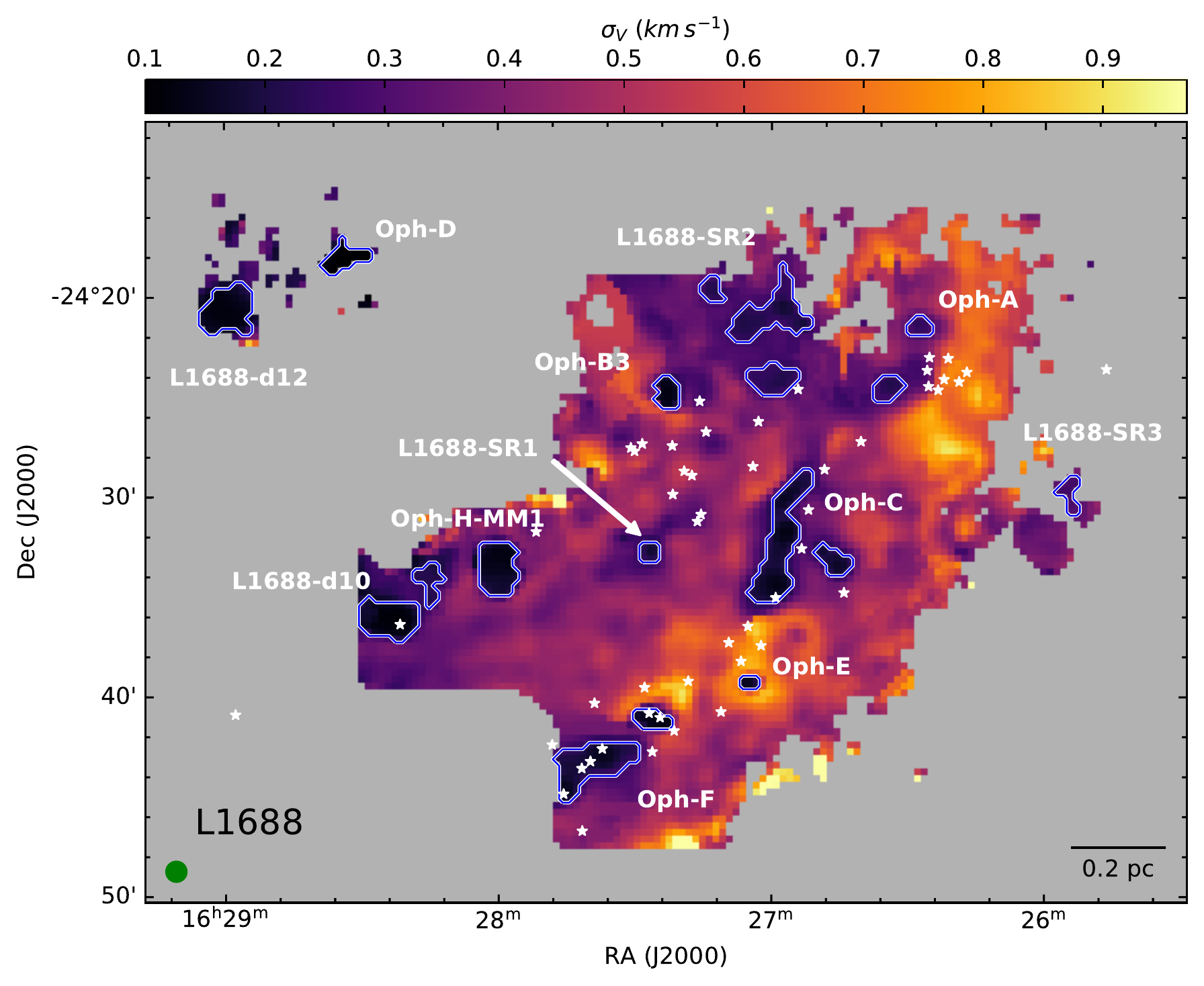}
    \caption{LSR velocity (top panel) and velocity dispersion (bottom panel) in L1688. 
    The solid black contours in top panel and solid blue contours in bottom panel show the coherent cores in the region (described later in Section \ref{sec_id_coh}).
    The black (top panel) and white (bottom panel) stars show the positions of Class 0/I and flat-spectrum protostars \citep{yso_l1688}. The beam is shown in green in the bottom left corner, and the scale bar is shown in the bottom right corner. \label{sigma-vlsr} }
    
\end{figure*}

\begin{figure*}[!ht]  
\centering
\includegraphics[width=0.76\textwidth]{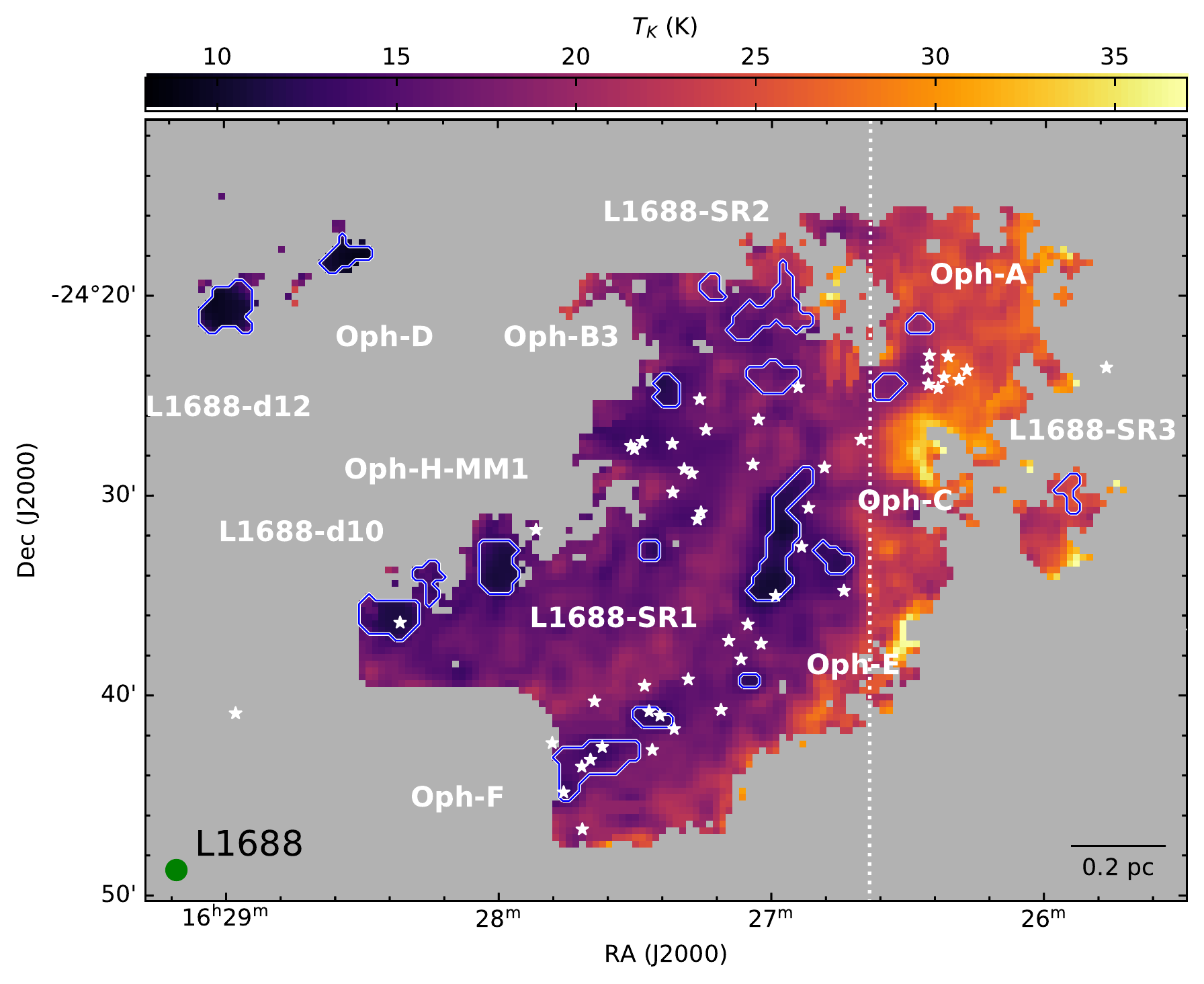}
  \caption{Kinetic temperature in L1688. 
  The solid blue contours show the coherent cores in the region (described later in Section \ref{sec_id_coh}).
  The white stars show the positions of Class 0/I and flat-spectrum protostars. To see the difference between the part of the cloud affected by the external radiation, and the dark cloud further away from the illumination source, we consider a vertical boundary, as shown. The beam is shown in green in the bottom left corner, and the scale bar is shown in the bottom right corner. \label{tk}}
\end{figure*}

\begin{figure*}[!ht]  
\centering
\includegraphics[width=0.76\textwidth]{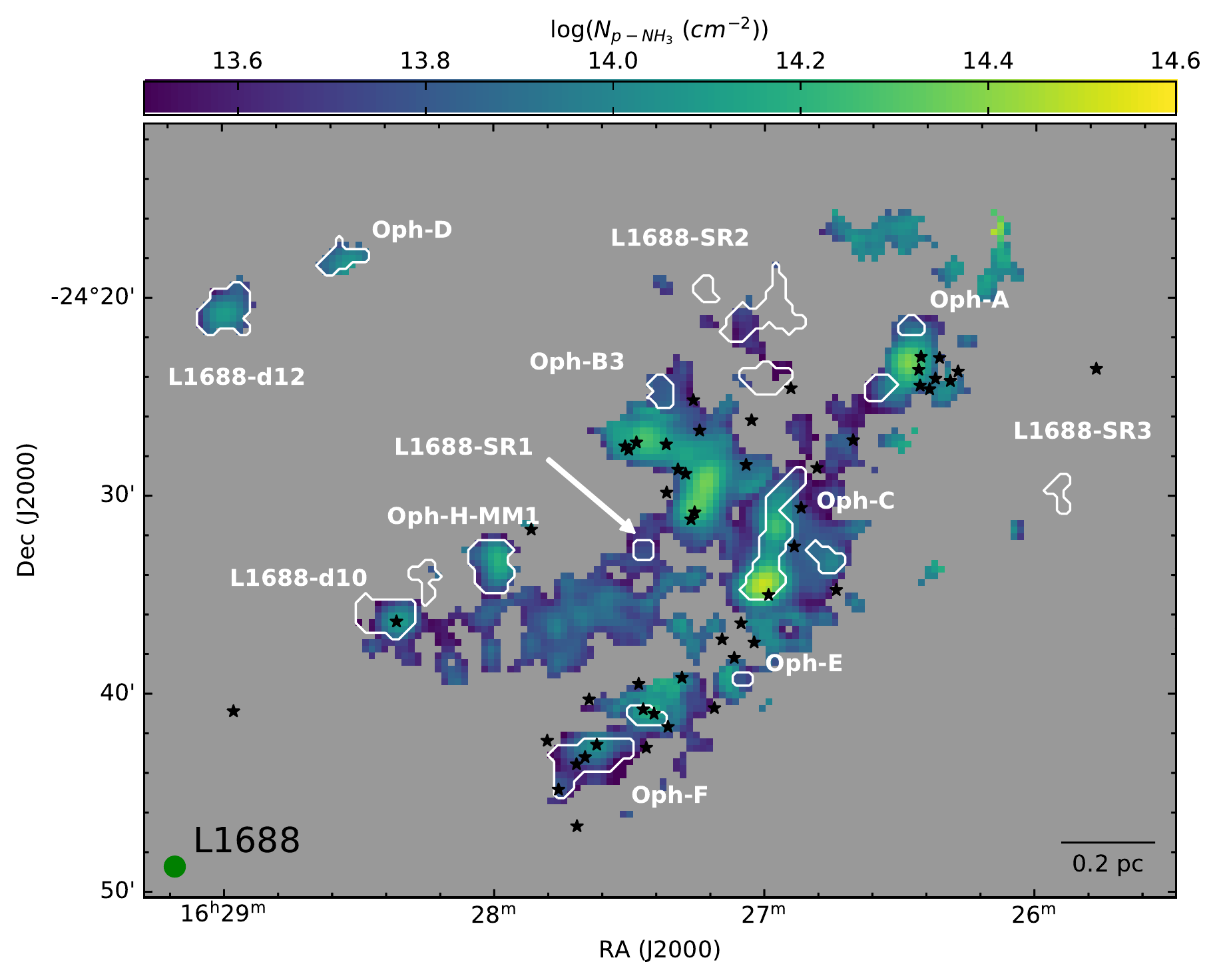}  
  \caption{p-\amm column density in L1688. 
  The solid white contours show the coherent cores in the region (described later in Section \ref{sec_id_coh}).
  The black stars show the positions of Class 0/I and flat-spectrum protostars. The beam is shown in green in the bottom left corner, and the scale bar is shown in the bottom right corner. \label{amm_col}}
\end{figure*}

In Figures \ref{sigma-vlsr}-\ref{amm_col}, we show the velocity, velocity dispersion, \amm kinetic temperature and p-\amm column density maps for L1688, obtained from our fits. We obtain much more extended maps for velocity dispersion and kinetic temperature (73.4\% and 65\% of the total observed area, respectively), compared to the original GAS DR1 maps \citep[Figures 13 and 17 in][19.2\% and 13\% of the total observed area respectively]{GASDR1}.  Also, for the first time, we are able to map the temperature in the dense cores, as well as the surrounding gas in the same density tracer.

In the less dense, outer region of the cloud, the \amm (1,1) line becomes optically thin, and so, the excitation temperature is harder to be satisfactorily determined\footnote{We observe that typically for $\tau$ {(1,1)} < 0.12, it is difficult to constrain \tex and therefore the ammonia column densities.}. 
Correspondingly, the p-\amm column density (Figure \ref{amm_col}) is also not well-determined away from some of the continuum cores \citep[identified in][]{motte1998}.

\subsection{Variation of temperature and velocity dispersion throughout the cloud}
\label{sec_tk_sig}

Figure \ref{tk} shows the kinetic temperature in L1688. \citet{motte1998} identify 7 dense cores, Oph-A, Oph-B1 and Oph-B2, Oph-C, Oph-D, Oph-E and Oph-F, from continuum observations in the cloud (these are shown in Figure \ref{h2_col}). It can be clearly seen from Figure \ref{tk}, that these continuum cores are at a lower temperature than the gas surrounding them. On average, the kinetic temperature in the N(\htw) peak inside the continuum cores is $\approx$ 3.7 K lower than the mean ambient cloud temperature (calculated in the area shown by the dashed rectangle in Figure \ref{h2_col}).

To the west of the cloud, there is a strong radiation field due to a young, massive B2-type star \citep{hd_spec}, HD147889 \citep[RA:16h25m24.31s, Dec:-24\degr$27\arcmin 56.57\arcsec$;][]{hd_coord}. The radiation field near the western edge is estimated to be $\sim 400\ G_0$ \footnote{Definition of $G_0$ from \citet{hollen_tielen_1999}: $G_0$ represents the unit of an average interstellar flux of 1.6 erg cm$^{-2}$ s$^{-1}$ in the energy range 6-13.6 eV \citep{habing1968}.}\citep{habart2003}.
Due to this external illumination, the kinetic temperature is $\approx$ 7.8 K higher by in the western edge of the cloud (considered as the region to the right of the vertical boundary shown in Figure \ref{tk}), compared to the mean cloud temperature. This increase in temperature can also be seen in the dust temperature map from $\emph{Herschel}$ (Figure \ref{dust_temp}).

In the velocity dispersion map in Figure \ref{sigma-vlsr}, we see that the continuum cores identified in \citet{motte1998} (shown in Figure \ref{h2_col}), except Oph-B1 and Oph-B2, show less turbulent gas, compared to the surrounding gas. The dispersion, averaged over one beam around the continuum peaks of Oph-C, Oph-D, Oph-E and Oph-F is $\leq$ 0.2 \kms, and around the continuum peak of Oph-A is $\approx$ 0.3 \kms. 
Oph-B1 and Oph-B2 contain a cluster of Class 0/I and flat spectrum protostars, similar to Oph-F in type, number and luminosity. However, Oph-F show gas in low dispersion, whereas the \sig observed in Oph-B is significantly higher($\approx$ 0.45 and 0.54 \kms, averaged over one beam around the continuum peaks of Oph-B1 and Oph-B2, respectively). A strong outflow is observed in the vicinity of Oph-B1 and Oph-B2 \citep{kamakazi2003,white2015}, and no such outflow is seen near Oph-F. Therefore, the higher \sig observed in Oph-B might be associated with presence of protostellar feedback.

\begin{figure*}[]
\centering
\includegraphics[width=0.5\textwidth]{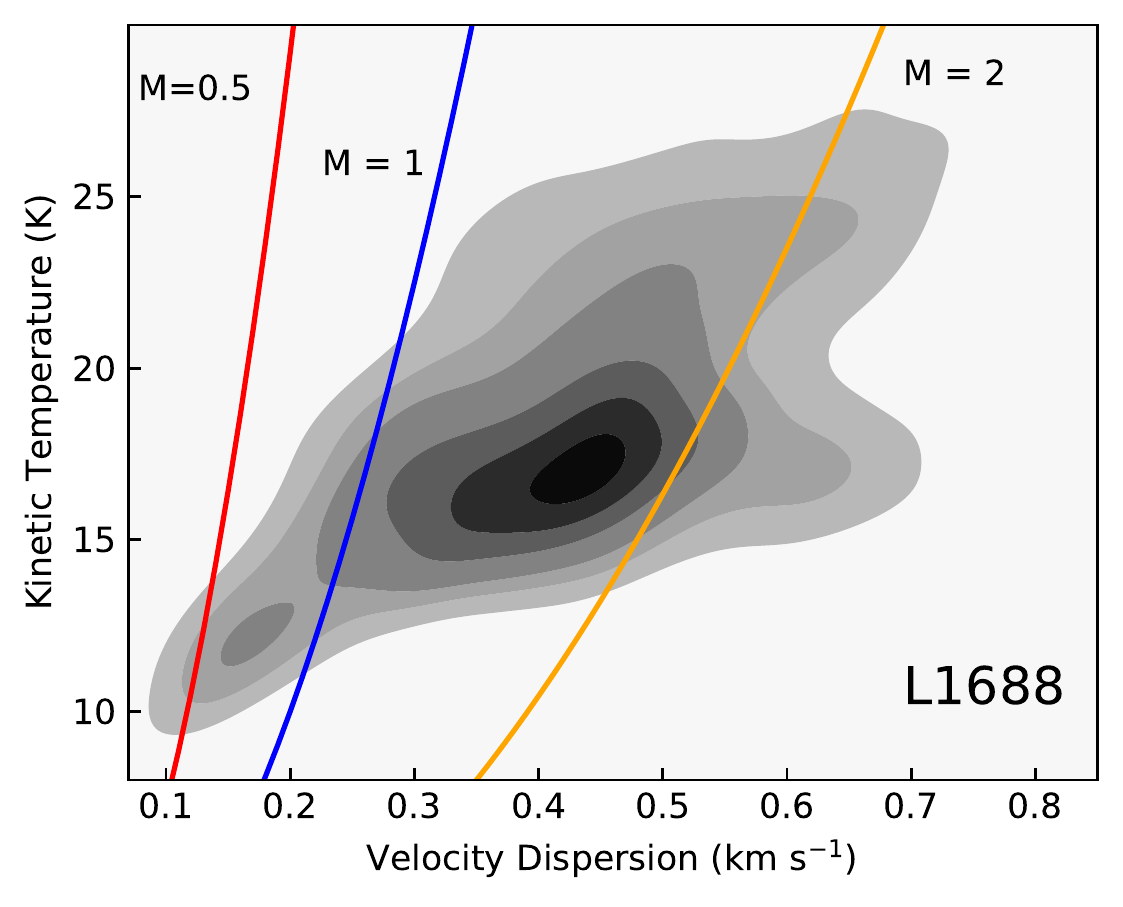}
\caption{
Kinetic temperature and velocity dispersion in L1688, shown here as a normalised kernel density distribution. The contour levels are 0.1,0.2,0.3,0.5,0.7 and 0.9. The red, blue and orange lines represent Mach number of 0.5, 1 and 2, respectively.}
\label{tk-sigma}
\end{figure*}

Figure \ref{tk-sigma} shows how the kinetic temperature of the gas changes with respect to velocity dispersion in L1688. Due to the large number of data points, we show a density distribution plot, instead of the actual data-set, to represent the distribution more clearly. Here, we use a Gaussian Kernel Density Estimator (KDE) from the package \verb+scipy+. KDE replaces each data point with a Gaussian of a constant width, and the sum of the individual Gaussians is used as an estimate for the density distribution of the data. The width of the Gaussian is determined using Scott's rule \citep{scott_1992_kde}, and depends on the number of data points.

\subsection{Identification of coherent cores }
\label{sec_id_coh}

In order to study the change in kinetic temperature and velocity dispersion from cores to the surrounding material, we first identify what we consider as ``coherent cores''. For this purpose, we use the sonic Mach number, $\mathcal{M_S}$, which is defined as the ratio of the non-thermal velocity dispersion to the sound speed in the medium :
\begin{equation}
    \mathcal{M_S} = \frac{\sigma_{\rm v,NT}}{c_S}~,
\end{equation}
where $c_S$ is the one-dimensional sound speed in the gas, given by :
\begin{equation}
    c_S = \sqrt{\frac{k_B \tk}{\mu_{\rm gas}}}~,
\end{equation}
where $\rm{k_B}$ is the Boltzmann's constant, \tk is the kinetic temperature in the region, and ${\mu_{\rm gas}}$ is the average molecular mass \citep[taken to be 2.37 amu, ][]{jens_2008_mu}. 
The non-thermal component of the velocity dispersion is calculated by removing the thermal dispersion for the observed molecule (${\sigma_{\rm T}}$) from the total observed velocity dispersion (\sig) :
\begin{equation}
    {\sigma_{\textrm{v,NT}}}^2 = {\sig}^2 - {\sigma_{\textrm{T,\amm}}}^2 -{\sigma_{\textrm{chan}}}^2~,
\end{equation}
where $\sigma_{\rm chan}$ is the contribution due to the width of the channel. The thermal component of the dispersion observed in \amm is: 
\begin{equation}
    \sigma_{T,\amm} = \sqrt{\frac{k_B T_K}{\mu_{\amm}}}~,
\end{equation}
with $\rm{\mu_{\amm}}$ = 17 amu, being the mass of the ammonia molecule. For \tk = 14\,K, the thermal component is $\sigma_{\rm T,\amm}$ is 0.082 \kms.

When a Hann-like kernel is used in the spectrometer, then the correction factor for the measured velocity dispersion is given by \citep[see][]{leroy_2016_chan-width, koch_2018_chan-width} :
\begin{equation}
    \sigma_{\rm chan} = \frac{\Delta}{\sqrt{2 \pi}} (1.0 + 1.18k + 10.4 k^2)~,
\end{equation}
where $\Delta$ is the spectral resolution and $k$ is dependent on the Hann-like function applied.
In the case of the VEGAS spectrometer used in the GAS observations, $k$ is 0.11, which corresponds to a correction term of:
\begin{equation}
    \sigma_{chan} = \frac{\Delta}{1.994} = 0.036\, \kms ~.
\end{equation}
After applying this small correction, we obtain a typical non-thermal component of the velocity dispersion of $\sigma_{\rm v,NT}=0.182$ \kms, for \sig = 0.2 \kms.

In Figure \ref{tk-sigma}, it can be observed that a group of points lies below the line depicting a Mach number of 1, and a much larger distribution lies in the supersonic part of the plot. This separation is taken as the basis for separating the subsonic cores from the extended, more turbulent cloud. We first identify a mask with the regions in L1688 having a sonic Mach number of 1 or lower  \citep[similar to][]{pineda2010,chen2019}. 
Regions smaller than the beam size are removed from the mask. 
All the remaining regions are then considered as the ``coherent cores'' in L1688.

Based on this definition, we identify 12 regions in L1688 as coherent cores \citepalias[same as][]{choudhury2020_letter}. These include:
\begin{itemize}
    \item Five continuum cores, Oph-A, Oph-C, Oph-D, Oph-E and Oph-F. We identify two islands in Oph-A as coherent cores (Oph-A North, and Oph-A South), and consider them together as Oph-A.
    \item One DCO$^+$ core, Oph-B3 \citep{loren_1990_b3, friesen_2009_b3}.
    \item Two `coherent droplets', L1688-d10 and L1688-d12, \citep{chen2019}. 
    \item One prestellar core Oph-H-MM1 \citep{johnstone_2004_hmm1}.
\end{itemize} 

\begin{figure*}[!ht]  
\centering

\includegraphics[width=0.85\textwidth]{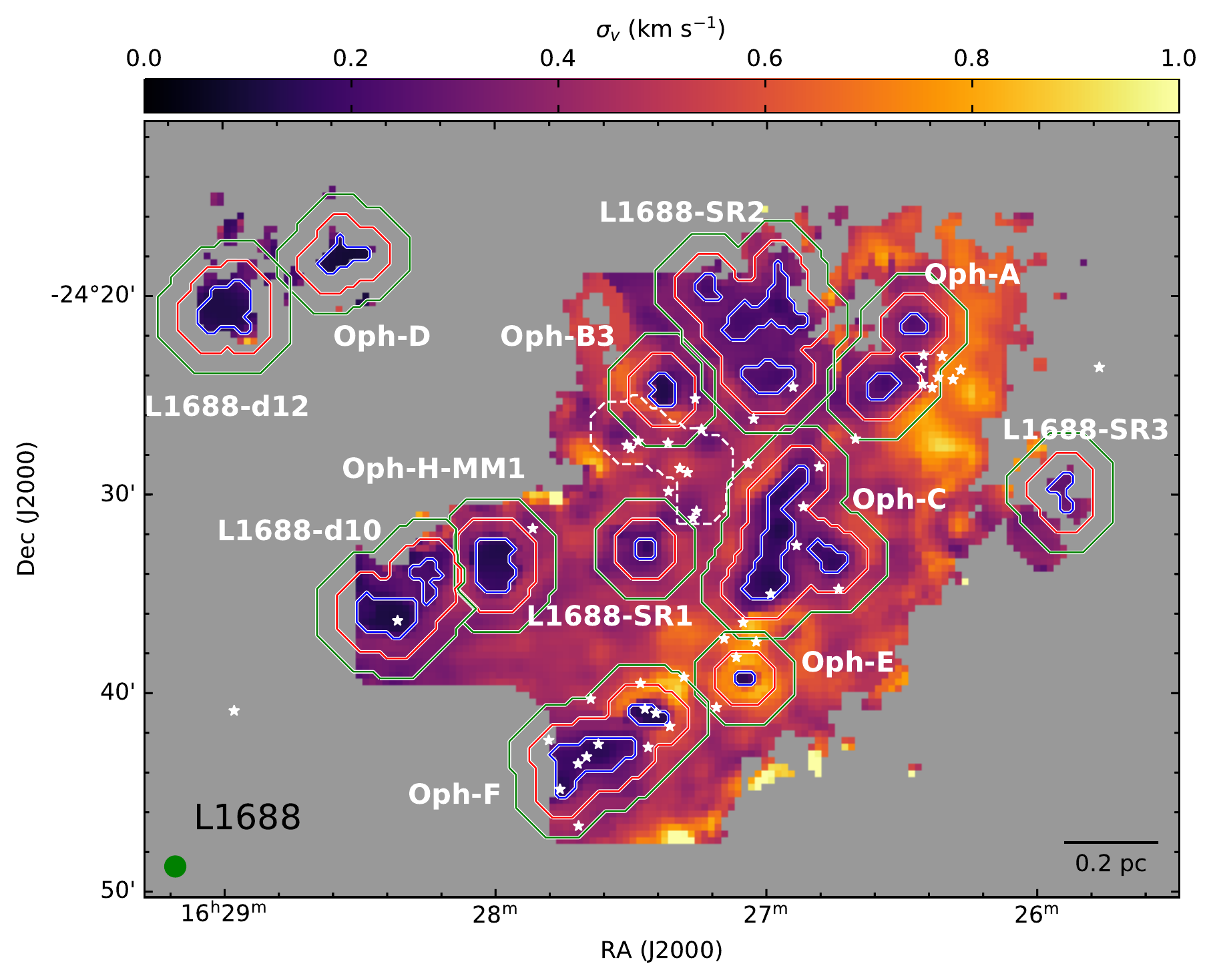}
\caption{The coherent cores and their immediate neighbourhood, as defined in Section \ref{sec_id_coh}, are shown in the velocity dispersion map. 
The coherent cores, shell-1, and shell-2 regions are shown with blue, red and green contours, respectively.
The white dashed contour shows the boundary considered for continuum cores Oph-B1 and Oph-B2 (See Section \ref{sec_id_coh} for details).
The white stars show the positions of Class 0/I and flat-spectrum protostars in the cloud. The beam and the scale bar are shown in the bottom left and bottom right corners, respectively.} 
\label{coh_cores}
\end{figure*}

Apart from these previously identified cores\footnote{Oph-B3, Oph-D and H-MM1 are also identified as L1688-c4, L1688-d11 and L1688-d9, respectively, in \citet{chen2019}. The two structures that we identify as Oph-A, are likely associated with Oph-A-N2 and Oph-A-N6, as identified in N$_2$H$^+$ by \citet{di_fran_2004_ophA_n}.}, we also identified three more subsonic regions :  L1688-SR1 (south of Oph-B1), L1688-SR2 (three islands east of Oph-A) and L1688-SR3 (near the western edge of the map). These cores are shown in Figure \ref{coh_cores}, along with the positions of Class 0/I and flat spectrum protostars in L1688 \citep{yso_l1688}. Following the classification of Class-0 and -I protostars in the region according to \citet{enoch_2009_str0-1}, we find no significant difference between the positions of Class 0/I protostars and those of flat spectrum protostars, with respect to the coherent cores. The continuum cores Oph-B1 and Oph-B2 are not subsonic, as previously discussed \citep[Section \ref{sec_tk_sig}, see also :][]{friesen_2009_b3}, and therefore, they are not considered as `coherent cores' by our definition. 

Due to using a larger beam (almost twice in size) compared to the GAS data, our definitions of the coherent boundaries are expected to vary slightly compared to other analyses done with the original data \citep[e.g.,][which also applied more stringent criteria\footnote{\citet{chen2019} had further requirements for the coherent regions to be identified as a `droplet' (such as associated local maxima in peak \amm (1,1) intensity and N(\htw), and relatively smooth \vel distribution inside), whereas we consider the sonic Mach number as the only criterion.} in their definition]{chen2019}. However, comparison with \citet{chen2019} shows that the overall coherent boundaries in the two works are consistent within one beam ($1\arcmin$).

\citet{ladje_2020_hgbs} identifies a prestellar core at the position of SR1, and an unbound starless core at the position of SR3. SR1 and SR3 are associated with a local peak in \amm integrated intensity. SR2 does not contain any bound core in $\emph{Herschel}$ data, nor does it show a local \amm peak. However, as can be seen from our later results (presented in Section \ref{sec_sec_comp}), the ambient cloud in the line of sight to the cores has a higher temperature compared to the gas in the core. However, in the computation of N(\htw) from $\emph{Herschel}$ data, a single temperature was assumed for the entire line of sight. Therefore, it is possible that small clumps of over-densities at the position of coherent cores, which have a temperature much lower than that of the ambient cloud, will not be seen in the N(\htw) maps. Without allowing for different temperatures in different parts of the cloud along the line of sight in the calculations of N(\htw), we cannot dismiss the possibility of an over-density being associated with SR2.

In order to study the environment surrounding the coherent cores, we define two shells around each core. Shell-1 is defined as all the pixels around each coherent core, within a distance of one beam (the smallest resolved scale) from the boundary. Shell-2 is defined as all the pixels around shell-1 within one beam. This allows us to study the environment surrounding the cores in the two consecutive layers. The regions defined\footnote{For most cases, the shells of different cores have little to no overlap. For L1688-d10 and H-MM1, there would have been some overlap with the shell definitions, but we avoid this overlap by restricting the boundaries of respective shells-2, as can be seen in Figure \ref{coh_cores}} are displayed in Figure \ref{coh_cores}.

As mentioned earlier, Oph-B1 and Oph-B2 are not coherent cores by our definition, while being continuum cores. Being close to these two cores, the shells of Oph-B3 and L1688-SR1 might contain some high column density gas with supersonic velocity dispersions, produced by protostellar feedback, which is
not representative of the cloud surrounding a coherent core. Therefore, we define a boundary roughly containing Oph-B1 and Oph-B2, based on column density (N(\htw)>2.1$\times 10^{22}\,$cm$^{-2}$, shown in white dashed contours in Figure \ref{coh_cores}), and remove the pixels inside this boundary from the shells of Oph-B3 and L1688-SR1, for all subsequent analyses.

\section{Discussion}
\label{sec_diss}

\subsection{Transition from coherent cores to their immediate neighbourhood: A distribution analysis}
\label{sec_transi_dist}

\begin{figure*}[!h]
\centering
\includegraphics[width=0.45\textwidth]{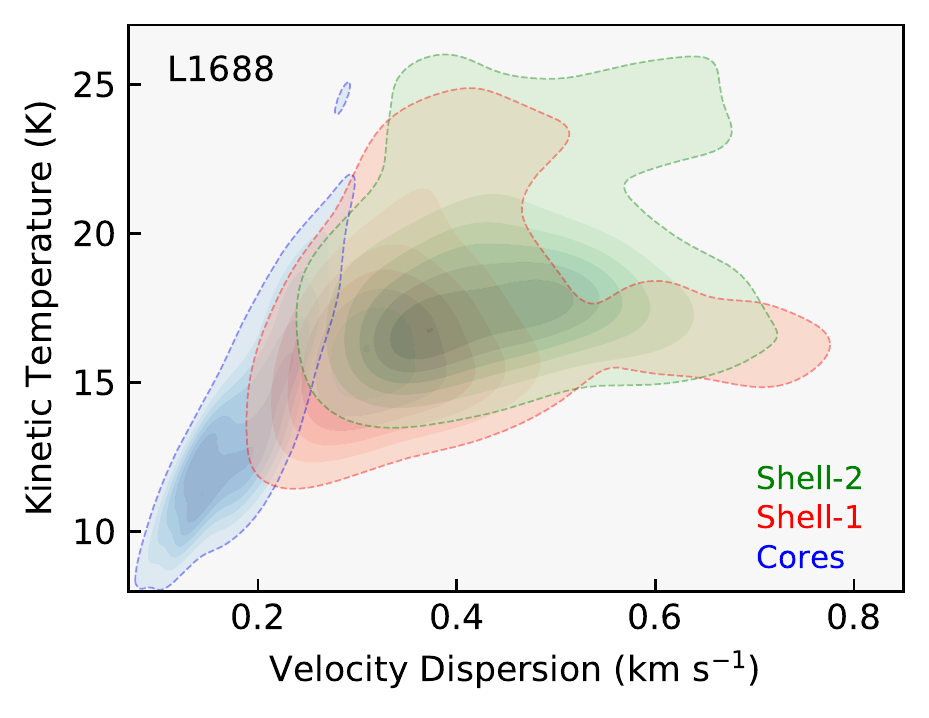}
\includegraphics[width=0.45\textwidth]{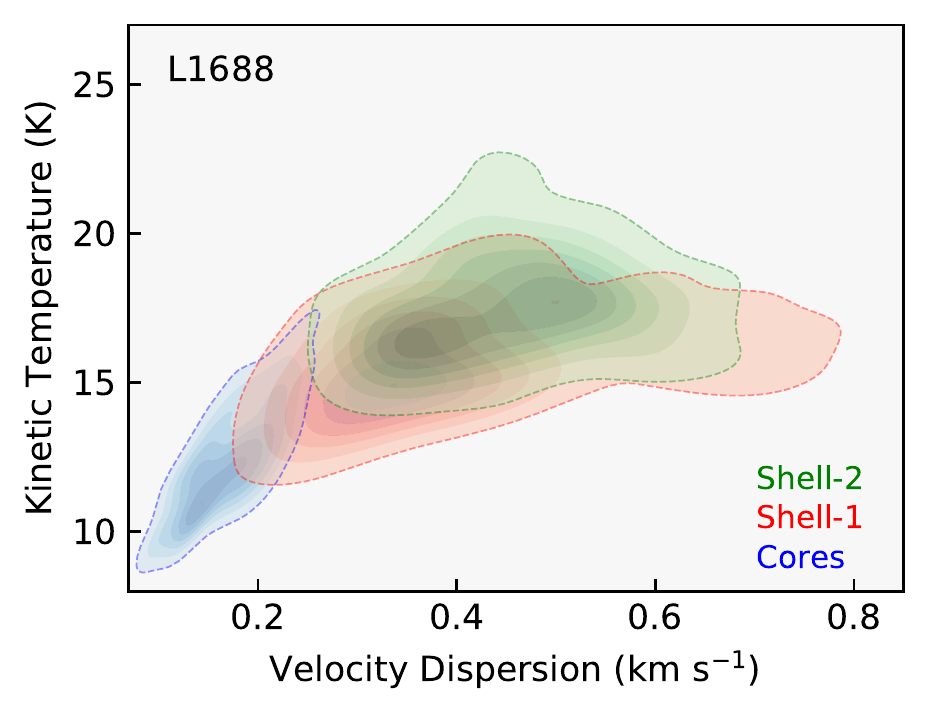}
\caption{Left : Distribution of kinetic temperature and velocity dispersion, for all the coherent cores and the shell-1 and shell-2 regions. 
Shells-1 and shells-2 are defined as two shells of width equal to one beam around the respective coherent cores.
Each kernel density distribution was normalised to have a peak density of 1. The contours show normalised KDE levels of 0.1, 0.25, 0.4, 0.55, 0.7 and 0.85. 
Right : Same as the left panel, but without the regions L1688-SR3, L1688-SR2 and Oph-A, and ignoring one beam at each star position. This is done to remove the effect of external heating, and possible contribution from prostellar feedback.}
\label{tk-sigma_coh_core}
\end{figure*}

We study the change in kinetic temperature and dispersion, observed between the coherent cores to their  surrounding environment (shell-1 and shell-2). 
The left panel of Figure \ref{tk-sigma_coh_core} shows the kernel density distribution of kinetic temperature with velocity dispersion for all the cores, and corresponding shell-1 and shell-2. 
The coherent cores are characterised by a kinetic temperature  3-7 K lower than shell-1 and shell-2. The velocity dispersion in the cores is also 0.2-0.4 \kms lower than in the shells.
Compared to the change in \tk and \sig from core to shell-1, the values of the two parameters in shell-1 and shell-2 do not change as much. The kinetic temperatures of the two shells are within $\approx$2 K of each other, and the difference in dispersion is $\leq$ 0.1 \kms.

The transition from the coherent cores to the surrounding molecular cloud in kinetic temperature is gradual, and there is no sudden jump, indicating that it changes smoothly from core to cloud at the angular resolution of the present work. By our definition of the cores, the transition in velocity dispersion at the core boundaries is sharp.
The average kinetic temperature and velocity dispersion for the coherent cores, are 12.107$\pm$0.009 K and 0.1495$\pm$0.0001\ \kms, respectively\footnote{All the averages reported in this section are weighted averages, with associated error on the weighted average.} with a standard deviation of 3.2$\,$K and 0.047 \kms. Similarly, for the first shell, the average values are $\tk$=15.80$\pm$0.01\ K (std. dev. = 3.12 K) and $\sig$=0.3175$\pm$0.0002\ \kms (std. dev. = 0.127 \kms); and for the second shell, $\tk$=18.11$\pm$0.02\ K (std. dev. = 3.5 K) and $\sig$=0.4176$\pm$0.0004\ \kms (std. dev. = 0.127 \kms).
We selected three roundish cores: Oph-E, Oph-H-MM1 and L1688-d12, and compared the radial profiles of velocity dispersion with those reported in \citet{chen2019}. We observed that \sig inside these coherent cores lie inside the 1$\sigma$ distribution of pixels in each distance bin equal to the beam size for the corresponding droplet in \citealt{chen2019}. Furthermore, we also observed that the radial profiles showed similar shape as compared to those in \citet{chen2019}, up to shell-2. Therefore, we see the similar radial profiles compared to \citet{chen2019}, even though we are using a larger beam.
On average, we see a kinetic temperature difference of $\approx$ 4 K between core and shell-2, i.e., $\approx$ 2\arcmin\ from core. \citet{harju_2017} observed a similar increase in kinetic temperature, $\approx$ 5 K, 2\arcmin\ away from the centre of the starless core H-MM1 \citep[see also][]{crapsi_2007_temp_drop_core}. This is consistent with the temperature structure of externally heated dense cores \citep[e.g.][]{evans_2001_core-temp,zucconi_2001_core-temp}. 

Assuming the gas to be completely molecular, so that $N_{\rm H} = 2 \times N(\htw)$, and the relation between hydrogen column density and optical extinction to be 
$N_{\rm H}\ ({\rm cm}^{-2}) \approx 2.21 \times 10^{21}\ {A_V}\ ({\rm mag})$ \citep{guever2009}, we find that the average extinction through the cores is $\approx$16 mag. Through shell-1 and shell-2, the average extinctions are ${A_V}\approx$ 13.16 and ${A_V}\approx$ 11.2 mag, respectively. Since we do not have the p-\amm column density in the extended cloud (away from the cores), we fit the average spectra (see Section \ref{sec_transi_avg}) of each core, and their shells, to get an idea of the average N(p-\amm) in the regions. From that analysis, we find that the average p-\amm column density in the cores is $\rm \approx (8.99 \pm 0.08) \times 10^{13} cm^{-2}$; and that in shell-1 and shell-2 are $\rm (5.2 \pm 0.1) \times 10^{13} cm^{-2}$ and $\rm (3.7 \pm 0.1) \times 10^{13} cm^{-2}$, respectively.

As mentioned in Section \ref{sec_tk_sig}, the west part of L1688 is affected by a strong external radiation field, the effect of which can be clearly seen in the kinetic temperature map (Figure \ref{tk}). 
The regions Oph-A, L1688-SR3 and L1688-SR2 lie in the affected region of the cloud. To get a clearer view of the behaviour of kinetic temperature and velocity dispersion in the embedded cores (without the effect of the outside illumination), we omit the regions L1688-SR3, L1688-SR2 and Oph-A, as well their neighbourhoods. 
In order to remove any possible contribution from prostellar feedback in the regions, we also mask one beam at the positions of known protostars \citep{yso_l1688}. 
The distribution of the remaining  cores and the shells is shown in the right panel of Figure \ref{tk-sigma_coh_core}. When the effect of external illumination, as well as possible contributions from protostellar feedback, is masked, the average kinetic temperature in the cores drops (11.66$\pm$0.01\ K, std. dev. = 1.71 K). The change in \sig is not as stark, the average for the new distribution being 0.1487$\pm$0.0001\ \kms (std. dev. = 0.038 \kms). Similarly, masking the heating by the external radiation removes the high kinetic temperature region in shells 1 and 2, reducing the average temperature to 14.68$\pm$0.02\ K (std. dev. = 1.9 K) and 17.37$\pm$0.02\ K (std. dev. = 2.1 K), respectively. By comparison, the spread in \sig remains almost constant, and its average value is unchanged (within errors). Therefore, our results suggest that the external irradiation is not accompanied by turbulence injection in the neighbourhood of the cores.

\subsection{Transition from coherent cores to their immediate neighbourhood: Analysis of Individual Cores}
\label{sec_transi_avg}

\begin{figure*}[!h]
\centering
\includegraphics[width=0.45\textwidth]{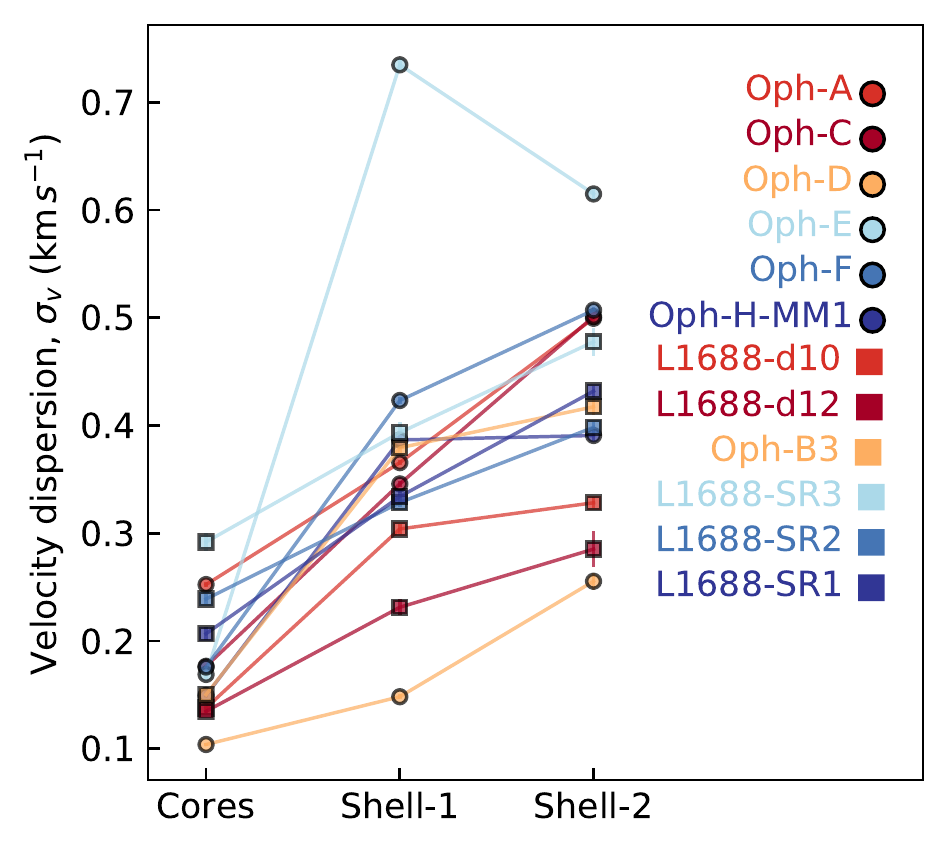}
\includegraphics[width=0.45\textwidth]{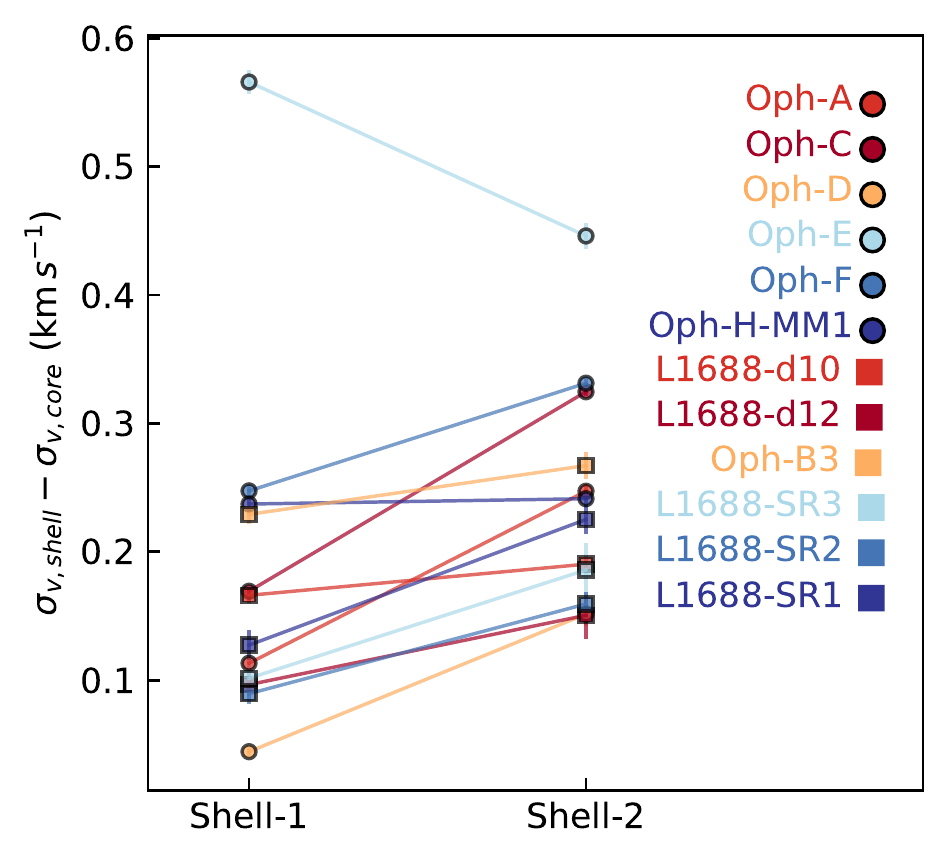}
\caption{
Left panel: velocity dispersion in core and the shells, determined from average spectra in the respective core/shell. 
Shells-1 and shells-2 are defined as two shells of width equal to one beam around the respective coherent cores.
Right panel : velocity dispersion in the shells relative to their respective cores.}
\label{sig_avg_spec}
\end{figure*}

\begin{figure*}[!h]
\centering
\includegraphics[width=0.45\textwidth]{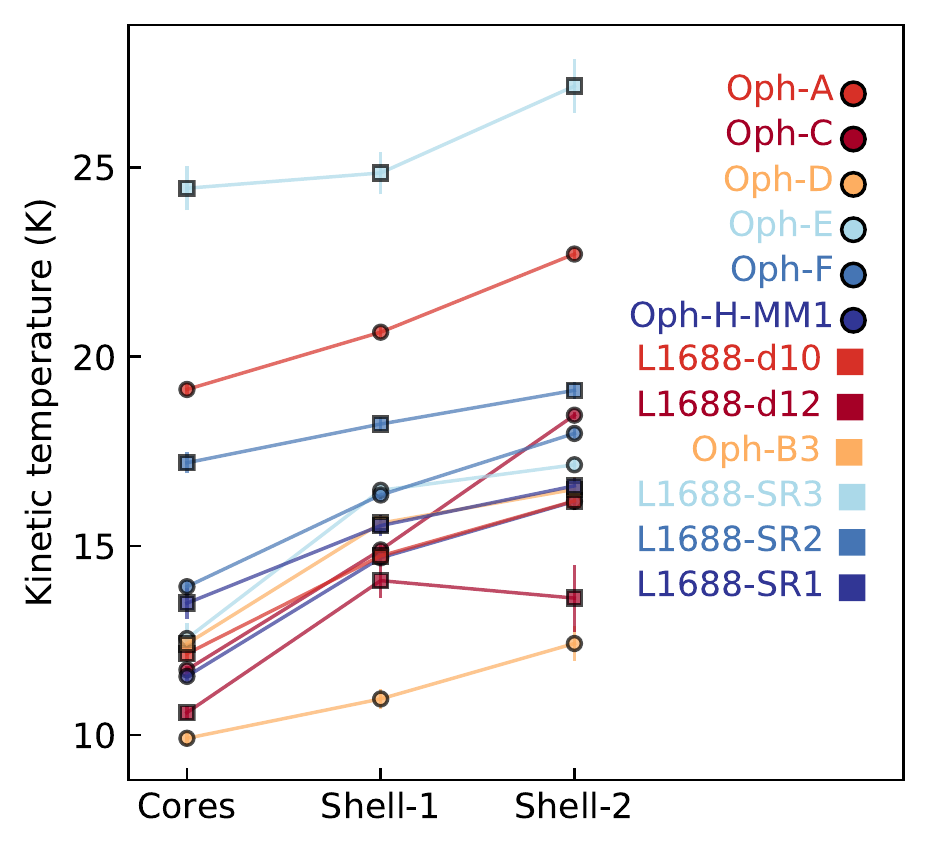}
\includegraphics[width=0.45\textwidth]{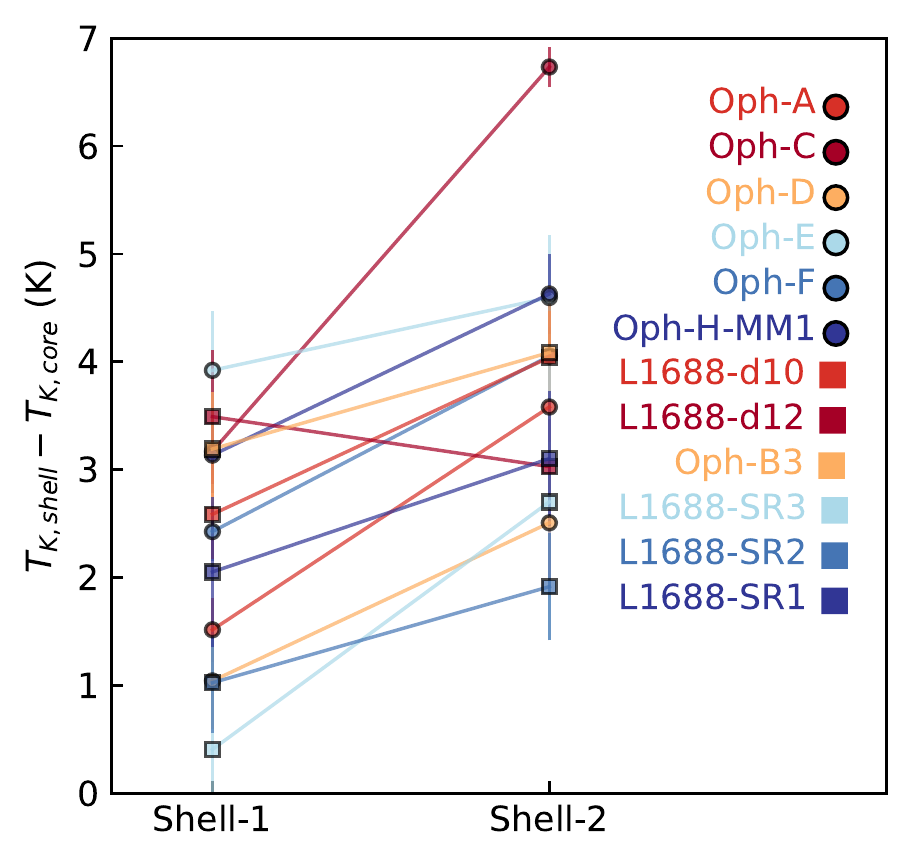}
\caption{
Left panel: kinetic temperature in core and the shells, determined from average spectra in the respective core/shell.
Shells-1 and shells-2 are defined as two shells of width equal to one beam around the respective coherent cores.
Right panel : kinetic temperature in the shells relative to their respective cores.}
\label{tk_avg_spec}
\end{figure*}

For the cores L1688-SR3, Oph-D and d12, we have the kinetic temperature information for very few points in the shells 1 and 2. The individual pixels in these shells do not have sufficient SNR for a good fit, and therefore, a direct determination of the kinetic temperature in the individual pixels is not possible. Moreover, \tk could not be determined in some pixels in the outer shells of d10, L1688-SR2, Oph-A and Oph-F, as well. Therefore, to get a fair comparison of the temperatures in the cores, shells-1 and shells-2, we average the spectra in each of these regions. Stacking the spectra for a large number of points results in significantly reduced noise levels, and we have sufficient SNR to be able to fit these spectra and obtain the kinetic temperature for each region. With the low noise, we are also able to look at minor details in the spectra towards each core and shells.
To avoid any possible line broadening due to averaging in a region with velocity gradients, we align the spectra at each pixel within a region before averaging. For this, we take the velocity at a pixel, determined from the single-component fit at that pixel, and using the \texttt{channelShift} function from module \texttt{gridregion} in the GAS pipeline\footnote{https://github.com/GBTAmmoniaSurvey/GAS/tree/master/GAS}, we shift the spectra at that pixel by the corresponding number of channels. Then, we average the resultant spectra from all pixels inside a region, now essentially aligned at v=0.

Figures \ref{sig_avg_spec} and \ref{tk_avg_spec} show the velocity dispersion and kinetic temperature, respectively, at each individual core and their respective neighbourhoods, determined from single-component fits to the averaged spectra in those regions. This gives us an idea of the change in the parameters as we move from core to shell-1, and then to shell-2. The left panel of Figure \ref{sig_avg_spec} shows the velocity dispersions of each individual core, shell-1 and shell-2. The right panel shows the \sig of the shells, relative to their corresponding cores. Similarly in Figure \ref{tk_avg_spec}, for kinetic temperature.

Figure \ref{sig_avg_spec} shows that, as expected, the velocity dispersion increases from core to shell-1, for all cores. For most cores, the dispersion steadily increases outwards to shell-2. Oph-E is an exception to this, where \sig decreases slightly in shell-2. 
It is clearly seen that for all the cores, the kinetic temperature steadily increases from the cores to shells 1 and 2. For d12, we see a slight drop in temperature from shell-1 to shell-2, but the difference is within the error margin, and therefore, not significant. It can be again seen that the highest kinetic temperatures for cores, shell-1 and shell-2 are for the regions affected by the outside illumination (Oph-A, L1688-SR2 and L1688-SR3). The temperature rise from shell-1 to shell-2 for Oph-C is very drastic, as the second shell includes part of the cloud heated by the external radiation (see Figure \ref{cores_on_tk} for reference).

We estimate the volume densities in the core, shell-1 and shell-2, with the \tex, \tk and $\tau$ measurements from the single-component fits, using the following relation described in \citet{foster_2009_nh2}:

\begin{equation}
    n = \frac{(J(\tex) - J(T_{cmb}) k J(\tk))}{h \nu_{(1,1)} (1-J(\tex))} \times n_{crit} \times \beta ~,
\end{equation}
where $T_{\rm cmb} = 2.73\,\rm K$, $n_{\rm crit} = 2 \times 10^4$, $k$ is the Boltzmann's constant, h is the Planck's constant, $\beta$ is the escape probability, estimated as $\beta = (1-e^{- \tau})/\tau$, and 

\begin{equation}
    J(T) = \frac{h \nu_{(1,1)}}{k (1 - e^{- \frac{h \nu_{(1,1)}}{k T}})}~.
\end{equation}

We find the cores to have a mean density of 1.66$\pm$0.09$\times$10$^5$ cm$^{-3}$. Shell-1 and shell-2 have similar densities, 
1.5$\pm$0.1$\times$10$^5$ cm$^{-3}$ and 1.4$\pm$0.2$\times$10$^5$ cm$^{-3}$, respectively\footnote{Non-weighted averages}. The densities in all of these three regions are more than an order of magnitude higher than the average density of the cloud traced by \amm, calculated using N(\htw) estimate from $\emph{Herschel}$ ($\sim 4 \times 10^3 \rm{cm}^{-3}$, see Section \ref{sec_amm_data}).
It should be noted, that density estimate from \citet{foster_2009_nh2} uses measurements with \amm, which traces a higher density than dust continuum. Therefore, by design, the \citet{foster_2009_nh2} estimate is sensitive mainly to higher density regions, whereas the $\emph{Herschel}$ continuum measurements refer to a significantly larger volume (as dust is much more extended than \amm).
Also, the $\emph{Herschel}$ column density maps are obtained assuming a single line of sight dust temperature, which might create discrepancies in the N(\htw) estimate, especially towards the cores, which are much colder than the surrounding gas. Therefore, a systematic difference in the two density estimates (using \amm, and from N(\htw) map) is expected.

\subsection{Comparison of kinetic temperature with dust temperature}
\label{sec_comp_tk_td}

\begin{figure*}[!h]  
\centering
\includegraphics[width=0.76\textwidth]{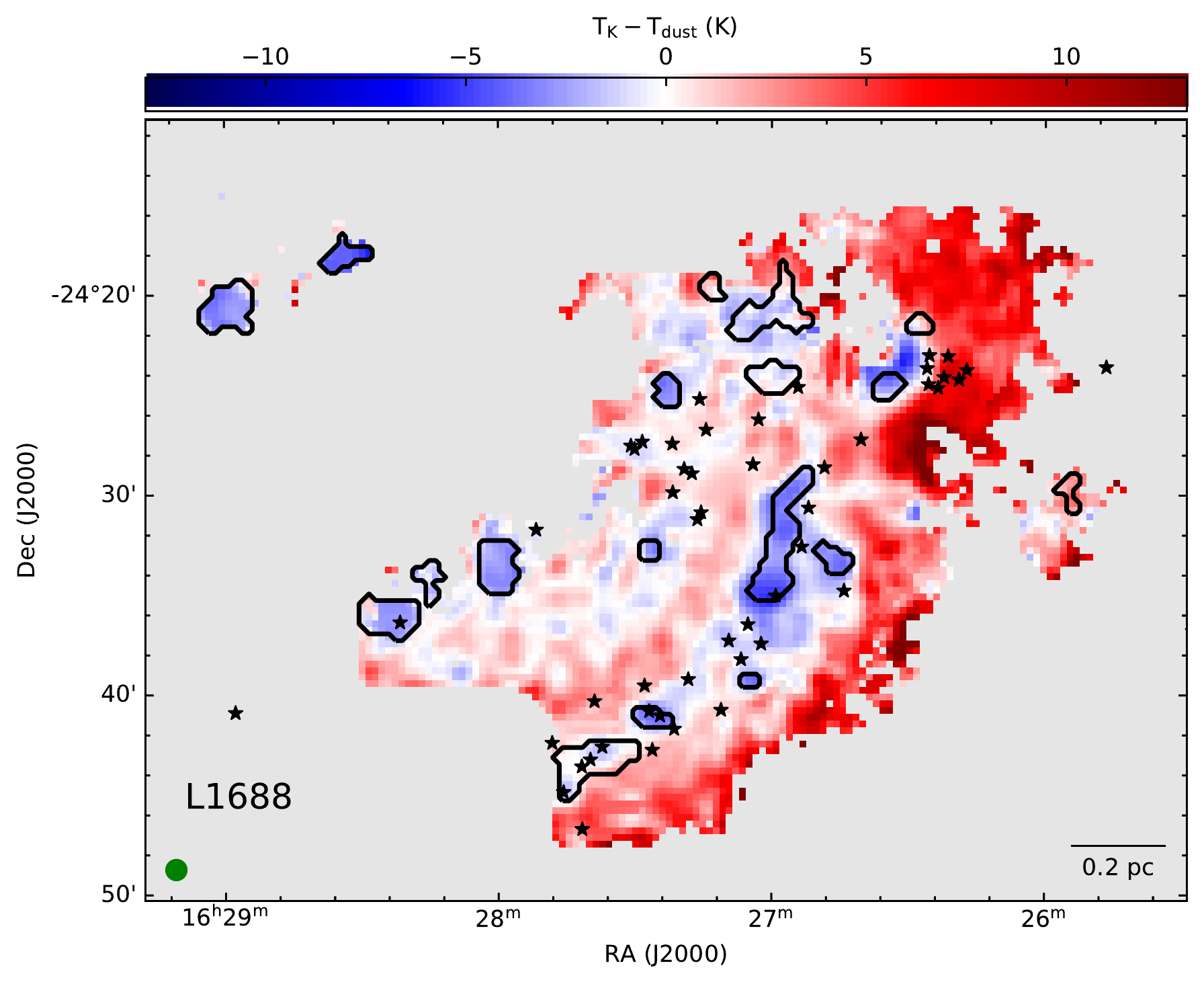}
\caption{
Map of the difference between kinetic temperature (from \amm) and dust temperature (from $\emph{Herschel}$). 
The black stars show the positions of known protostars in the region. 
The solid black contours show the coherent cores in the region (see Section \ref{sec_id_coh}).
The beam and the scale bar are shown in the bottom left and bottom right corners, respectively.}
\label{tk_dust_comp}
\end{figure*}

In Figure \ref{tk_dust_comp}, we show the map of the difference between the kinetic temperature derived using \amm and the dust temperature derived from $\emph{Herschel}$. The blue region in the map shows the gas with kinetic temperature less than dust temperature, and the red part shows the region in L1688 with kinetic temperature higher than dust temperature. Gas and dust are expected to be effectively coupled only at densities above $\rm{10^{4.5}} cm^{-3}$ \citep{goldsmith_2001_dust_gas_coup,galli_2002_dust_gas_coup}. 
The average density in L1688 is lower than $10^{4}$\,cm$^{-3}$ (Section \ref{sec_amm_data}), hence gas-dust coupling is not achieved in this cloud (except in the core regions and their neighbourhoods).

Although \amm emission in L1688 is extended, the comparison with the dust temperature indicates that \amm is not tracing the larger scale molecular cloud traced by the dust emission (or by low-density tracers such as CO). Instead, it is likely tracing the inner region of the cloud (at $A_v$ > 10 mag), where \amm is abundant and its inversion transitions can be excited. 
The coherent cores have a lower kinetic temperature than dust temperatures, even though their identification is done only using the kinematical information (\sig). 
This is expected, because the \amm emission towards the cores is dominated by the inner dense, cold core material; whereas the dust temperature is representative of the temperature of the foreground, outer part of the cloud.
On average, the kinetic temperature in the coherent cores is 1.8 K lower than dust temperature.
In the two shells surrounding the cores, the situation is reversed. In shell-1, the kinetic temperature is marginally higher ($\approx$ 0.3 K) than the dust temperature, and in shell-2, the difference is even stronger, \tk - \td $\approx$ 1.8 K.
The gas towards the western edge, which is affected by the strong external radiation (defined as the region to the right of the vertical dotted line shown in Figure \ref{dust_temp}), show ammonia at a temperature $\approx$5\ K higher than the dust temperature. 
This region is illuminated by a far ultraviolet field of $\sim$ 400 $G_0$ \citep{habart2003}. Therefore, the gas temperature in this region is significantly higher than the dust temperature, as expected in a dense region with high external illumination \citep{tk-td_pdr_koumpia2015}, and as gas-dust coupling is not achieved at those densities.

\subsection{Ammonia Abundance}
\label{sec_amm_abun}

To see how the ammonia abundance varies going from the cores to the surrounding cloud, we separate the points in the coherent cores and  the shells. It can be seen from Figure \ref{amm_col}, that we are not able to obtain a much extended map for the p-\amm column density. Consequently, we have too few points in shell-2 of the cores for a meaningful analysis. Therefore, we limit our analysis for \amm abundance to only the cores and shell-1. As mentioned in Section \ref{sec_line_fit}, in this work we focus on the column density of only para-ammonia. We do not attempt to convert this into the total \amm column density because of unknown ortho-to-para ratio.

\begin{figure*}[!h]  
\centering
\includegraphics[width=0.8\textwidth]{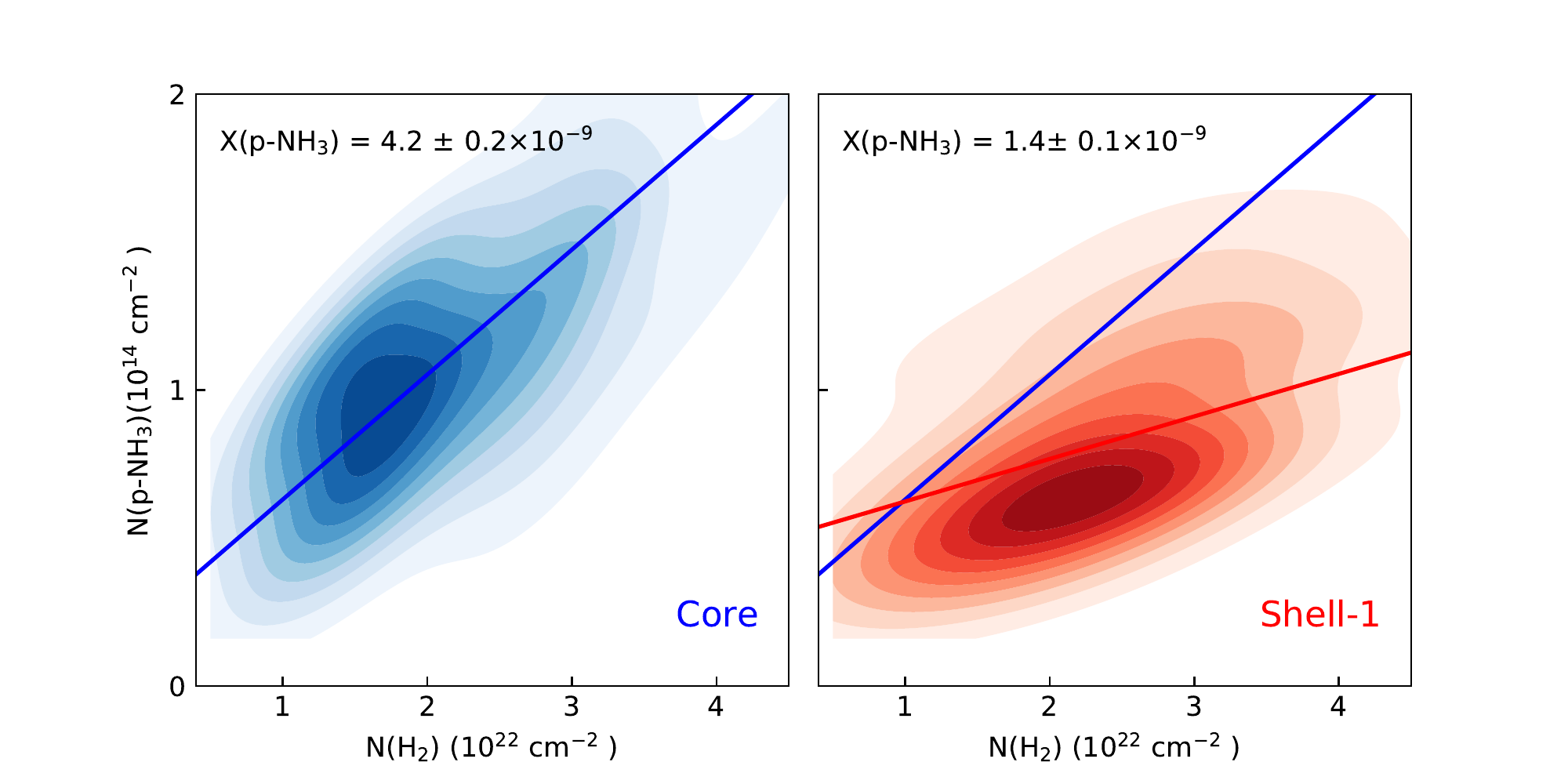}
 \caption{KDE representation of the distribution of  p-\amm column density with N(\htw) in the cores (in red) and shell-1 (in blue), as defined in Section \ref{sec_id_coh} and shown in Figure \ref{coh_cores}. The straight lines show the best linear fit to the data in the corresponding region. For comparison between the regions, the linear fit to the data in the cores (red line) is also shown in the plot for shell-1. The slopes of the linear fits indicate the fractional p-\amm abundance (with respect to \htw) in the region, which is shown in the top left corner.}
    \label{amm_abun}
\end{figure*}

Figure \ref{amm_abun} shows the distribution of N(p-\amm) with N(\htw), inside the cores (left panel) and shell-1 (right panel). Note that the plots only show the pixels, for which a good determination of N(p-\amm) was possible.
To clearly show the distribution of a large number of points, we plot the Kernel Density Estimate (KDE), instead of the individual data points in the plots (the calculation of KDE distribution from the data is explained in Section \ref{sec_tk_sig}).  The error in p-\amm column density is considered in the calculation of the KDE.

To get an estimate of the para-ammonia abundance in the cores and shell-1, we fit a straight line of the form \texttt{y = mx + c}, to each data-set. The slope of the line, \texttt{m}, gives us the fractional para-ammonia abundance, $X(p\text{-}NH_3) = {N(p\text{-}NH_3)}/{N(H_2)}$, in the region. We use \verb+curve_fit+ from the python package \verb+scipy+ to obtain the best linear fit. We take into account the error in our calculation of N(p-\amm) in fitting this data. However, as the error associated with the \htw column density was not available in the $\emph{Herschel}$ public archive, we do not incorporate any error in the x-axis for the fit. 

The model in the core is a good fit to the data. The fit in shell-1 is slightly offset from the peak due to the presence of a small number of high N(p-\amm) points. However, the fit still provides a good constraint on the slope, which is the parameter of interest. From this analysis, we report an average para-ammonia fractional abundance (with respect to \htw) of 4.2$\pm$0.2$\times$10$^{-9}$ in the coherent cores, and 1.4$\pm$0.1$\times$10$^{-9}$ in shell-1. Therefore, the para-ammonia abundance drops by a factor of 3, going from cores to their immediate surrounding.
The p-\amm abundance within the coherent cores reported here is comparable to that in L1544, as found by \citet{tafalla_2002, crapsi_2007_temp_drop_core, caselli_2017_amm_abun}, which is $\approx$ 4$\times$10$^{-9}$. \citet{crapsi_2007_temp_drop_core} also report a similar drop in abundance for a similar distance outside the core in L1544. They report an abundance of $\approx$2$\times$10$^{-9}$, at a distance $\sim$10\,000 au. This is comparable to the abundance we report for shell-1, which is at a similar distance ($\sim$10\,000 au) from the cores in our study\footnote{These studies report the total ammonia abundance (X(\amm)). We convert this to X(p-\amm) for comparison, using ortho-para ratio of unity, as assumed in these studies.}.

\subsection{Presence of a second component : comparison with \citetalias{choudhury2020_letter} results}
\label{sec_sec_comp}

From the fits and corresponding residuals to the average spectra towards all the cores and most of the shells (Figures \ref{avg_spec_ophA} to \ref{avg_spec_west}), we see that the single component fits do not recover all the flux. This indicates the presence of a second component in these spectra. In \citetalias{choudhury2020_letter}, we analysed the dual-component nature of the spectra towards the cores. 
It was shown that a faint supersonic component is present along with the narrow core component, towards all cores. 
We suggested that the narrow component is representative of the subsonic core, and the broad component traces the foreground cloud next to the cores. Here, we extend that analysis to shell-1 and shell-2 around the coherent cores, and study how the two components change going from cores to their shells. 

Similar to \citetalias{choudhury2020_letter}, we use the Akaike Information Criterion (AIC), which determines if the quality of the model improves significantly, considering the increase in the number of parameters used. We find that for all the regions, the two-component fit is a better model to the spectra. 
However, with a closer look at the spectra, we see that for shell-2 of d12, and shell-1 of SR3, one of the components fit by the model has very large \sig (> 1 \kms), and is very faint (peak $\rm{T_{MB, (1,1)}} \approx 10\,mK$, which is comparable to, or lower than the noise in the spectrum). Therefore, we do not have reliable constraints for the fits to these components. So, we do not consider them for the further analysis.

Extending the definition of the two components used in \citetalias{choudhury2020_letter}, we refer the two components as `narrow' and `broad', based on their velocity dispersion. However, it should be noted that unlike towards the cores (in \citetalias{choudhury2020_letter}), the narrow components in the shells, especially shell-2, are not always subsonic. In this paper, our distinction of the components are merely based on the velocity dispersion, in order to be able to track the transition of each component from core to shell-2.

\begin{figure*}[!h]
\centering
\includegraphics[width=0.9\textwidth]{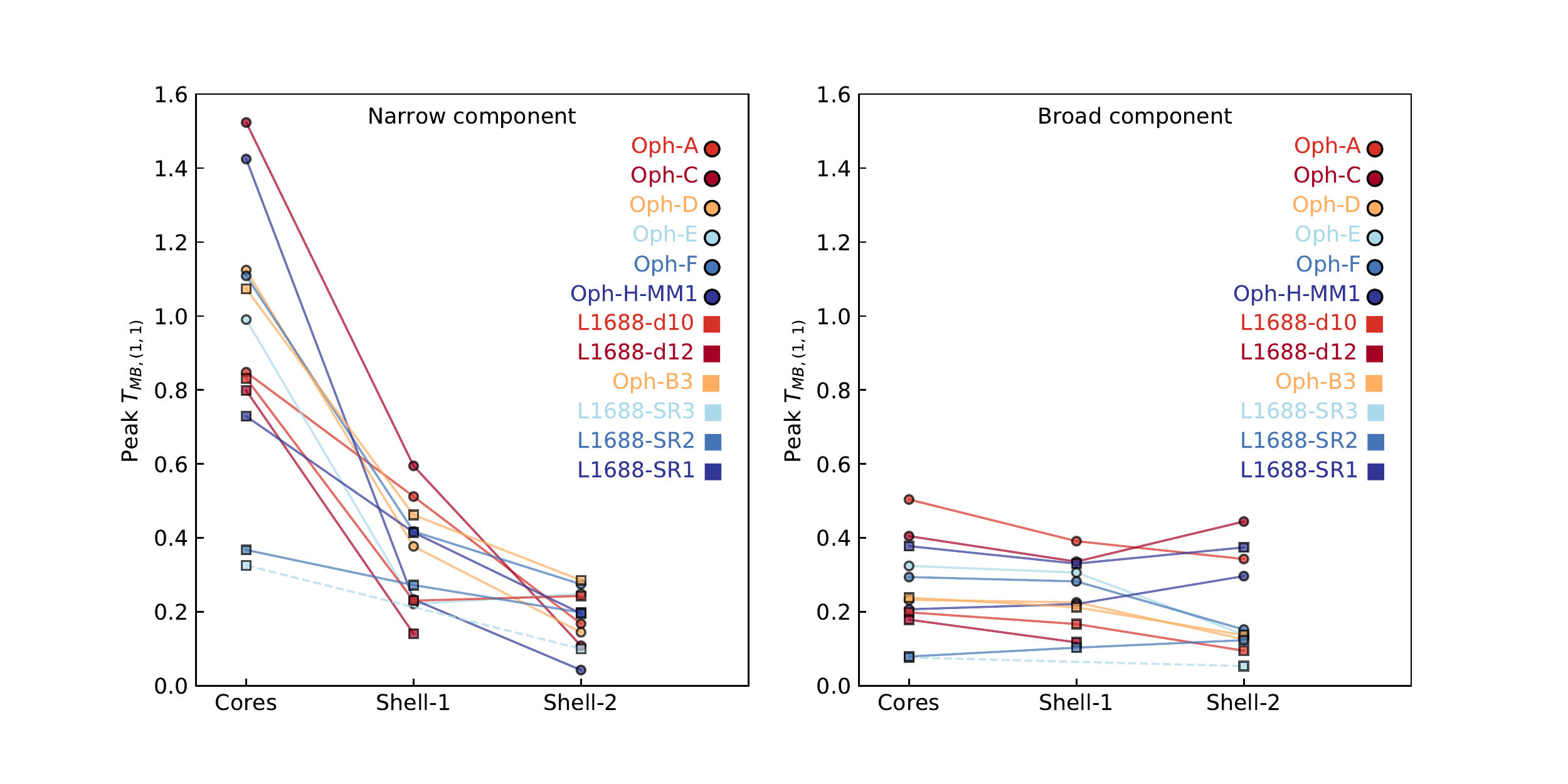}
\caption{
Left panel: Peak main beam brightness temperature of (1,1), for the narrow component in the cores, shells-1 and shells-2. 
Shells-1 and shells-2 are defined as two shells of width equal to one beam around the respective coherent cores.
Right panel : Same variance, but for the broad component. \textit{Note} : As we do not consider the two-component fit to shell-1 of SR3 (see Section \ref{sec_sec_comp}), the values for the core and shell-2 are connected by a dashed line.}
\label{tmb_comp_2-cmp}
\end{figure*}

\begin{figure*}[!h]
\centering
\includegraphics[width=0.9\textwidth]{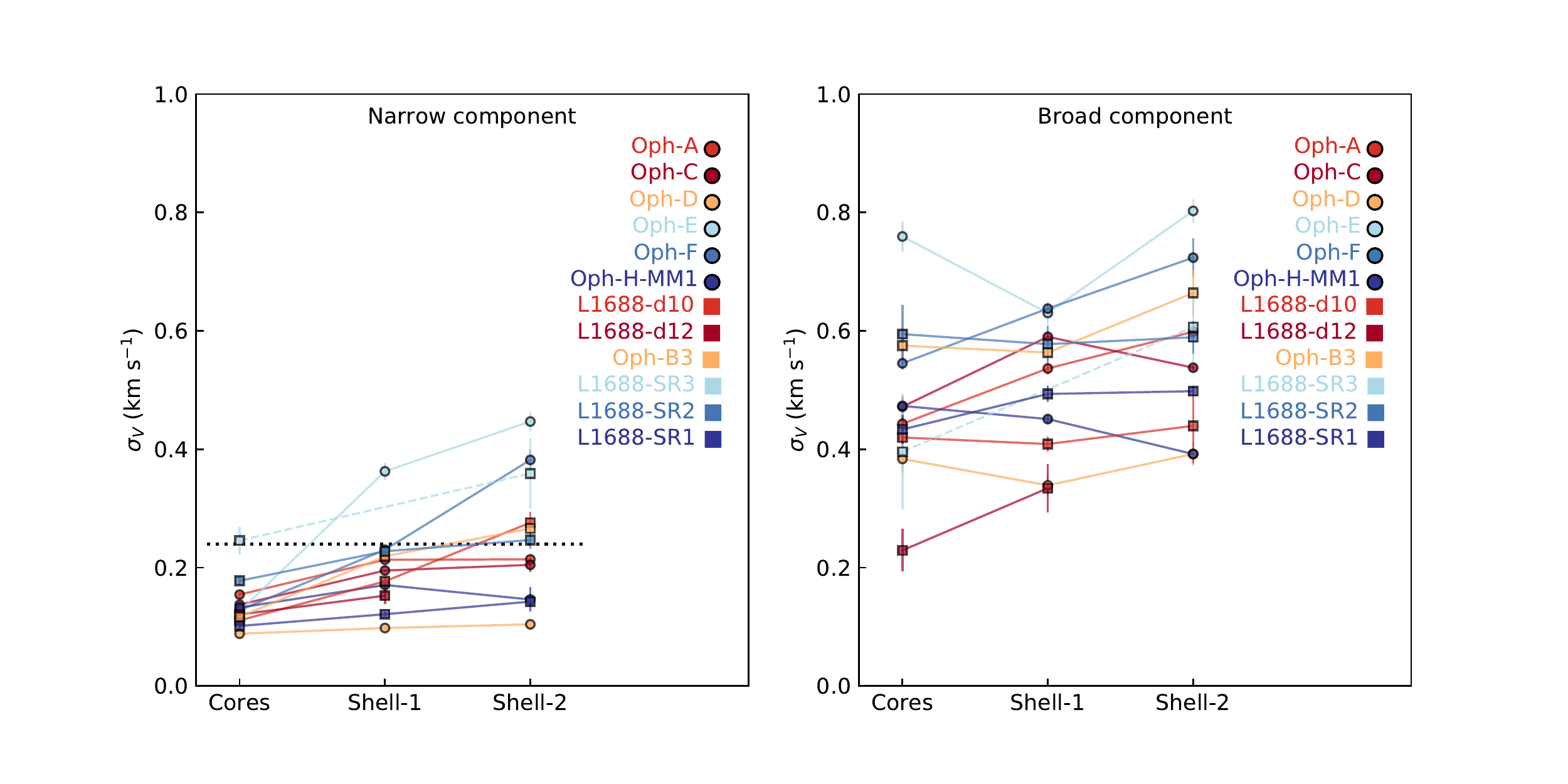}
\caption{
Left panel: Velocity dispersion of the narrow component in the cores, shells-1 and shells-2. 
Shells-1 and shells-2 are defined as two shells of width equal to one beam around the respective coherent cores.
The black-dotted line shows the velocity dispersion with $\mathcal{M_S}=1$ at typical core temperature of \tk=$\rm 10\,K$.
Right panel : Same, but for the broad component.
\textit{Note} : As we do not consider the two-component fit to shell-1 of SR3 (see Section \ref{sec_sec_comp}), the values for the core and shell-2 are connected by a dashed line.}
\label{sig_comp_2-cmp}
\end{figure*}

\begin{figure*}[!h]
\centering
\includegraphics[width=0.9\textwidth]{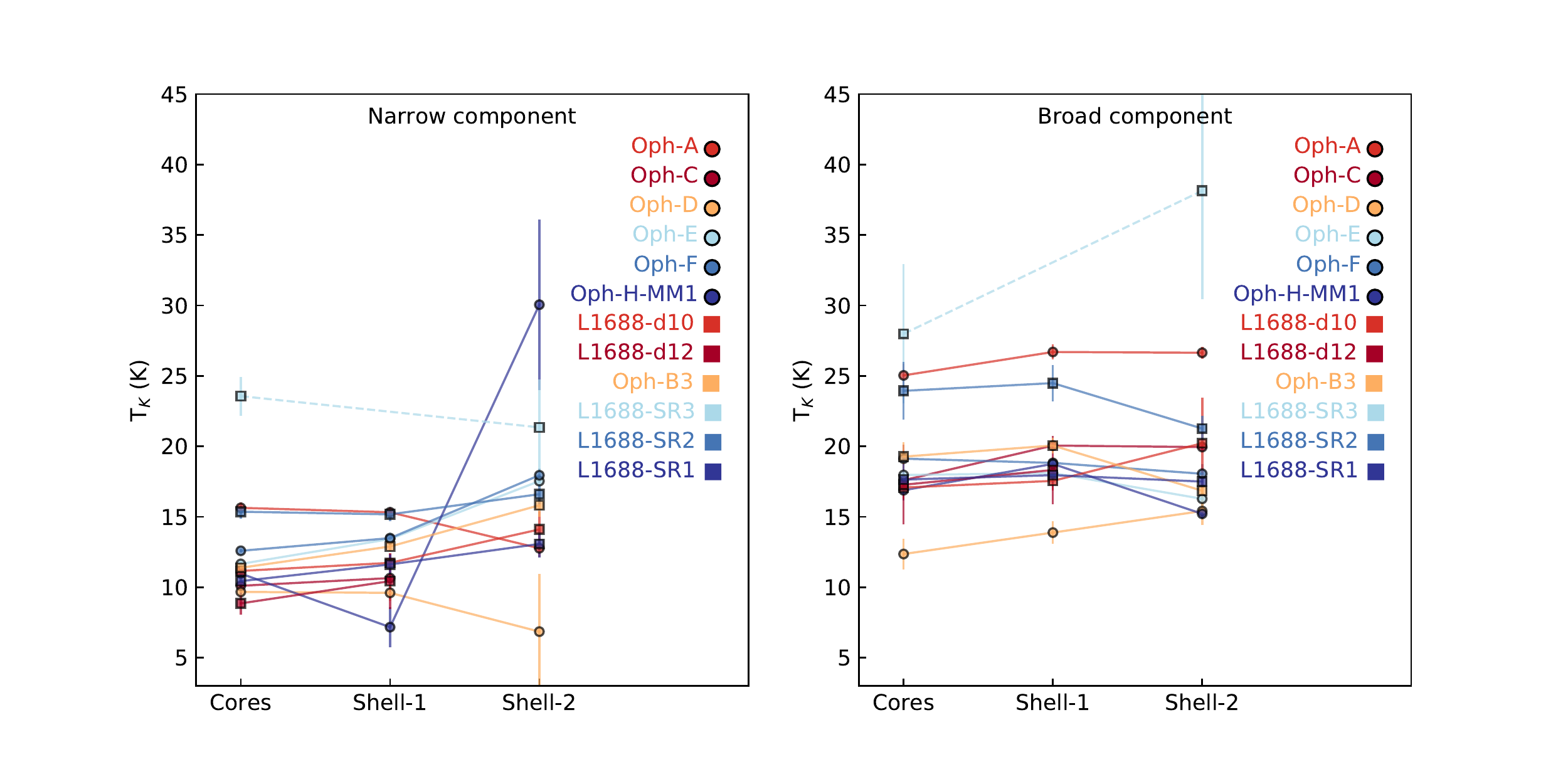}
\caption{
Left panel: Kinetic temperature of the narrow component in the cores, shells-1 and shells-2. 
Shells-1 and shells-2 are defined as two shells of width equal to one beam around the respective coherent cores.
Right panel : Same variation, but for the broad component.
\textit{Note} : As we do not consider the two-component fit to shell-1 of SR3 (see Section \ref{sec_sec_comp}), the values for the core and shell-2 are connected by a dashed line.}
\label{tk_comp_2-cmp}
\end{figure*}

Figure \ref{tmb_comp_2-cmp} shows the variation of the 
peak main beam brightness temperature 
of the (1,1) across cores and shells, for the narrow component (left panel) and the broad component (right panel). It can be seen that the intensity of the narrow component decreases sharply from core to shell-1, and less so from shell-1 to shell-2. In comparison, the broad component's intensity remains largely constant for all three regions. This result agrees with the conclusions of \citetalias{choudhury2020_letter}, that the narrow component is the core component, and the broad component is a single cloud component, representative of the surrounding cloud. This figure also shows that the single component fit results would be skewed towards the narrow component in the core. The flux is dominated by the broad components in shell-2 because of larger width, as the intensities are similar to those of the narrow components. For shell-1, neither component is clearly dominant.

The variation of velocity dispersion and kinetic temperature, from the cores to the shells, for the two components, is shown in Figures \ref{sig_comp_2-cmp} and \ref{tk_comp_2-cmp}, respectively. It can clearly be seen from Figure \ref{sig_comp_2-cmp} that the narrow component gets more turbulent, as we go outwards from the cores. 
The broad cloud component does not show any clear trend, and is mostly invariant. The narrow component, even in shell-2, has lower \sig than the broad component towards cores. This again agrees with the interpretation of \citetalias{choudhury2020_letter}, of the narrow component tracing the subsonic core, and the broad component tracing the more turbulent surrounding cloud.

For the kinetic temperature, we do not see a clear trend in either of the two components (Figure \ref{tk_comp_2-cmp}). It can be seen that the narrow component is slightly colder towards the cores compared to the shells. In general, the narrow component is at a lower temperature than the broad component. In shell-2 of Oph-C, the (2,2) line could not be fit for the broad component, and therefore, the kinetic temperature of that component is not well-defined. Hence, the shell-2 of Oph-C is omitted from Figure \ref{tk_comp_2-cmp} (for the broad component).

{Figure \ref{tk_comp_2-cmp} shows that for H-MM1, the kinetic temperature of the narrow component show an unusual trend, in that \tk in shell-2 seem to be $\approx 20\,$K higher than that in the core. A look at the spectra (Figure \ref{avg_spec_hmm1}) show that the relative positions of the narrow and broad component seem to switch from core and shell-1, to shell-2. The narrow component in shell-2 seem to be brighter in (2,2) than in shell-1. These unexpected behaviours could be explained by the relatively small change in AIC in the 2-component fit, from the 1-component fit (see \ref{tab_fit_para_all}). As mentioned in Appendix E of \citetalias{choudhury2020_letter}, the constraints in the fit-derived physical properties are relatively low, for small $\Delta_{\rm AIC}$ ($ = \rm AIC_{1-comp} - AIC_{2-comp}$). In particular, for H-MM1, we find that more restrictions (e.g., not allowing the two components to switch their relative positions) results in a different 2-component fit, with only a slightly smaller $\Delta_{\rm AIC}$ (=27). This alternate fit (Figure \ref{avg_spec_hmm1_alt}) indicate that there is no subsonic component in shell-2, but rather, two supersonic components (\sig $\approx$ 0.28 \kms for both). Therefore, as already mentioned in \citetalias{choudhury2020_letter}, we need to be cautious while considering the regions with low $\Delta_{\rm AIC}$ values, and need to inspect the fits to the spectra carefully.}

In \citetalias{choudhury2020_letter}, we reported that the ambient cloud in L1688 (considered to be represented by the rectangular box shown in Figure \ref{h2_col}) shows two supersonic components, with a small relative velocity. We suggested that a collision between these two cloud components might result in a local density increase, where the merging of the two broad components locally produces the narrow feature, following a corresponding dissipation of turbulence; thus creating the observed coherent cores with subsonic linewidths. The gradual decrease in the dispersion of the narrow component, from shell-2 to core is congruent with this hypothesis. 

\begin{figure}[!ht]
\includegraphics[width=0.45\textwidth]{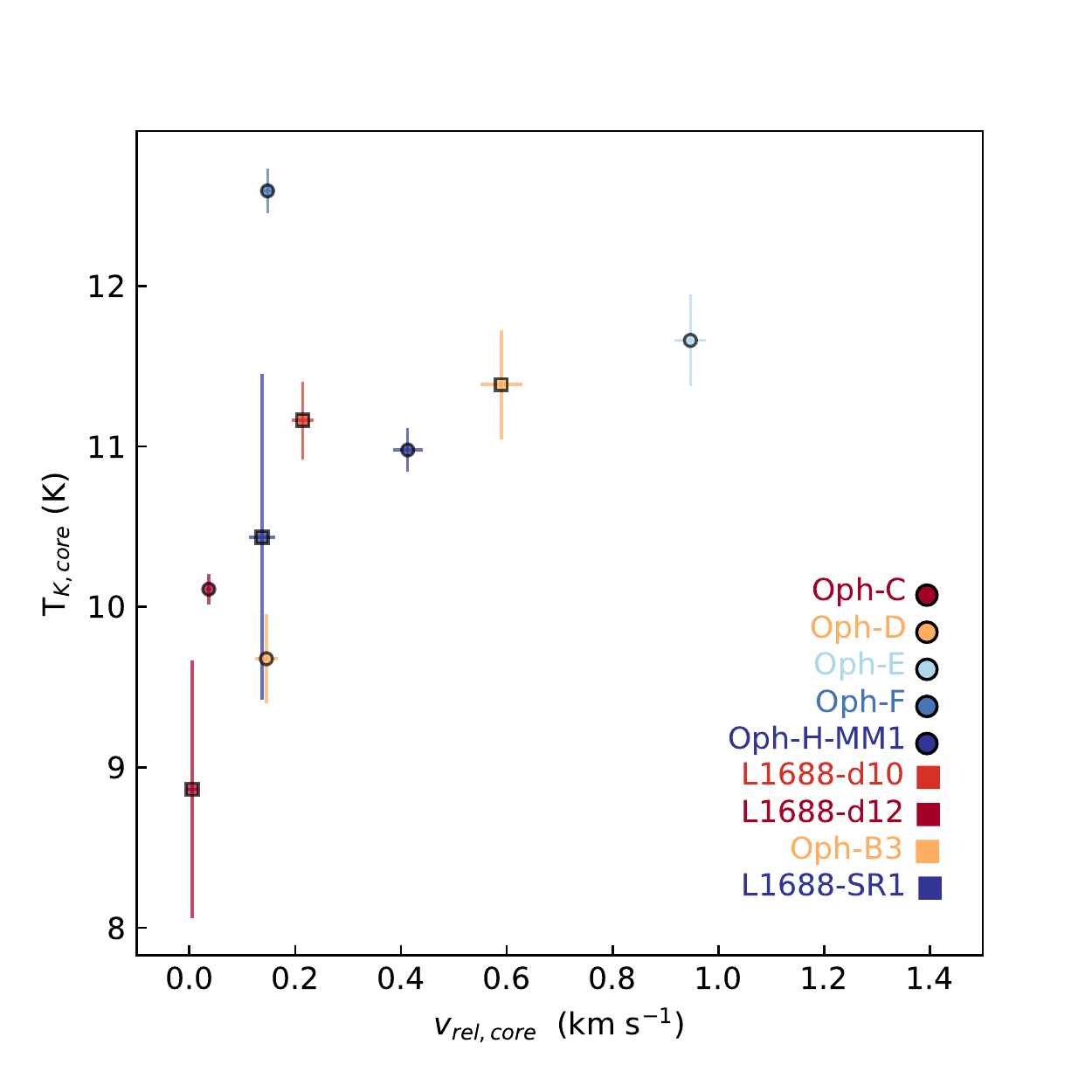}
\caption{Kinetic temperature of the core (narrow) component, against the relative velocity between the core (narrow) component and the cloud (broad) component, towards each core. Oph-A, SR2 and SR3 are affected by the external illumination, and therefore, are not considered in this analysis.}
\label{rel-vel_tk-n}
\end{figure}

Figure \ref{rel-vel_tk-n} shows the temperature of each  coherent core (narrow component) as a function of the relative velocity between the two components towards the core. Here, we have not considered Oph-A, SR2 and SR3, so as to remove any possible contribution of the external radiation from the west. It can be seen that there is a general trend of the core temperature increasing with larger relative velocity of the two components. This might point to possible collision between the two components being responsible for a local temperature increase. However, owing to our relatively large beam, we cannot conclusively comment on the presence of local shocks \citep[e.g.][]{pon_2012_shock}, and would require higher resolution data in the region and observations in shock tracers, to confirm or rule out this possibility.

\subsection{Evidence of subsonic component beyond previously identified coherent zones}

The subsonic component seen in the cores is also detected in shell-1 for ten out of the twelve coherent cores (except Oph-E, where there is no subsonic component in shell-1, and SR3, where the two-component fit in shell-1 of SR3 is not well-constrained; see Section \ref{sec_sec_comp}). 
Furthermore, in five regions (Oph-A, Oph-D, H-MM1, SR1 and SR2), the component with subsonic turbulence can be detected even in shell-2. 
This suggests that the subsonic component extends well beyond the typical boundary of the coherent cores, albeit being much fainter in intensity ($\approx$1/5th the peak intensity in \amm (1,1)) compared to that towards the cores. 
This suggests that the transition to coherence is not sharp, as previously reported \citep[e.g.][]{pineda2010, GASDR1, chen2019}, rather, gradual. 
The sharp transition to coherence observed by \citet{pineda2010} was within a 0.04 pc scale, which is very similar to our spatial resolution (1$\arcmin$ at $\approx$138 pc). It is likely that the transition observed by \citet{pineda2010} in B5 was the transition between the narrow and the broad component. Due to the poorer sensitivity of the data, the broad component towards the core and the subsonic component outside the core boundary were not detected in their work. 
With averaged spectra, we are able to detect the two components towards the cores and the shells, and we observe that the coherent cores in Ophiuchus are indeed more extended than previously found with single-component fit.

It should be noted that due to the lack of sensitivity, even with the smoothed data, we are not able to detect the broad component across the entire region; and the two-component analysis is only possible with the stacked spectra in the coherent cores and the shells. Therefore, with our current sensitivity, we cannot use the narrow component to define the coherent core, which would be the ideal case, as we do not have a map of the narrow (or broad) component. So, we are limited by the sensitivity to use the single component fit, to define the coherent boundaries, as is the case in the existing literature. Our present results indicate that with better sensitivity data, which allow for a two-component fit, the coherent boundary could be improved.

We also observe that the core component smoothly broadens up to a supersonic velocity dispersion towards some cores, but the dispersion still does not reach that of the broad component at the corresponding position (see Figure \ref{sig_comp_2-cmp}).
Although the value of the broad component could be  interpreted as the level of turbulence outside the core, it is more likely that the larger velocity dispersion is due to change in the scale traced along the line of sight, which could include large scale velocity structure.

\subsection{Turbulence-size relation}

We observe the subsonic component extending outside the typical boundary of the coherent cores. We also observe this component gradually broadening outwards from the cores (Figure \ref{sig_comp_2-cmp}). 
Therefore, we looked at the relation of the turbulence (non-thermal velocity dispersion) in the component with size of the regions. 
We define the equivalent radius, $r_{\rm eq}$ for each region as $A = \pi r_{\rm eq}^2$, where $A$ is the area inside that region (core, shell-1 or shell-2). 

\begin{figure}[!ht]
\includegraphics[width=0.5\textwidth]{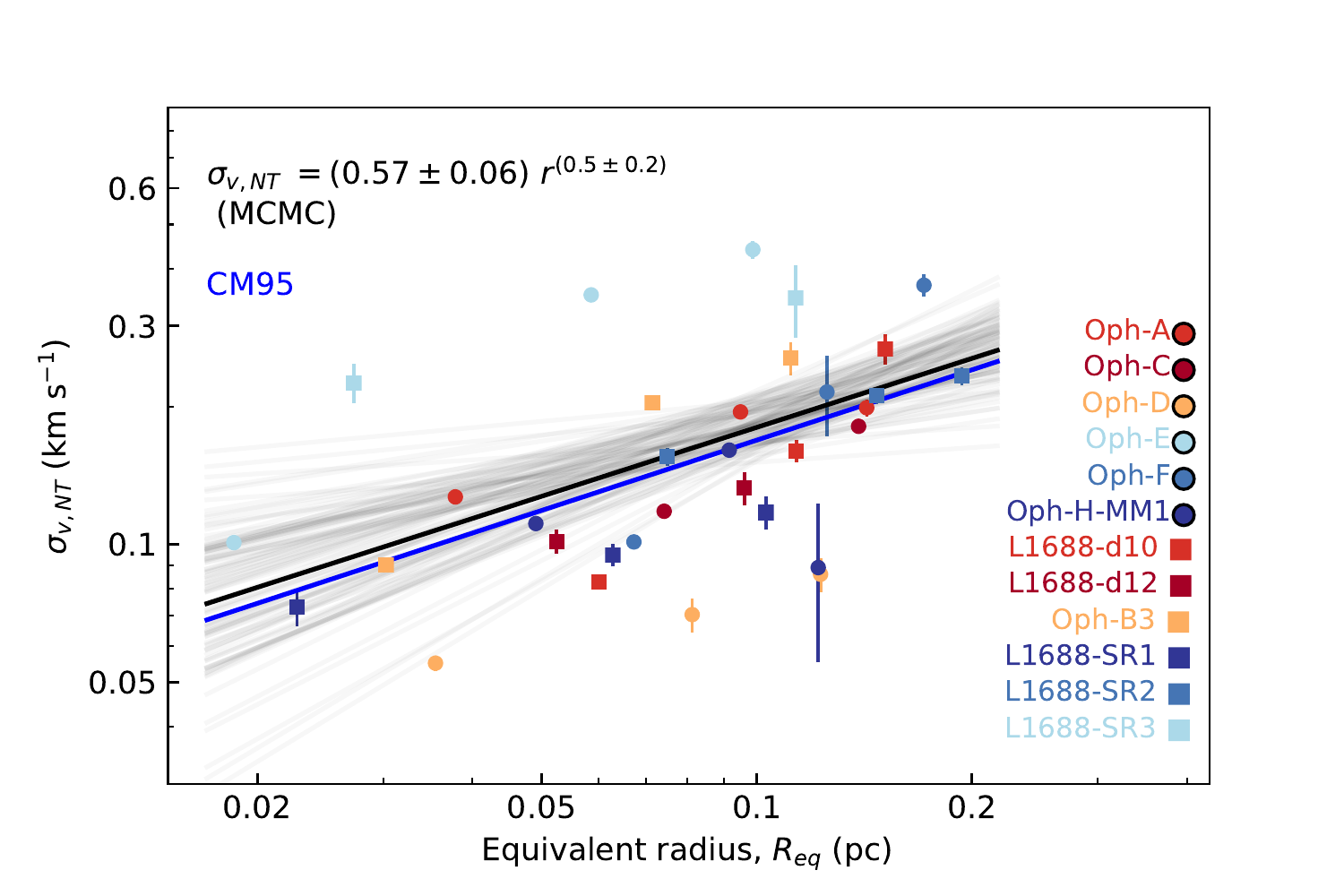}
\caption{Non-thermal velocity dispersions in the narrow component, for each core, shell-1 and shell-2, as a function of their equivalent radii. 
The grey lines show the power law fits for the range of parameters obtained from a MCMC fit, and the black line shows the best fit model. The blue line show the turbulence-size relation from \citep{CM95_lw_sz} for low-mass cores.}
\label{turb_sz}
\end{figure}

As we could only detect the subsonic component outside the typical coherent core using stacked spectra in the shells with our current data, for each core, we only have three data-points to fit for the turbulence-size relation. 
Moreover, the $\sigma_{\rm v,NT}$ for shell-2 often has relatively large errors (see Table \ref{tab_fit_para_all}). 
Therefore, with the present sensitivity, we cannot obtain a reliable turbulence-size relation for each core separately, and so, we only fit an average relation considering all cores together. 
Figure \ref{turb_sz} show the turbulence in the narrow component towards each coherent core, shell-1 and shell-2, as a function of the respective equivalent radii. 

We fit a power-law, $\sigma_{\rm v,NT} = a\,r^b$, to the whole sample using the uncertainty in the derived non-thermal velocity dispersion with \texttt{emcee} \citep{emcee2013}. Since we fit for an average relation for all the cores, we allow for a Gaussian intrinsic scatter term (with variance $V$) to the model, to account for the spread due to considering multiple cores. In the calculation of the likelihood parameter, the uncertainty in the model, $\sigma$, is then replaced by 

\begin{equation}
    \sigma_{\rm eff} = \sqrt{\sigma^2 + V} ~.
\end{equation}

We use uniform priors for the exponent ($b$) and the constant ($a$) with a range that includes both tails of the individual posterior probability distributions. We then sample the distribution using \texttt{emcee} with 32 walkers in a 2D Gaussian around the maximum likelihood result, and run 5000 steps of MCMC (Markov Chain Monte Carlo). With this, we obtain an auto-correlation time of $\approx 40$ steps. We discard the first 200 steps to avoid effects of initialisation, and trim the distribution by half the auto-correlation time (20 steps). From this analysis, we determine the best fit to the data as 
\begin{equation}
    \sigma_{\rm v,NT} = (0.57\pm0.06)\, r_{\rm eq}^{(0.5\pm0.2)} ~.
    \label{eq_lw-sz}
\end{equation}

The derived exponent is consistent with the one reported by \citet{CM95_lw_sz} within the error bar. However, in that work, they use a sample of mixed tracers, including $^{13}$CO. Our results focus this relation to a narrow range of small scale structures, down to $\approx 0.02$ pc. The similarity of the exponent with the results from \citet{CM95_lw_sz} suggest that the turbulence-size relation remains approximately the same in scales closer to the core centre. \citet{chen_2020_core-form} also observed similar exponents in linewidth-size relations for Phase-I and -II cores in synthetic observations.\footnote{It should be noted that \citet{chen_2020_core-form} considered \sig in their relation. For comparison with their work, we calculated the relation \ref{eq_lw-sz} for \sig.}

\section{Conclusions}
We present new analysis on the GAS DR1 data with GBT towards L1688 smoothed to 1$\arcmin$ resolution. 
Our results can be summarised as follows:

   \begin{enumerate}
    \item For the first time, we obtain substantially extended kinetic temperature and velocity dispersion maps, covering 65\% and 73.4\% of the observed area, respectively, including dense cores and the surrounding molecular cloud, using the same density tracer. This ensures continuity in the physical properties from core to cloud.
    
    \item We identify 12 coherent cores in L1688 \citepalias[same as][]{choudhury2020_letter}, including 3 previously unidentified subsonic regions, SR1, SR2 and SR3. 
    
    \item Using single-component fit to the data, we observe that both the kinetic temperature and the velocity dispersion gradually increases outwards from the coherent cores. On average, the kinetic temperature 1$\arcmin$ ($\approx 8000$ au) and 2$\arcmin$ ($\approx 16000$ au) away from the core boundary is approximately 4 K and 6 K higher than the core temperature. Similarly, the velocity dispersion in these regions are 0.15 \kms and 0.25 \kms higher than that in the core.

    \item We find that the external illumination at the western edge of the cloud is not accompanied by turbulence injection.
    
    \item The kinetic temperature towards the coherent cores is, on average, $\approx$ 1.8$\,$K lower than the dust temperature. Outside the cores, the kinetic temperature of the gas is higher than the dust temperature.
    
    \item We find an average para-ammonia fractional abundance (with respect to \htw) of 4.2$\pm$0.2$\times 10^{-9}$ in the coherent cores, and 1.4$\pm$0.1$\times 10^{-9}$, at 1$\arcmin$ from the core. Previous works report a similar abundance within the core, and a similar drop in abundance for a similar distance outside the core, for L1544. 
    
    \item By stacking the spectra towards the cores and their neighbourhoods, we are able to detect two velocity components, one narrow and one broad, superposed in velocity. For most cores, we observe that the component with subsonic turbulence is extended beyond the previously identified coherent regions. This suggests that the transition to coherence is gradual, in contrast to previous results. We observe that the subsonic component towards the cores becomes fainter and broader outwards, with a turbulence-size relation of $\sigma_{\rm v,NT} \propto r_{eq}^b$, with $b=0.5\pm0.2$, similar to what was found in other low-mass dense cores, using multiple molecular tracers.
    
    \item In contrast, the broad component shows near-constant intensity and dispersion towards core and cloud. This supports our conclusions from \citetalias{choudhury2020_letter} of the broad component tracing material across the larger scale cloud seen with ammonia.
    
    \item We observe that on average the cores with larger velocities relative to the surrounding cloud show higher temperatures. With higher resolution maps of the region in adequate sensitivity, it would be possible to determine whether there is any presence of local shocks around the coherent cores.

   \end{enumerate}

\begin{acknowledgements}
SC, JEP, and PC acknowledge the support by the Max Planck Society.
This material is based upon work supported by the Green Bank Observatory which is a major facility funded by the National Science Foundation operated by Associated Universities, Inc. AC-T acknowledges the support from MINECO projects AYA2016-79006-P and PID2019-108765GB-I00. AP acknowledges the support from the Russian Ministry of Science and Higher Education via the State Assignment Project FEUZ-2020-0038. AP is a member of the Max Planck Partner Group at the Ural Federal University.
\end{acknowledgements}

\appendix

\section{Selection of cores, and shells}
\label{app_cores_on_tk}

\begin{figure*}[!t]  
\centering
\includegraphics[width=0.85\textwidth]{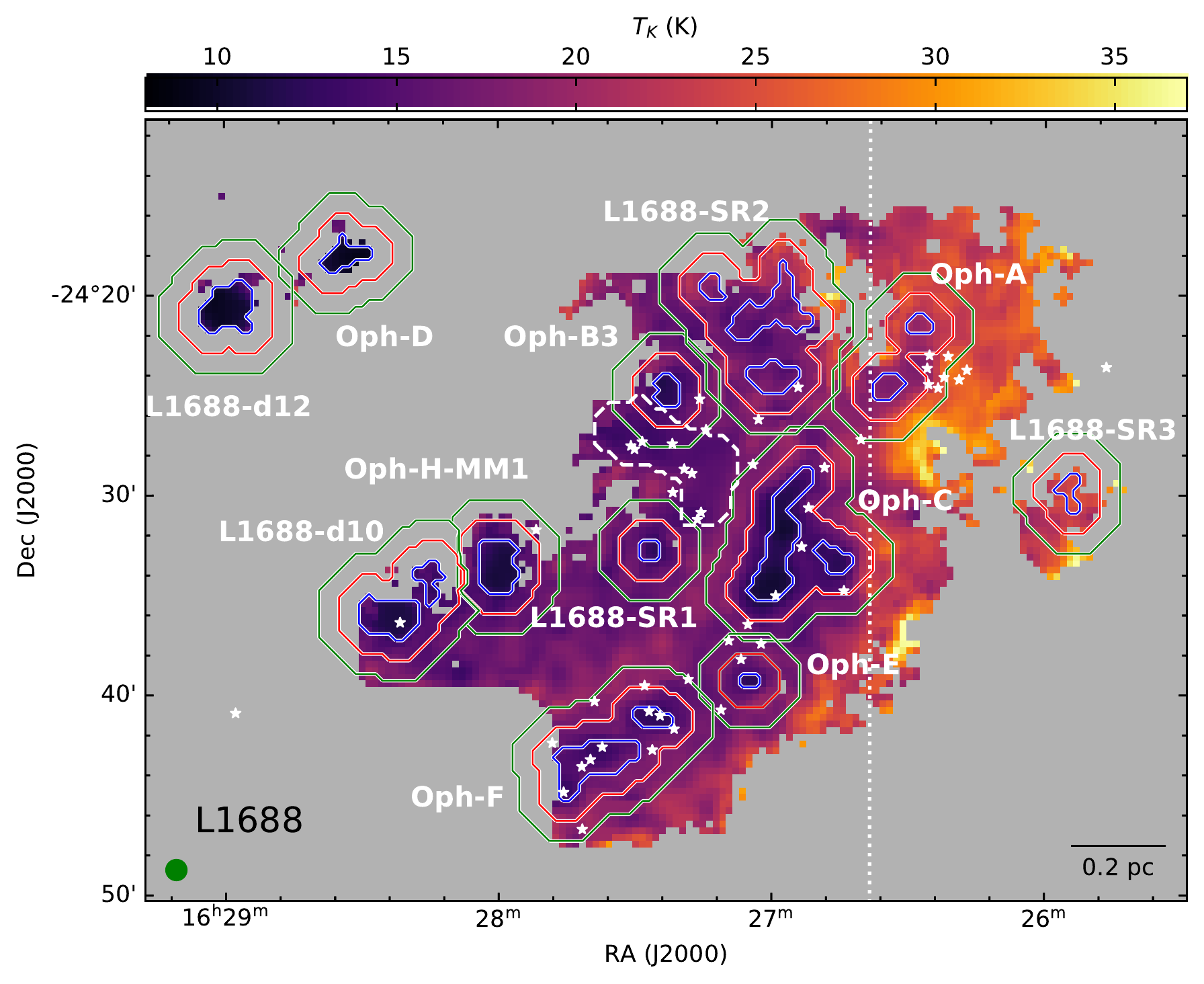}
  \caption{The coherent cores, shells-1 and -2, as defined in Section \ref{sec_id_coh}, are shown on the kinetic temperature map, with blue, red and green contours, respectively. The white stars show the positions of Class 0/I and flat-spectrum protostars in the cloud. The white dashed contour shows a rough boundary for Oph-B1 and Oph-B2 (See Section \ref{sec_id_coh}). The white dotted line roughly separates the dark cloud to the left from the molecular material affected by the external illumination, to the right.}
     \label{cores_on_tk} 
\end{figure*}

\begin{figure*}[!t]  
\centering
\includegraphics[width=0.85\textwidth]{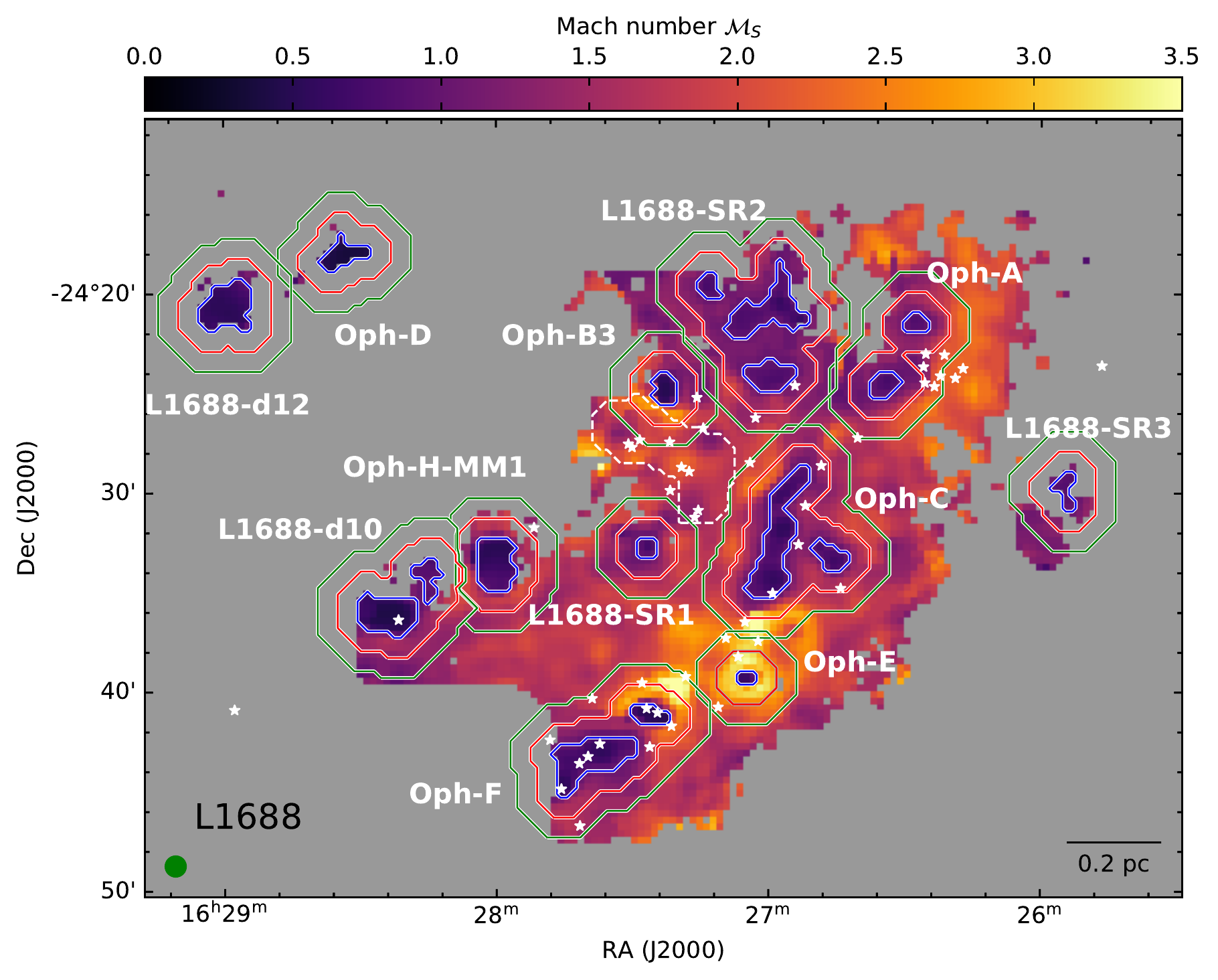}
  \caption{Similar to Figure \ref{cores_on_tk}, but with sonic Mach number as the background colour map.}
     \label{cores_on_mach} 
\end{figure*}

Figures \ref{cores_on_tk} and \ref{cores_on_mach} show the cores, as well as the shells 1 and 2 around each of them, similar to Figure \ref{coh_cores}, but with kinetic temperature and sonic Mach number as the background colour-scale. 

\section{Averaged spectra in the cores, shell-1 and shell-2}

\begin{figure*}[!ht]  
\centering
\includegraphics[width=1\textwidth]{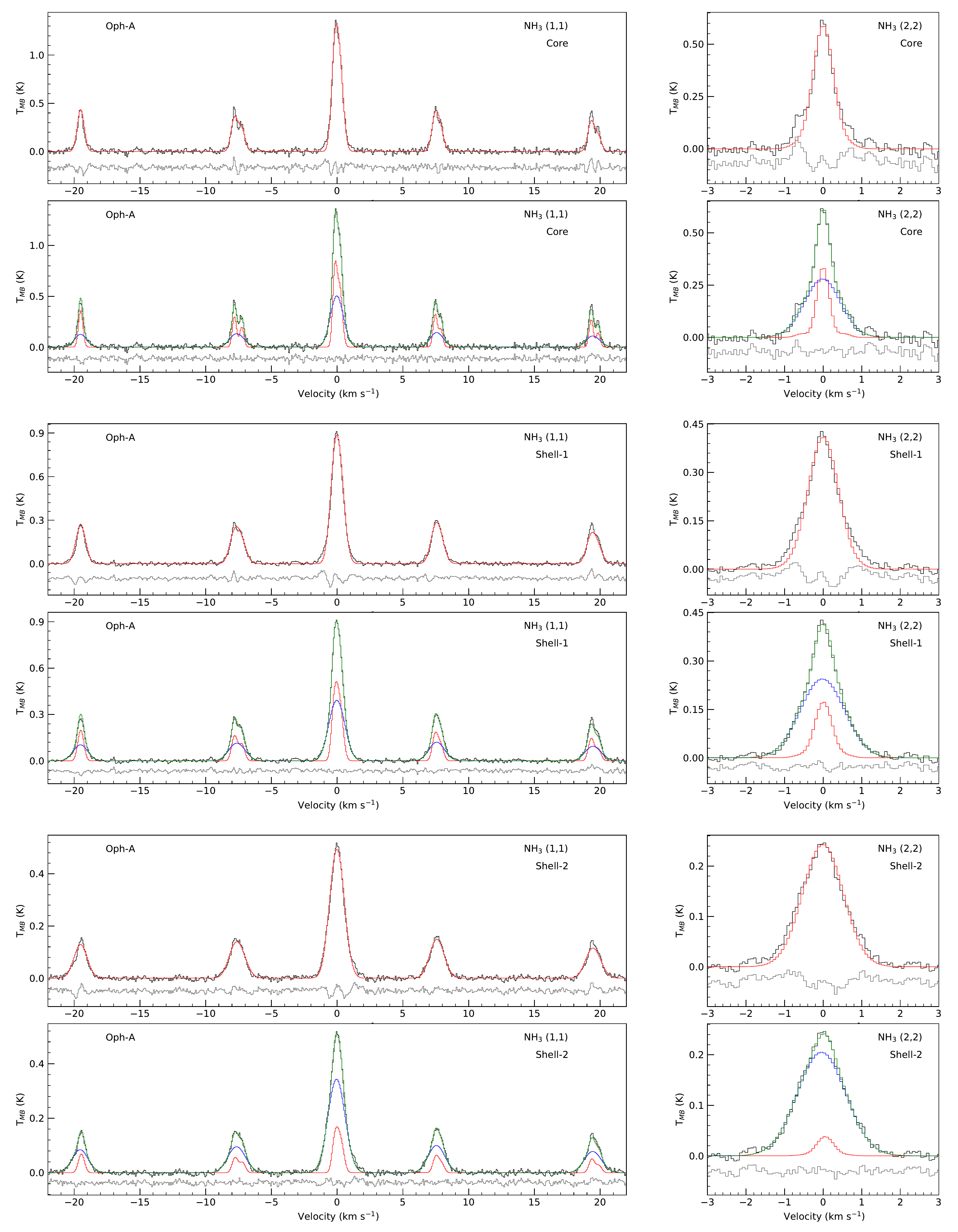}
  \caption{Top panels: Average \amm (1,1) and (2,2) spectra of Oph-A core, shell-1 and shell-2, with a single-component fit. Bottom panels: Same spectra, with two-component fit (green). The narrow (red) and broad (blue) components are also shown separately.}
     \label{avg_spec_ophA} 
\end{figure*}

\begin{figure*}[!ht]  
\centering
\includegraphics[width=1\textwidth]{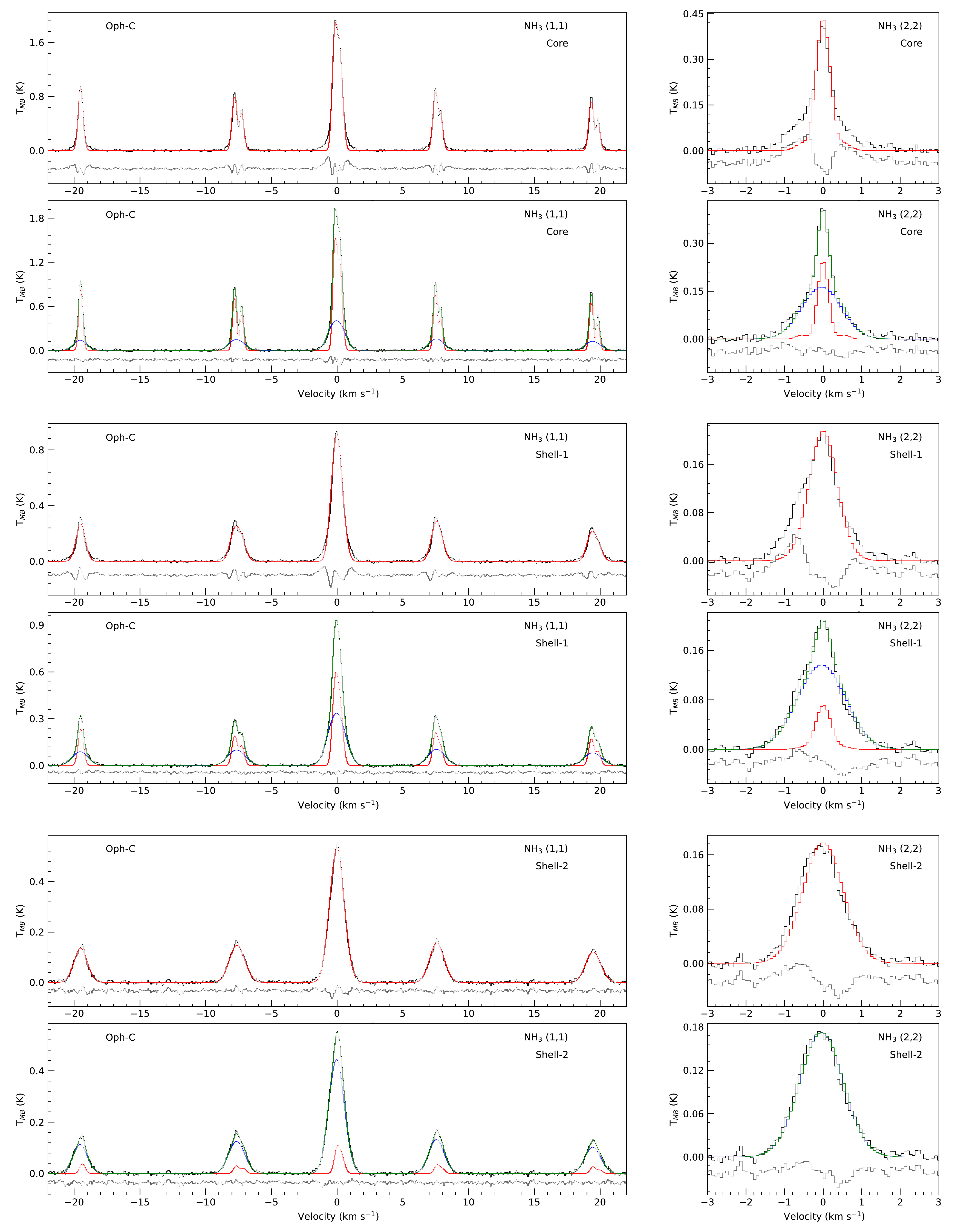}
  \caption{Same as Figure \ref{avg_spec_ophA}, but for Oph-C}
     \label{avg_spec_ophC} 
\end{figure*}

\begin{figure*}[!ht]  
\centering
\includegraphics[width=1\textwidth]{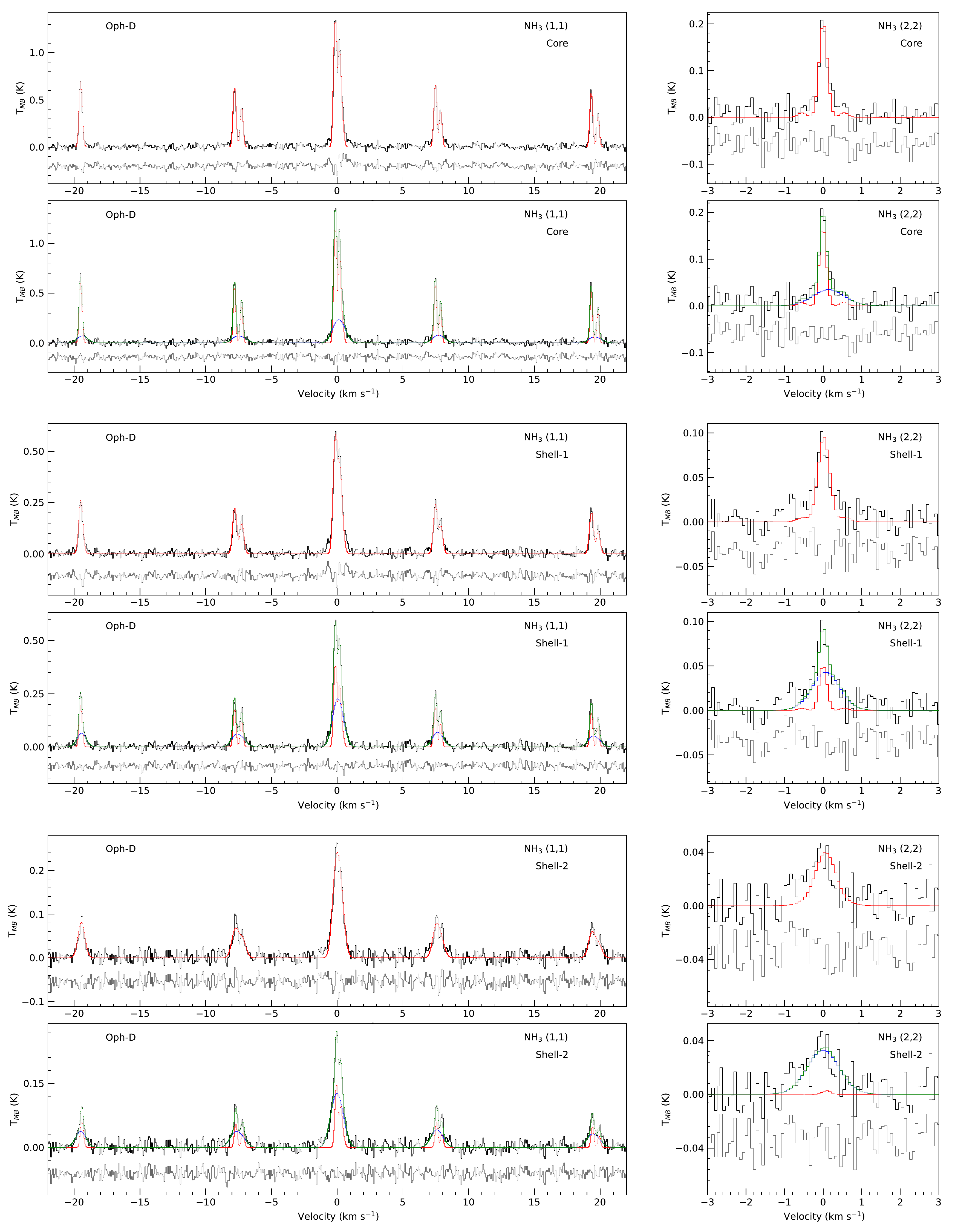}
  \caption{Same as Figure \ref{avg_spec_ophA}, but for Oph-D}
     \label{avg_spec_ophD} 
\end{figure*}

\begin{figure*}[!ht]  
\centering
\includegraphics[width=1\textwidth]{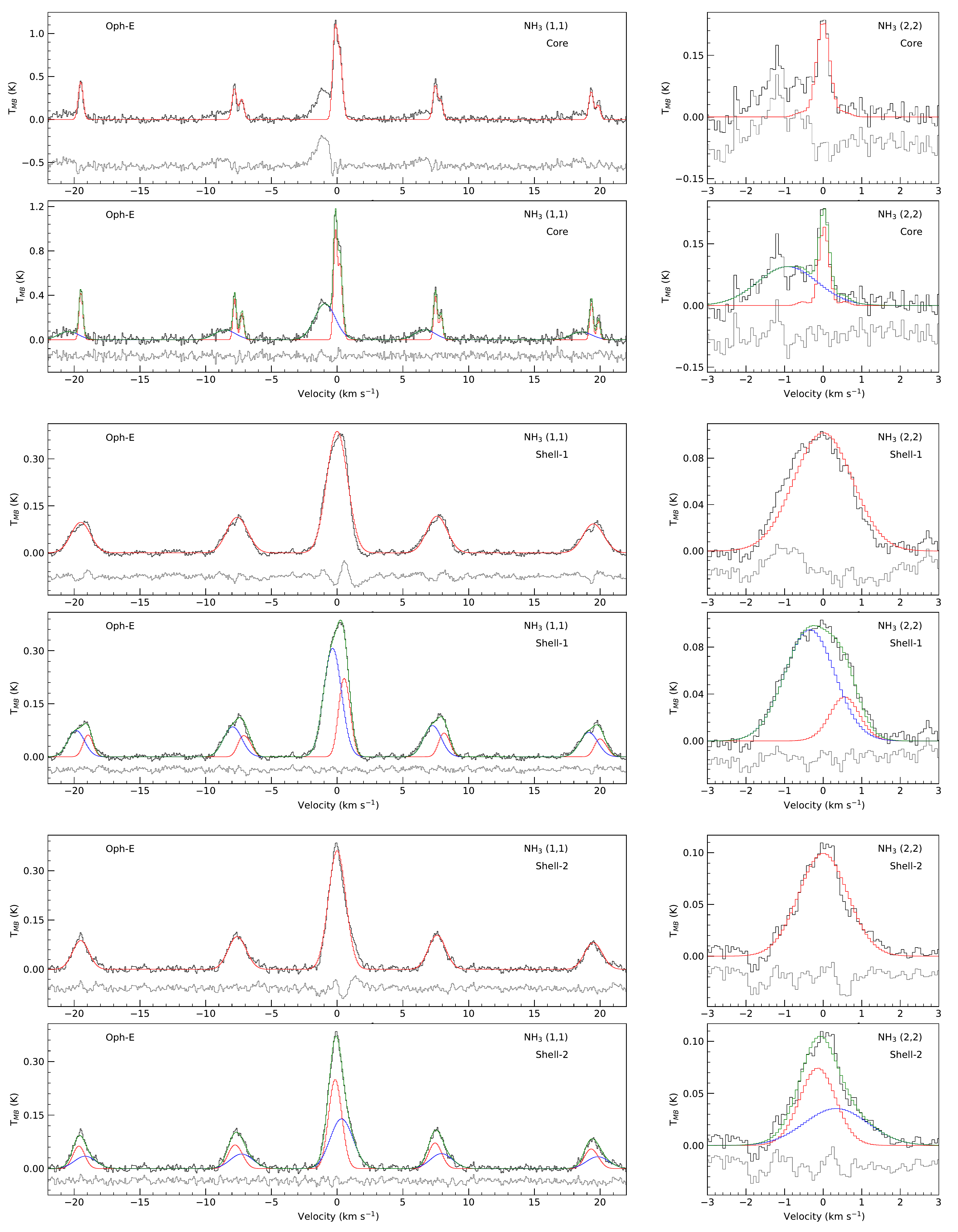}
  \caption{Same as Figure \ref{avg_spec_ophA}, but for Oph-E}
     \label{avg_spec_ophE} 
\end{figure*}

\begin{figure*}[!ht]  
\centering
\includegraphics[width=1\textwidth]{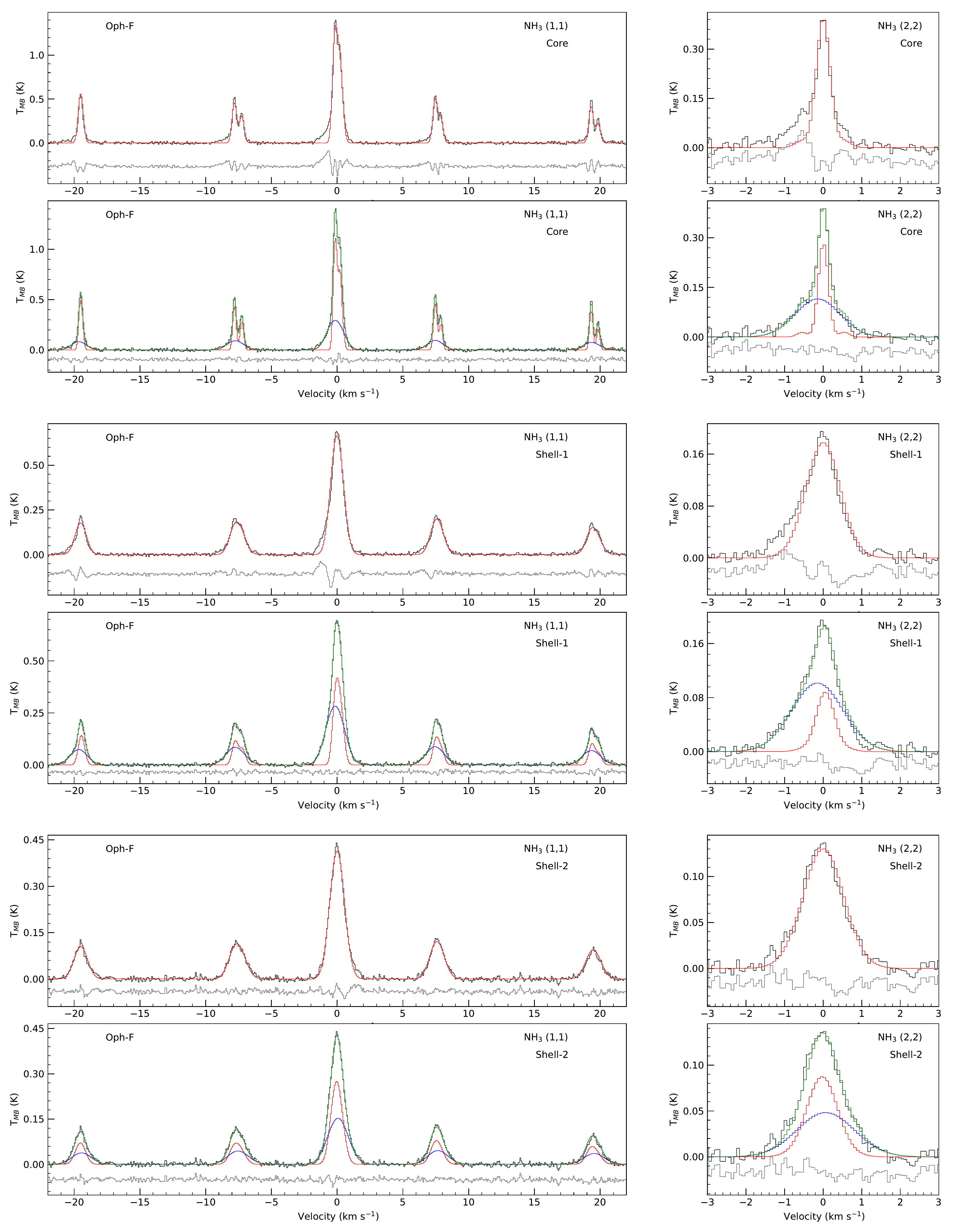}
  \caption{Same as Figure \ref{avg_spec_ophA}, but for Oph-F}
     \label{avg_spec_ophF} 
\end{figure*}

\begin{figure*}[!ht]  
\centering
\includegraphics[width=1\textwidth]{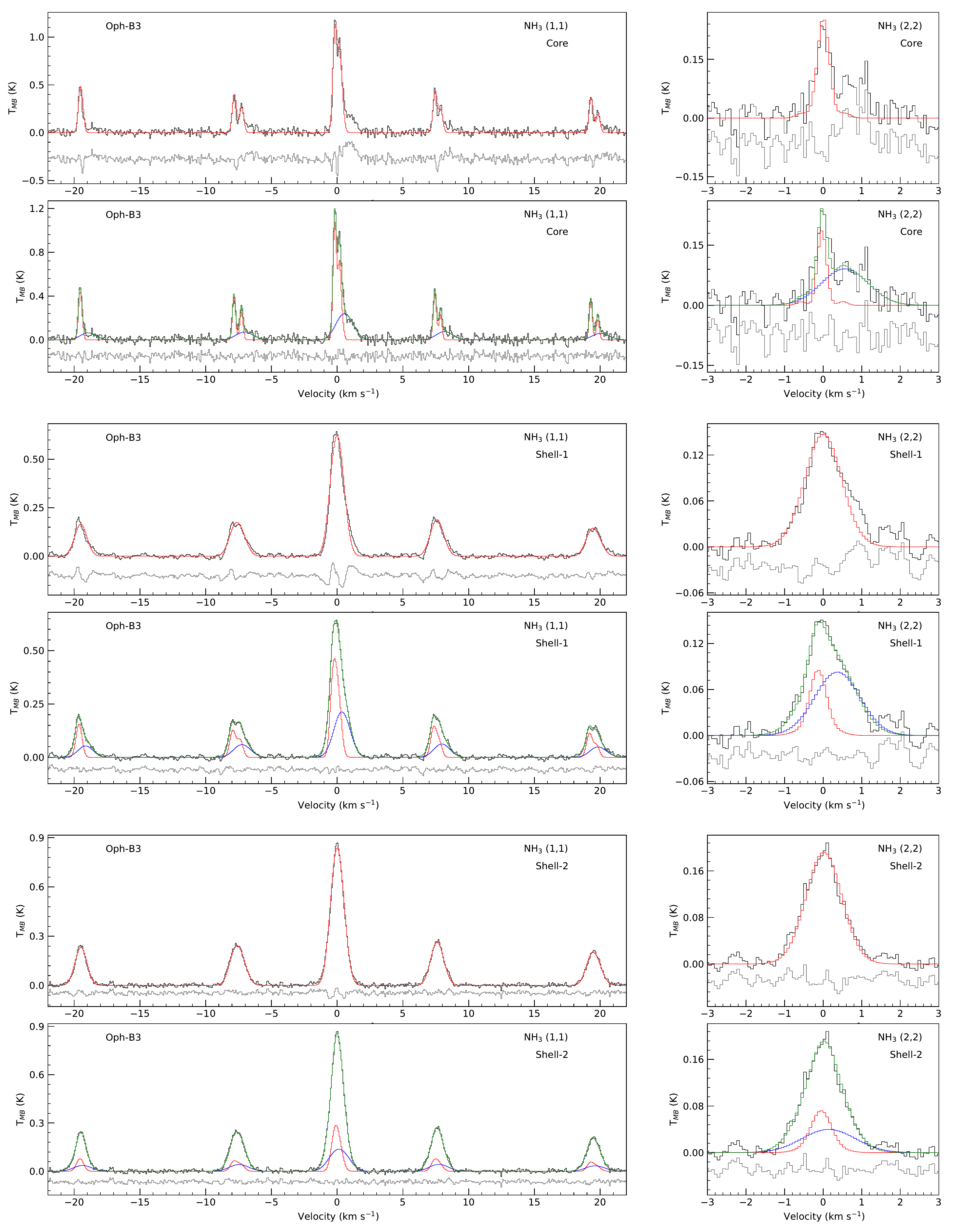}
  \caption{Same as Figure \ref{avg_spec_ophA}, but for Oph-B3}
     \label{avg_spec_c4} 
\end{figure*}

\begin{figure*}[!ht]  
\centering
\includegraphics[width=1\textwidth]{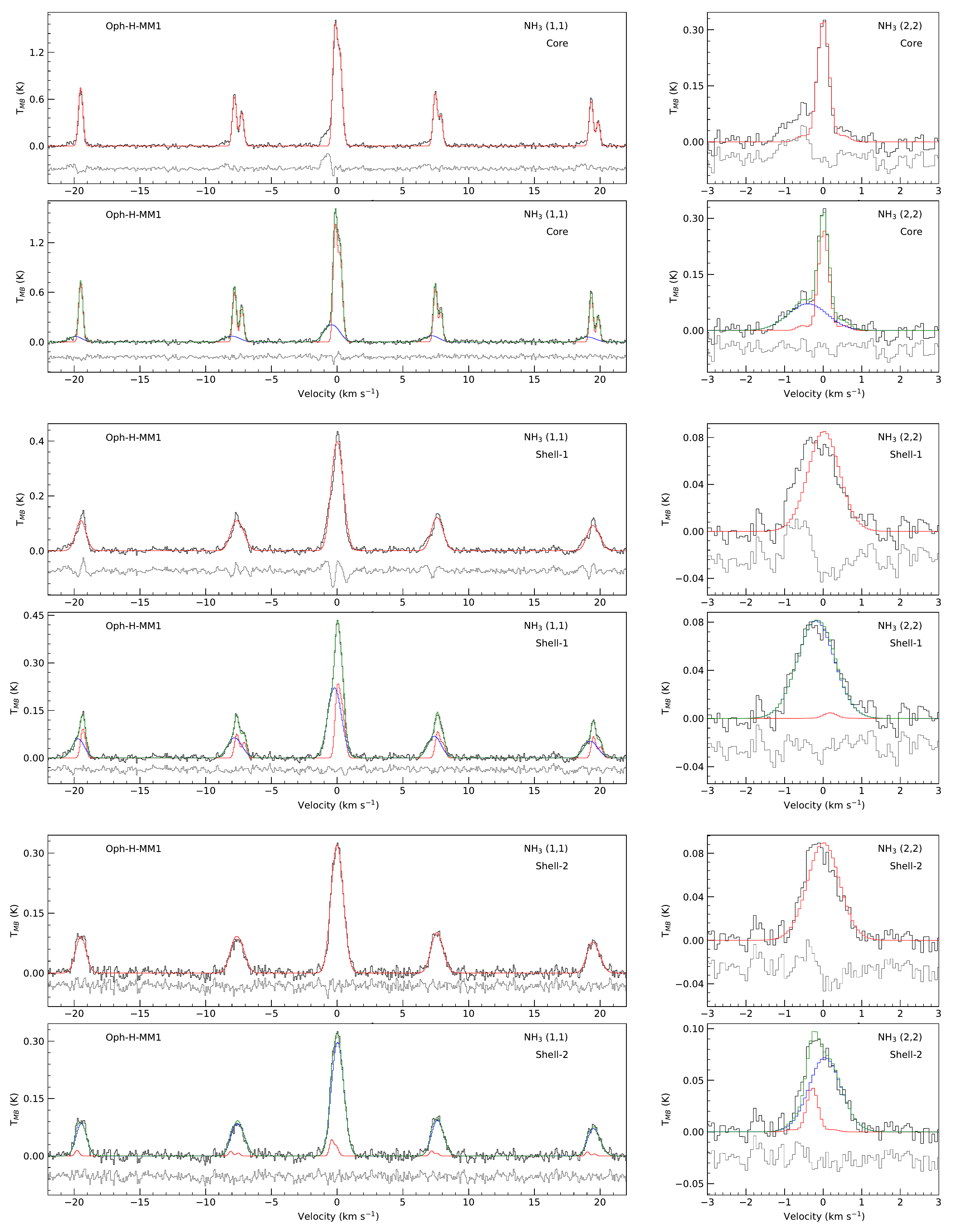}
  \caption{Same as Figure \ref{avg_spec_ophA}, but for Oph-H-MM1}
     \label{avg_spec_hmm1} 
\end{figure*}

\begin{figure*}[!ht]  
\centering
\includegraphics[width=1\textwidth]{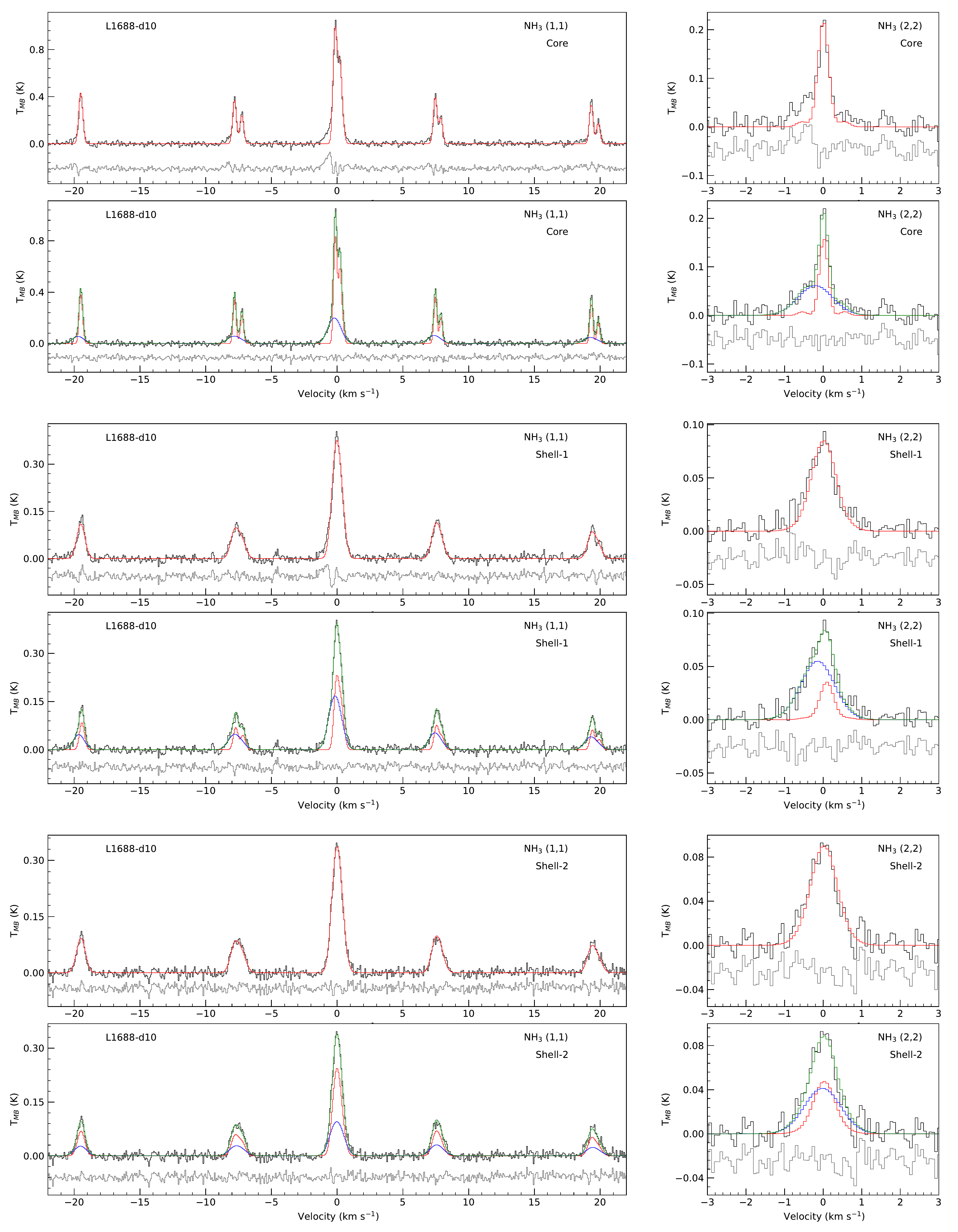}
  \caption{Same as Figure \ref{avg_spec_ophA}, but for L1688-d10}
     \label{avg_spec_d10} 
\end{figure*}

\begin{figure*}[!ht]  
\centering
\includegraphics[width=1\textwidth]{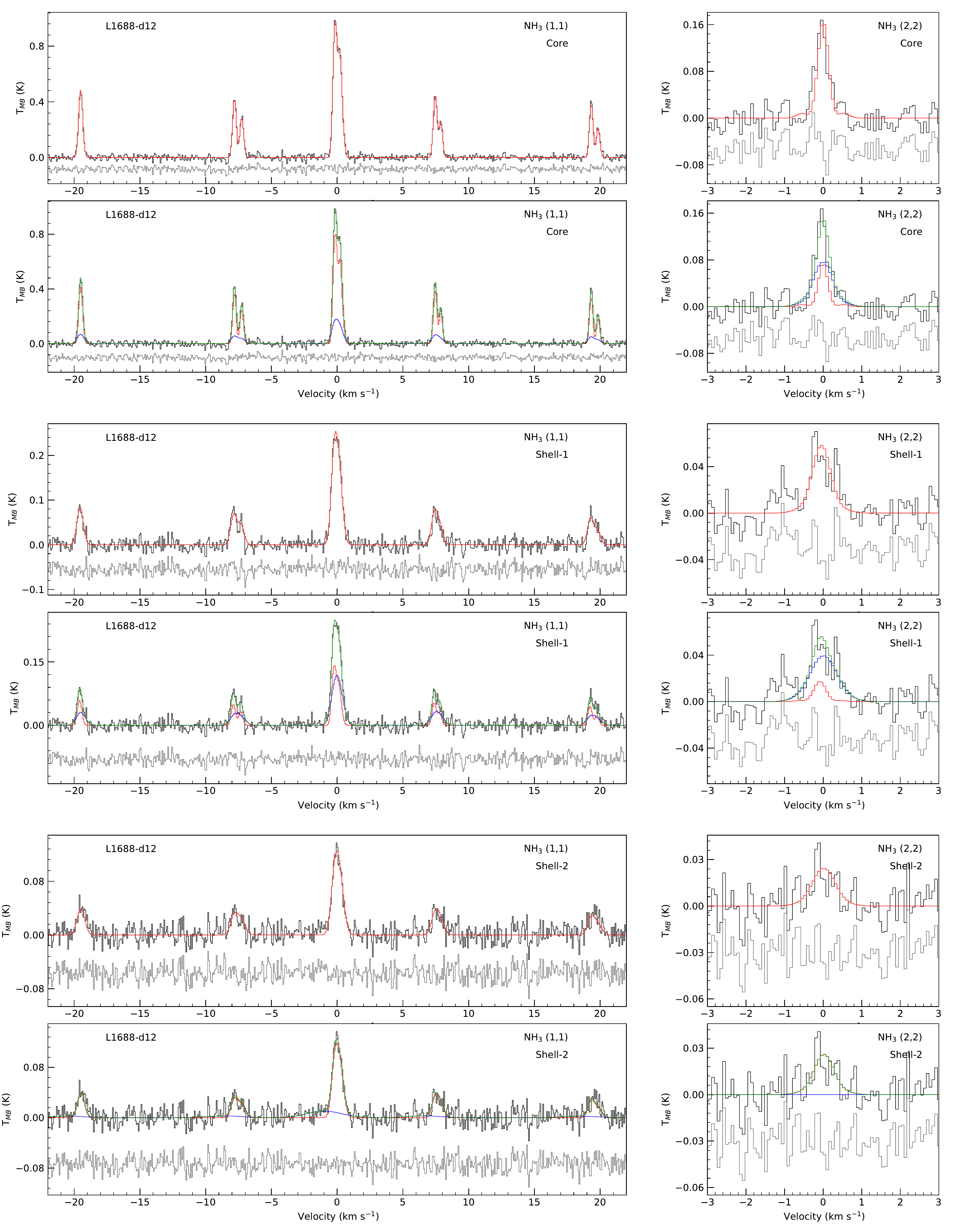}
  \caption{Same as Figure \ref{avg_spec_ophA}, but for L1688-d12}
     \label{avg_spec_d12} 
\end{figure*}

\begin{figure*}[!ht]  
\centering
\includegraphics[width=1\textwidth]{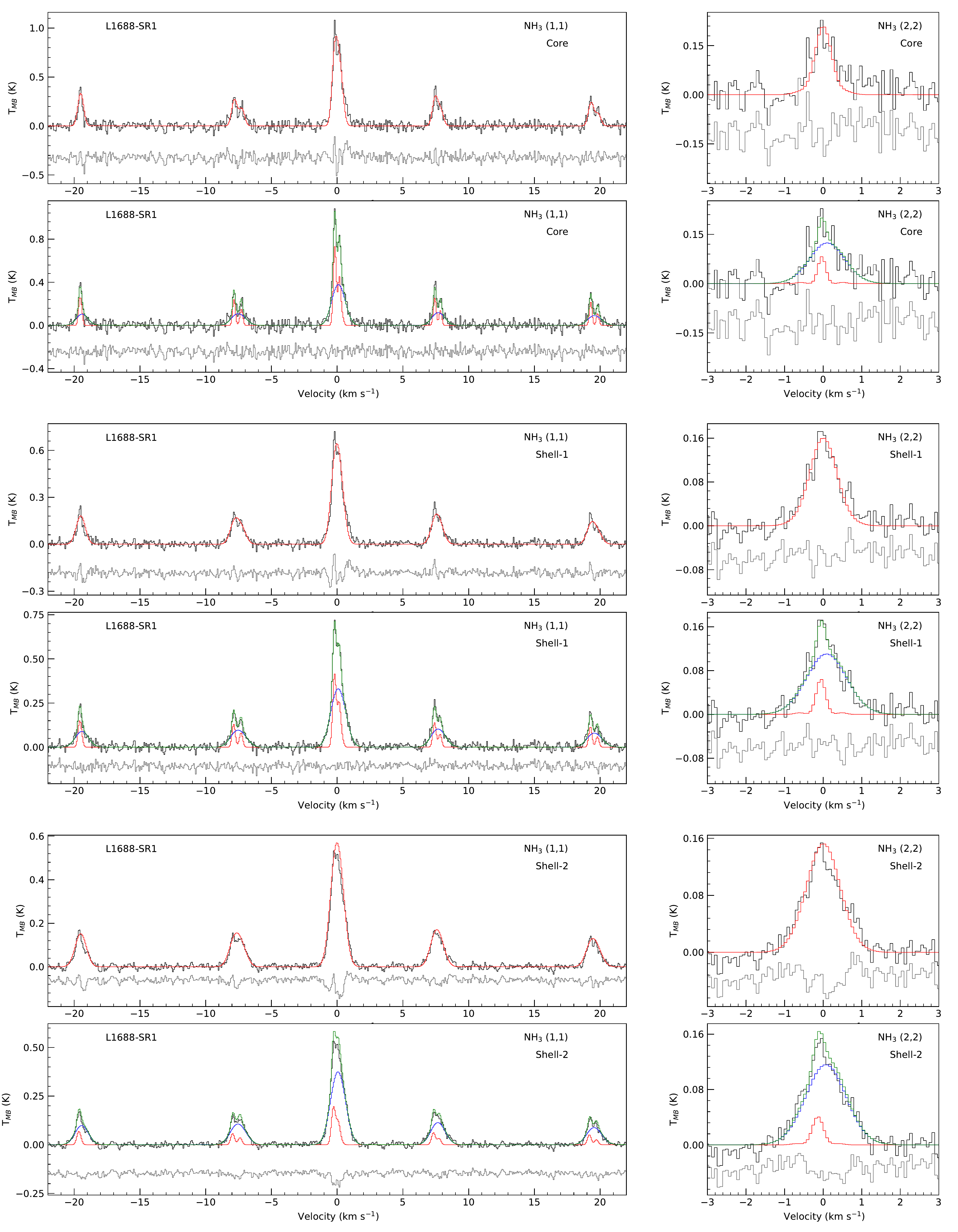}
  \caption{Same as Figure \ref{avg_spec_ophA}, but for L1688-SR1}
     \label{avg_spec_bsou} 
\end{figure*}

\begin{figure*}[!ht]  
\centering
\includegraphics[width=1\textwidth]{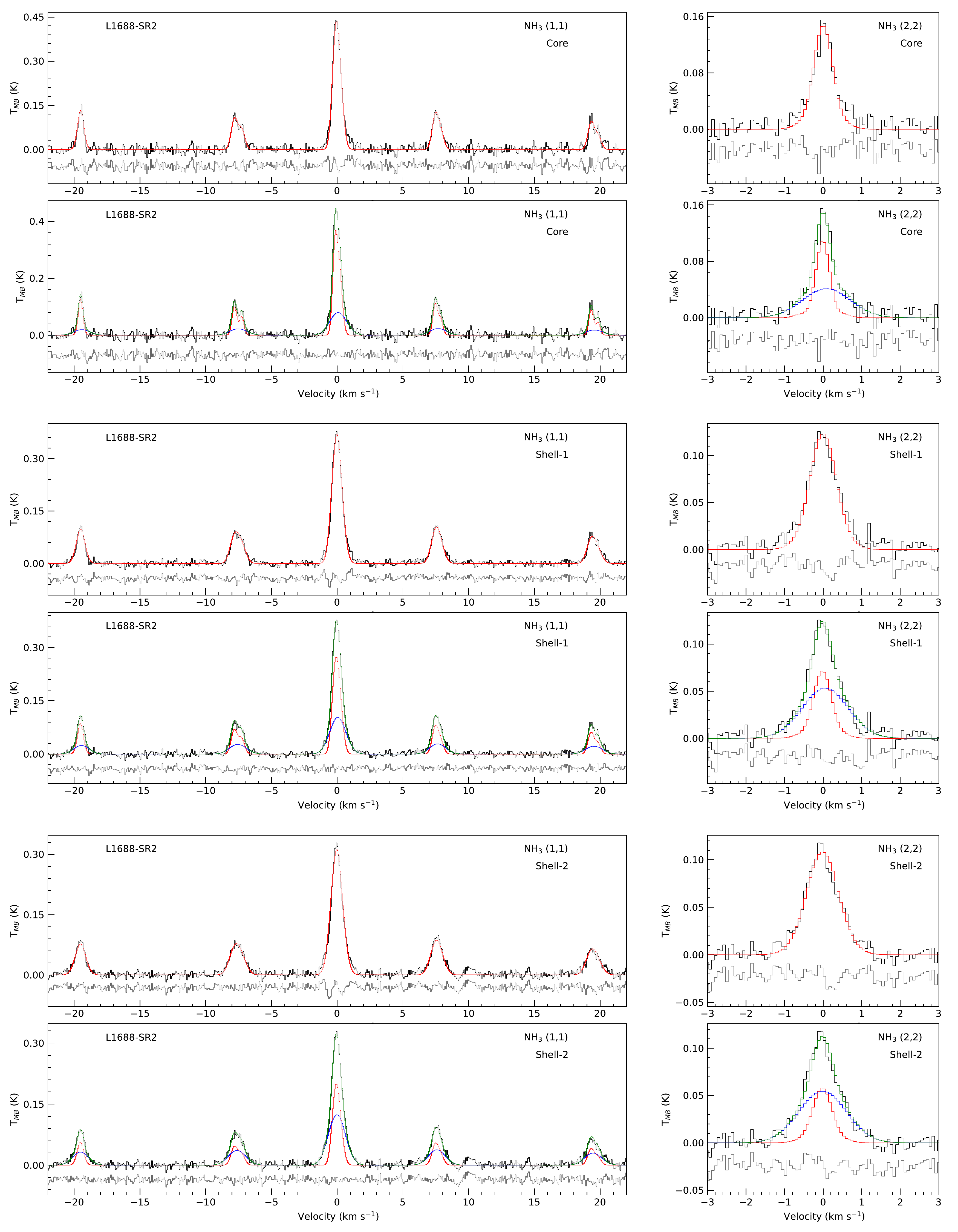}
  \caption{Same as Figure \ref{avg_spec_ophA}, but for L1688-SR2}
     \label{avg_spec_top} 
\end{figure*}

\begin{figure*}[!ht]  
\centering
\includegraphics[width=1\textwidth]{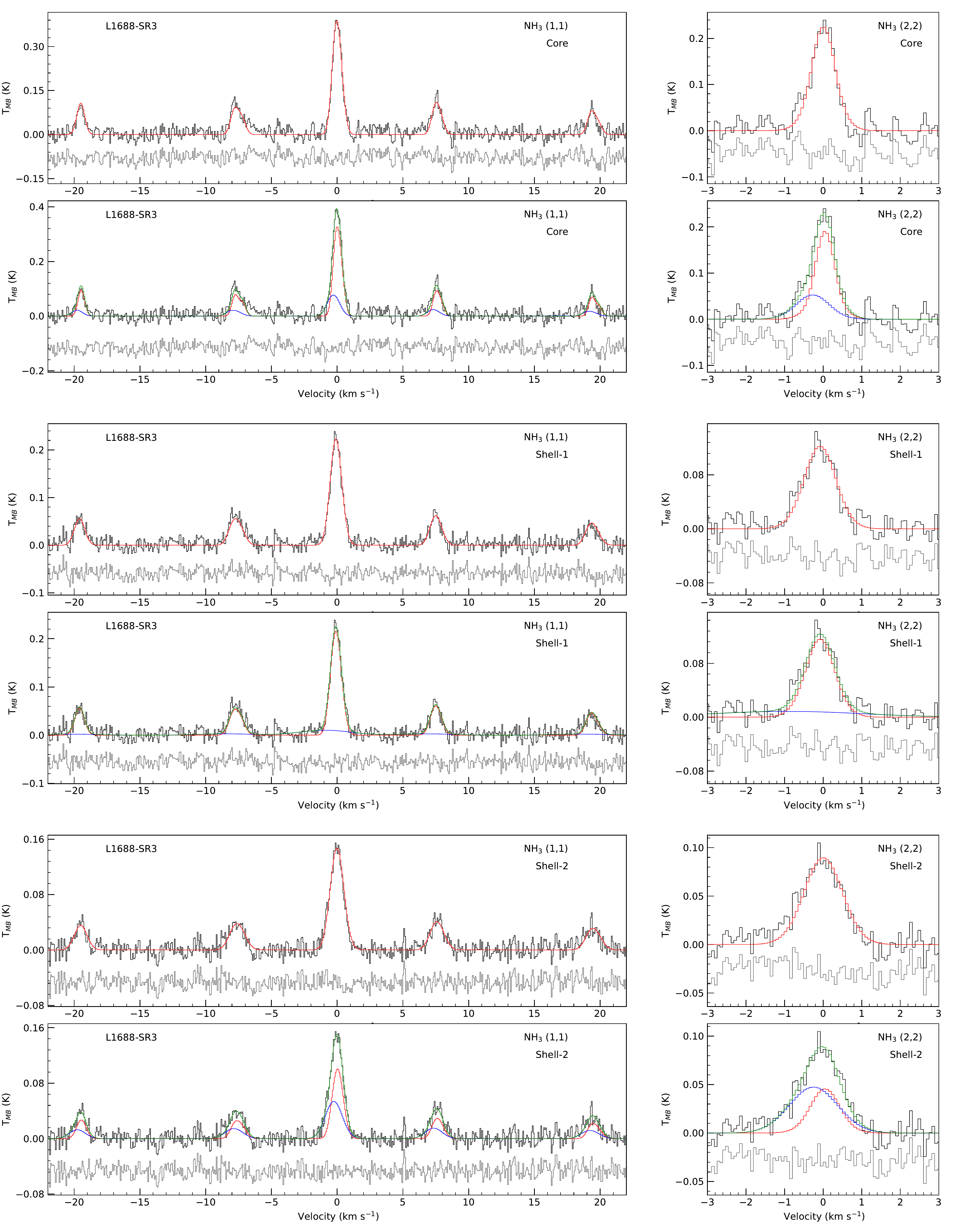}
  \caption{Same as Figure \ref{avg_spec_ophA}, but for L1688-SR3}
     \label{avg_spec_west} 
\end{figure*}

Figures \ref{avg_spec_ophA} to \ref{avg_spec_west} shows the average spectra in the respective coherent cores, used in Section \ref{sec_transi_avg}. The final model determined by the fit is overlaid (in green) on the spectra.

\begin{figure*}[!ht]  
\centering
\includegraphics[width=1\textwidth]{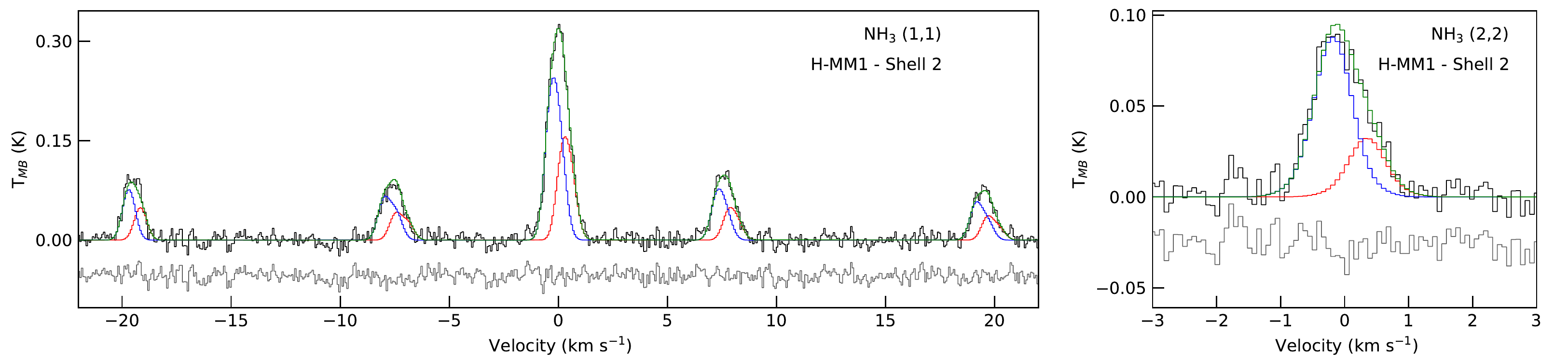}
  \caption{Alternate two-component fit to the average spectra in shell-2 of H-MM1 (see Section \ref{sec_sec_comp}).}
     \label{avg_spec_hmm1_alt} 
\end{figure*}

Figure \ref{avg_spec_hmm1_alt} show an alternate 2-component fit to the shell-2 of H-MM1. Here, we have restricted the relative positions of the narrow and the broad components from switching, i.e., the narrow component to be at a higher velocity than the broad component (Section \ref{sec_sec_comp}).

\section{Fit parameters for average spectra}
\label{app_tab_para}

In Table \ref{tab_fit_para_all}, we present the kinetic temperature, velocity dispersion and p-\amm column densities for the cores and shells 1 \& 2, from single-component fits and two-component fits to the average spectra in the respective regions. We also show the improvement in the fit (as change in AIC value) from single- to two-component fit, and the noise level, for each spectrum.

\begingroup
\onecolumn
\begin{longtable}{ccccccc|cc} 
\label{tab_fit_para_all} \\

\caption[c]{ Best-fit parameters for single and two-component fits in cores and shells.} \\

\hline \hline
& Component \tablefootmark{a} & \tk & $\log_{10}(N(p-\amm)/{\rm cm}^{-2})$ & \sig \tablefootmark{b} & $\rm v_{rel}$ \tablefootmark{c} & $\mathcal{M}_S$ & noise \tablefootmark{d} & $\Delta_{AIC}$\\
& & (K) & & (km s$^{-1}$) & (km s$^{-1}$) & & (mK) &  \\
                    \hline
\endfirsthead
\caption[]{(continued)} \\
\hline \hline
& Component \tablefootmark{a} & \tk & $\log_{10}(N(p-\amm)/{\rm cm}^{-2})$ & \sig \tablefootmark{b} & $\rm v_{rel}$ \tablefootmark{c} & $\mathcal{M}_S$ & noise \tablefootmark{d} & $\Delta_{AIC}$\\
& & (K) & & (km s$^{-1}$) & (km s$^{-1}$) & & (mK) &  \\
                    \hline
\endhead

 & Single & 19.1(2) & 13.86(2) & 0.253(3) & 0.0 \tablefootmark{f} & 0.9 &  &  \\ 
Oph-A : core & Narrow & 15.8(4) & 13.74(4) & 0.156(4) & -0.001(2) & 0.6 & 17 & 761 \\ 
 & Broad & 24.9(8) & -- \tablefootmark{e} & 0.44(1) & -0.014(6) & 1.5 &  &  \\ 
\hdashline
 & Single & 20.7(1) & 13.88(1) & 0.366(2) & 0.001(2) & 1.3 &  &  \\ 
Oph-A : shell-1 & Narrow & 15.3(3) & 13.81(3) & 0.213(5) & 0.009(3) & 0.8 &  8, 7 & 1404 \\ 
 & Broad & 26.7(5) & 13.8(7) & 0.54(1) & -0.019(5) &1.7 &  &  \\ 
\hdashline
 & Single & 22.7(2) & 13.72(3) & 0.5(3) & -0.008(3) & 1.7 &  &  \\ 
Oph-A : shell-2 & Narrow & 12.8(6) & -- \tablefootmark{e} & 0.214(8) & 0.054(6) & 0.9 & 6 & 671 \\ 
 & Broad & 26.6(4) & 13.6(7) & 0.598(8) & -0.042(5) &1.9 &  &  \\ 
\hline

 & Single & 11.72(9) & 14.168(5) & 0.177(1) & 0.0 \tablefootmark{f} & 0.8 &  &  \\ 
Oph-C : core & Narrow & 10.13(9) & 14.088(6) & 0.1376(8) & -0.0015(4) & 0.6 & 9 & 6724 \\ 
 & Broad & 17.7(3) & 14.13(3) & 0.477(7) & -0.033(5) &1.9 &  &  \\ 
\hdashline
 & Single & 14.9(1) & 13.81(1) & 0.346(2) & 0.0(9) & 1.5 &  &  \\ 
Oph-C : shell-1 & Narrow & 10.7(2) & 13.74(2) & 0.195(2) & 0.008(2) & 0.9 & 5 & 5664 \\ 
 & Broad & 20.1(2) & 13.79(5) & 0.59(7) & -0.038(4) &2.2 &  &  \\ 
\hdashline
 & Single & 18.5(1) & 13.63(2) & 0.501(2) & -0.003(2) & 1.9 &  &  \\ 
Oph-C : shell-2 & Narrow & --\tablefootmark{g} & 13.4(2) & 0.2(1) & 0.139(9) & --\tablefootmark{g} & 5 & 427 \\ 
 & Broad & 20.0(2) & 13.66(4) & 0.538(4) & -0.04(4) &2.0 &  &  \\ 
\hline
 & Single & 9.9(2) & 13.94(1) & 0.104(1) & 0.0005(5) & 0.4 &  &  \\ 
 
Oph-D : core & Narrow & 9.7(3) & 13.89(2) & 0.088(1) & -0.006(1) & 0.3 & 22, 21 & 383 \\ 
 & Broad & 12.0(1) & 13.8(1) & 0.39(2) & 0.14(2) & 1.8 &  &  \\ 
\hdashline
 & Single & 11.0(3) & 13.8(2) & 0.148(2) & 0.016(3) & 0.7 &  &  \\ 
Oph-D : shell-1 & Narrow & 9.6(6) & 13.77(4) & 0.098(4) & -0.001(2) & 0.4 & 13 & 351 \\ 
 & Broad & 13.9(8) & 13.5(2) & 0.34(2) & 0.07(1) &1.5 &  &  \\ 
\hdashline
 & Single & 12.4(5) & 13.62(7) & 0.255(8) & 0.058(6) & 1.2 &  &  \\ 
Oph-D : shell-2 & Narrow & 7.0(4) & -- \tablefootmark{e} & 0.104(6) & 0.091(6) & 0.6 & 11 & 123 \\ 
 & Broad & 15.0(1) & 13.7(1) & 0.39(2) & 0.0(3) &1.6 &  &  \\ 
\hline
 & Single & 12.5(4) & 13.77(4) & 0.168(4) & 0.002(4) & 0.7 &  &  \\ 
Oph-E : core & Narrow & 11.7(3) & 13.73(3) & 0.126(2) & 0.023(2) & 0.5 & 24 & 3698 \\ 
 & Broad & 17.9(6) & -- \tablefootmark{e} & 0.76(2) & -0.92(3) &3.0 &  &  \\ 
\hdashline
 & Single & 16.5(1) & 13.79(3) & 0.735(5) & 0.0(1) & 3.0 &  &  \\ 
Oph-E : shell-1 & Narrow & 13.4(5) & 13.48(6) & 0.36(1) & 0.56(1) & 1.6 &  5, 6 & 1574 \\ 
 & Broad & 18.1(2) & 13.49(7) & 0.63(2) & -0.34(4) &2.5 &  &  \\ 
\hdashline
 & Single & 17.1(2) & 13.52(7) & 0.615(5) & 0.0(9) & 2.5 &  &  \\ 
Oph-E : shell-2 & Narrow & 17.5(4) & 13.3(1) & 0.45(2) & -0.14(1) & 1.8 & 6 & 634 \\ 
 & Broad & 16.3(6) & 13.7(1) & 0.8(2) & 0.34(5) &3.3 &  &  \\ 
\hline
 & Single & 13.9(1) & 13.91(1) & 0.174(2) & -0.0003(5) & 0.7 &  &  \\ 
Oph-F : core & Narrow & 12.6(1) & 13.81(1) & 0.128(1) & 0.006(1) & 0.5 &  9, 11 & 5244 \\ 
 & Broad & 19.1(4) & 13.87(7) & 0.55(1) & -0.144(9) &2.1 &  &  \\ 
\hdashline
 & Single & 16.3(1) & 13.67(3) & 0.423(3) & -0.004(3) & 1.7 &  &  \\ 
Oph-F : shell-1 & Narrow & 13.5(2) & 13.59(3) & 0.23(4) & 0.058(3) & 1.0 &  5, 6 & 3536 \\ 
 & Broad & 18.8(3) & 13.79(5) & 0.638(8) & -0.148(7) &2.5 &  &  \\ 
\hdashline
 & Single & 18.0(1) & 13.57(4) & 0.507(3) & 0.002(4) & 2.0 &  &  \\ 
Oph-F : shell-2 & Narrow & 18.0(5) & -- \tablefootmark{e} & 0.38(2) & -0.025(6) & 1.5 & 6 & 220 \\ 
 & Broad & 18.1(7) & 13.6(2) & 0.72(3) & 0.06(2) &2.9 &  &  \\ 
\hline
 & Single & 11.6(1) & 14.0(9) & 0.15(1) & 0.0008(5) & 0.6 &  &  \\ 
Oph-H-MM1 : core & Narrow & 11.0(1) & 13.971(9) & 0.12(1) & 0.012(1) & 0.6 & 16 & 1489 \\ 
 & Broad & 16.9(7) & 13.99(8) & 0.47(2) & -0.4(3) &1.9 &  &  \\ 
\hdashline
 & Single & 14.7(2) & 13.6(5) & 0.387(5) & 0.015(4) & 1.7 &  &  \\ 
Oph-H-MM1 : shell-1 & Narrow & 7.0(1) & 13.54(5) & 0.171(5) & 0.182(5) & 1.0 &  6, 7 & 1513 \\ 
 & Broad & 18.8(4) & 13.65(7) & 0.451(8) & -0.19(1) &1.7 &  &  \\ 
\hdashline
 & Single & 16.2(2) & 13.68(4) & 0.391(5) & 0.007(4) & 1.6 &  &  \\ 
Oph-H-MM1 : shell-2 & Narrow & 30.0(6) & -- \tablefootmark{e} & 0.15(2) & -0.28(2) & 0.3 & 9 & 41 \\ 
 & Broad & 15.2(3) & 13.66(5) & 0.392(6) & 0.05(1) &1.7 &  &  \\ 
\hline

\pagebreak

 & Single & 12.1(2) & 13.79(2) & 0.137(2) & 0.005(1) & 0.6 &  &  \\ 
L1688-d10 : core & Narrow & 11.2(2) & 13.76(2) & 0.111(2) & 0.015(2) & 0.4 & 14 & 826 \\ 
 & Broad & 17.1(8) & 13.6(2) & 0.42(2) & -0.2(2) &1.7 &  &  \\
\hdashline

 
 & Single & 14.7(2) & 13.51(5) & 0.304(5) & 0.011(4) & 1.3 &  &  \\ 
L1688-d10 : shell-1 & Narrow & 11.7(7) & 13.4(1) & 0.177(8) & 0.097(7) & 0.8 & 8 & 303 \\ 
 & Broad & 17.6(7) & 13.6(1) & 0.41(1) & -0.14(2) &1.6 &  &  \\ 
\hdashline

 & Single & 16.2(2) & 13.39(7) & 0.328(5) & 0.017(4) & 1.3 &  &  \\ 
L1688-d10 : shell-2 & Narrow & 14.0(1) & -- \tablefootmark{e} & 0.28(2) & 0.03(1) & 1.2 &  9, 8 & 4 \\ 
 & Broad & 20.0(3) & -- \tablefootmark{e} & 0.44(6) & -0.02(3) &1.6 &  &  \\ 
\hline
 & Single & 10.6(2) & 13.92(1) & 0.135(1) & -0.0004(4) & 0.6 &  &  \\ 
L1688-d12 : core & Narrow & 9.4(8) & 13.91(3) & 0.123(4) & -0.0009(7) & 0.6 & 16 & 30 \\ 
 & Broad & 18.0(3) & 13.6(4) & 0.26(5) & 0.0(1) & 1 &  &  \\ 
\hdashline
 & Single & 14.1(5) & 13.49(9) & 0.231(7) & -0.046(6) & 1.0 &  &  \\ 
L1688-d12 : shell-1 & Narrow & 10.0(2) & -- \tablefootmark{e} & 0.15(1) & -0.08(1) & 0.7 & 11,12 & 23 \\ 
 & Broad & 18.0(2) & -- \tablefootmark{e} & 0.33(4) & 0.01(3) &1.3 &  &  \\ 
\hdashline
 & Single & 13.6(9) & 13.5(2) & 0.29(2) & 0.01(1) & 1.3 &  &  \\ 
L1688-d12 : shell-2 & Narrow\tablefootmark{h} & -- & -- & -- & -- & -- & 12, 11 & 9 \\ 
 & Broad\tablefootmark{h} & -- & --  & -- & -- & -- &  &  \\ 
\hline
 & Single & 12.4(3) & 13.81(3) & 0.149(3) & -0.002(2) & 0.6 &  &  \\ 
Oph-B3 : core & Narrow & 11.4(3) & 13.67(3) & 0.117(2) & -0.028(2) & 0.5 & 26, 28 & 1083 \\ 
 & Broad & 19.0(1) & 13.8(2) & 0.58(3) & 0.56(4) & 2.2 &  &  \\ 
\hdashline
 & Single & 15.6(2) & 13.59(5) & 0.379(5) & -0.012(4) & 1.6 &  &  \\ 
Oph-B3 : shell-1 & Narrow & 12.9(3) & 13.54(4) & 0.219(6) & -0.131(4) & 1.0 &  8, 9 & 1586 \\ 
 & Broad & 20.1(6) & 13.5(2) & 0.56(2) & 0.37(3) &2.1 &  &  \\ 
\hdashline
 & Single & 16.5(3) & -- \tablefootmark{e} & 0.417(7) & -0.005(6) & 1.7 &  &  \\ 
Oph-B3 : shell-2 & Narrow & 15.8(8) & -- \tablefootmark{e} & 0.27(2) & -0.05(9) & 1.1 &  8, 9 & 266 \\ 
 & Broad & 17.0(1) & 13.8(2) & 0.66(4) & 0.14(3) &2.7 &  &  \\ 
\hline
 & Single & 13.5(4) & 13.75(5) & 0.209(6) & 0.000(1) & 0.9 &  &  \\ 
L1688-SR1 : core & Narrow & 10(1) & 13.3(1) & 0.102(5) & -0.033(4) & 0.4 & 35 & 308 \\ 
 & Broad & 18.0(1) & 13.7(2) & 0.43(2) & 0.11(2) &1.7 &  &  \\ 
\hdashline
 & Single & 15.5(3) & 13.61(5) & 0.334(6) & 0.002(6) & 1.4 &  &  \\ 
L1688-SR1 : shell-1 & Narrow & 11.6(7) & 13.2(1) & 0.121(4) & -0.058(4) & 0.5 & 16 & 582 \\ 
 & Broad & 18.0(6) & 13.7(1) & 0.49(1) & 0.09(1) &1.9 &  &  \\ 
\hdashline
 & Single & 16.6(2) & 13.6(5) & 0.432(5) & 0.002(7) & 1.8 &  &  \\ 
L1688-SR1 : shell-2 & Narrow & 13.1(9) & 13.1(3) & 0.142(8) & -0.136(7) & 0.6 & 10, 12 & 373 \\ 
 & Broad & 17.5(4) & 13.67(7) & 0.498(9) & 0.072(8) &2.0 &  &  \\ 
\hline
 & Single & 17.2(3) & 13.35(8) & 0.239(4) & -0.000(2) & 0.9 &  &  \\ 
L1688-SR2 : core & Narrow & 15.4(5) & 13.3(1) & 0.178(6) & -0.009(4) & 0.7 & 11 & 164 \\ 
 & Broad & 24.0(2) & 13.0(4) & 0.6(5) & 0.08(3) & 2.0 &  &  \\ 
\hdashline
 & Single & 18.2(2) & 13.24(8) & 0.328(4) & -0.002(2) & 1.2 &  &  \\ 
L1688-SR2 : shell-1 & Narrow & 15.2(5) & 13.3(1) & 0.228(8) & -0.018(4) & 0.9 &  6, 7 & 251 \\ 
 & Broad & 24.0(1) & -- \tablefootmark{e} & 0.58(3) & 0.05(2) &1.9 &  &  \\ 
\hdashline
 & Single & 19.1(2) & -- \tablefootmark{e} & 0.398(5) & -0.017(4) & 1.5 &  &  \\ 
L1688-SR2 : shell-2 & Narrow & 16.6(7) & -- \tablefootmark{e} & 0.25(1) & -0.02(6) & 1.0 & 6 & 219 \\ 
 & Broad & 21.3(9) & 13.6(2) & 0.59(3) & -0.01(1) &2.1 &  &  \\ 
\hline
 & Single & 24.6(6) & -- \tablefootmark{e} & 0.295(8) & 0.0 \tablefootmark{f} & 0.9 &  &  \\ 
L1688-SR3 : core & Narrow & 24(1) & -- \tablefootmark{e} & 0.24(2) & 0.05(2) & 0.8 & 18 & 12 \\ 
 & Broad & 27.0(5) & -- \tablefootmark{e} & 0.41(7) & -0.2(2) & 1.3 &  &  \\ 
\hdashline
 & Single & 24.9(6) & -- \tablefootmark{e} & 0.39(1) & -0.084(7) & 1.3 &  &  \\ 
L1688-SR3 : shell-1 & Narrow \tablefootmark{h} & -- & --  & -- & -- & --&  9 & 54 \\ 
 & Broad \tablefootmark{h} & -- & --  & -- & -- & -- &  &  \\ 
\hdashline
 & Single & 27.2(7) & -- \tablefootmark{e} & 0.48(1) & 0.0(0) \tablefootmark{f} & 1.5 &  &  \\ 
L1688-SR3 : shell-2 & Narrow & 21.0(3) & -- \tablefootmark{e} & 0.36(6) & 0.05(4) & 1.3 & 9 & 40 \\ 
 & Broad & 38.0(8) & -- \tablefootmark{e} & 0.61(6) & -0.2(1) &1.6 &  &  \\ 
\hline

\end{longtable}
\endgroup

{ \textbf{Notes. }
Kinetic temperatures, p-\amm column densities, velocity dispersions and velocities, derived from single-component, and two-component fits, in the coherent cores and the shells. 
The values in parentheses represent the fit-determined error in the final decimal place of the corresponding parameter. These uncertainties do not include the calibration uncertainty, which is $\sim$10 \%. 
Also shown are the Mach numbers for each component, the rms noise in the averaged spectra, and the decrease in AIC parameter, from a 1-component fit to 2-component fit ($\rm \Delta_{AIC} = AIC_{1-comp.} - AIC_{2-comp.}$).\\
\tablefoottext{a}{single-component fit, or the individual components of the two-component fit.} \\
\tablefoottext{b}{The channel response, $\sigma_{chan}$ (see Section \ref{sec_id_coh}), is not removed from the \sig value reported here(The contribution from $\sigma_{chan}$ is very small, changes only in the third decimal place in \sig).} \\
\tablefoottext{c}{Velocity from the fit. Since we align the spectrum in the core by the velocity at each pixel (determined from single-component fit), the velocities reported in this Table are relative to the mean velocity in the corresponding core or shell.} \\
\tablefoottext{d}{Noise level estimated for both \amm (1,1) and (2,2). In cases where the noise in \amm (1,1) and (2,2) are not the equal, the noise in both the line are shown}. \\
\tablefoottext{e}{Excitation temperature could not be well-constrained from the fit (fit determined error > 30\%), and therefore, the column density estimate is not very reliable.} \\
\tablefoottext{f}{Value and error smaller than 10$^{-4}$ \kms.}
\tablefoottext{g}{Kinetic temperature not could not be determined. See Section \ref{sec_sec_comp}}
\tablefoottext{h}{Two-component fit not reliable, see Section \ref{sec_sec_comp}}}

\bibliographystyle{aa}
\bibliography{biblio}

\begin{thebibliography}{51}
\expandafter\ifx\csname natexlab\endcsname\relax\def\natexlab#1{#1}\fi

\bibitem[{{Andr{\'e}} {et~al.}(2010){Andr{\'e}}, {Men'shchikov}, {Bontemps},
  {K{\"o}nyves}, {Motte}, {Schneider}, {Didelon}, {Minier}, {Saraceno},
  {Ward-Thompson}, {di Francesco}, {White}, {Molinari}, {Testi}, {Abergel},
  {Griffin}, {Henning}, {Royer}, {Mer{\'\i}n}, {Vavrek}, {Attard},
  {Arzoumanian}, {Wilson}, {Ade}, {Aussel}, {Baluteau}, {Benedettini},
  {Bernard}, {Blommaert}, {Cambr{\'e}sy}, {Cox}, {di Giorgio}, {Hargrave},
  {Hennemann}, {Huang}, {Kirk}, {Krause}, {Launhardt}, {Leeks}, {Le Pennec},
  {Li}, {Martin}, {Maury}, {Olofsson}, {Omont}, {Peretto}, {Pezzuto}, {Prusti},
  {Roussel}, {Russeil}, {Sauvage}, {Sibthorpe}, {Sicilia-Aguilar}, {Spinoglio},
  {Waelkens}, {Woodcraft}, \& {Zavagno}}]{hgbs_andre2010}
{Andr{\'e}}, P., {Men'shchikov}, A., {Bontemps}, S., {et~al.} 2010, \aap, 518,
  L102

\bibitem[{{Barranco} \& {Goodman}(1998)}]{coh_core_barranco_goodman_1998}
{Barranco}, J.~A. \& {Goodman}, A.~A. 1998, \apj, 504, 207

\bibitem[{{Bergin} \& {Langer}(1997)}]{bergin_langer_97_depletion}
{Bergin}, E.~A. \& {Langer}, W.~D. 1997, \apj, 486, 316

\bibitem[{{Caselli} {et~al.}(2017){Caselli}, {Bizzocchi}, {Keto}, {Sipil{\"a}},
  {Tafalla}, {Pagani}, {Kristensen}, {van der Tak}, {Walmsley}, {Codella},
  {Nisini}, {Aikawa}, {Faure}, \& {van Dishoeck}}]{caselli_2017_amm_abun}
{Caselli}, P., {Bizzocchi}, L., {Keto}, E., {et~al.} 2017, \aap, 603, L1

\bibitem[{{Caselli} \& {Myers}(1995)}]{CM95_lw_sz}
{Caselli}, P. \& {Myers}, P.~C. 1995, \apj, 446, 665

\bibitem[{{Chen} {et~al.}(2020){Chen}, {Offner}, {Pineda}, {Goodman},
  {Burkert}, {Ginsburg}, \& {Choudhury}}]{chen_2020_core-form}
{Chen}, H. H.-H., {Offner}, S. S.~R., {Pineda}, J.~E., {et~al.} 2020, arXiv
  e-prints, arXiv:2006.07325

\bibitem[{{Chen} {et~al.}(2019){Chen}, {Pineda}, {Goodman}, {Burkert},
  {Offner}, {Friesen}, {Myers}, {Alves}, {Arce}, \& {Caselli}}]{chen2019}
{Chen}, H. H.-H., {Pineda}, J.~E., {Goodman}, A.~A., {et~al.} 2019, \apj, 877,
  93

\bibitem[{{Choudhury} {et~al.}(2020){Choudhury}, {Pineda}, {Caselli},
  {Ginsburg}, {Offner}, {Rosolowsky}, {Friesen}, {Alves}, {Chac{\'o}n-Tanarro},
  {Punanova}, {Redaelli}, {Kirk}, {Myers}, {Martin}, {Shirley}, {Chun-Yuan
  Chen}, {Goodman}, \& {Di Francesco}}]{choudhury2020_letter}
{Choudhury}, S., {Pineda}, J.~E., {Caselli}, P., {et~al.} 2020, \aap, 640, L6

\bibitem[{{Crapsi} {et~al.}(2007){Crapsi}, {Caselli}, {Walmsley}, \&
  {Tafalla}}]{crapsi_2007_temp_drop_core}
{Crapsi}, A., {Caselli}, P., {Walmsley}, M.~C., \& {Tafalla}, M. 2007, \aap,
  470, 221

\bibitem[{{Di Francesco} {et~al.}(2004){Di Francesco}, {Andr{\'e}}, \&
  {Myers}}]{di_fran_2004_ophA_n}
{Di Francesco}, J., {Andr{\'e}}, P., \& {Myers}, P.~C. 2004, \apj, 617, 425

\bibitem[{{Dunham} {et~al.}(2015){Dunham}, {Allen}, {Evans},
  {Broekhoven-Fiene}, {Cieza}, {Di Francesco}, {Gutermuth}, {Harvey},
  {Hatchell}, {Heiderman}, {Huard}, {Johnstone}, {Kirk}, {Matthews}, {Miller},
  {Peterson}, \& {Young}}]{yso_l1688}
{Dunham}, M.~M., {Allen}, L.~E., {Evans}, Neal~J., I., {et~al.} 2015, \apjs,
  220, 11

\bibitem[{{Enoch} {et~al.}(2009){Enoch}, {Evans}, {Sargent}, \&
  {Glenn}}]{enoch_2009_str0-1}
{Enoch}, M.~L., {Evans}, Neal~J., I., {Sargent}, A.~I., \& {Glenn}, J. 2009,
  \apj, 692, 973

\bibitem[{{Evans} {et~al.}(2001){Evans}, {Rawlings}, {Shirley}, \&
  {Mundy}}]{evans_2001_core-temp}
{Evans}, Neal~J., I., {Rawlings}, J. M.~C., {Shirley}, Y.~L., \& {Mundy}, L.~G.
  2001, \apj, 557, 193

\bibitem[{{Foreman-Mackey} {et~al.}(2013){Foreman-Mackey}, {Hogg}, {Lang}, \&
  {Goodman}}]{emcee2013}
{Foreman-Mackey}, D., {Hogg}, D.~W., {Lang}, D., \& {Goodman}, J. 2013, \pasp,
  125, 306

\bibitem[{{Foster} {et~al.}(2009){Foster}, {Rosolowsky}, {Kauffmann}, {Pineda},
  {Borkin}, {Caselli}, {Myers}, \& {Goodman}}]{foster_2009_nh2}
{Foster}, J.~B., {Rosolowsky}, E.~W., {Kauffmann}, J., {et~al.} 2009, \apj,
  696, 298

\bibitem[{{Friesen} {et~al.}(2009){Friesen}, {Di Francesco}, {Shirley}, \&
  {Myers}}]{friesen_2009_b3}
{Friesen}, R.~K., {Di Francesco}, J., {Shirley}, Y.~L., \& {Myers}, P.~C. 2009,
  \apj, 697, 1457

\bibitem[{{Friesen} {et~al.}(2017){Friesen}, {Pineda}, {co-PIs}, {Rosolowsky},
  {Alves}, {Chac{\'o}n-Tanarro}, {How-Huan Chen}, {Chun-Yuan Chen}, {Di
  Francesco}, {Keown}, {Kirk}, {Punanova}, {Seo}, {Shirley}, {Ginsburg},
  {Hall}, {Offner}, {Singh}, {Arce}, {Caselli}, {Goodman}, {Martin}, {Matzner},
  {Myers}, {Redaelli}, \& {GAS Collaboration}}]{GASDR1}
{Friesen}, R.~K., {Pineda}, J.~E., {co-PIs}, {et~al.} 2017, \apj, 843, 63

\bibitem[{{Gaia Collaboration}(2018)}]{hd_coord}
{Gaia Collaboration}. 2018, VizieR Online Data Catalog, I/345

\bibitem[{{Galli} {et~al.}(2002){Galli}, {Walmsley}, \&
  {Gon{\c{c}}alves}}]{galli_2002_dust_gas_coup}
{Galli}, D., {Walmsley}, M., \& {Gon{\c{c}}alves}, J. 2002, \aap, 394, 275

\bibitem[{{Ginsburg} \& {Mirocha}(2011)}]{pyspeckit}
{Ginsburg}, A. \& {Mirocha}, J. 2011, {PySpecKit: Python Spectroscopic Toolkit}

\bibitem[{{Goldsmith}(2001)}]{goldsmith_2001_dust_gas_coup}
{Goldsmith}, P.~F. 2001, \apj, 557, 736

\bibitem[{{Goodman} {et~al.}(1998){Goodman}, {Barranco}, {Wilner}, \&
  {Heyer}}]{coh_core_goodman_barranco_1998}
{Goodman}, A.~A., {Barranco}, J.~A., {Wilner}, D.~J., \& {Heyer}, M.~H. 1998,
  \apj, 504, 223

\bibitem[{{G{\"u}ver} \& {{\"O}zel}(2009)}]{guever2009}
{G{\"u}ver}, T. \& {{\"O}zel}, F. 2009, \mnras, 400, 2050

\bibitem[{{Habart} {et~al.}(2003){Habart}, {Boulanger}, {Verstraete}, {Pineau
  des For{\^e}ts}, {Falgarone}, \& {Abergel}}]{habart2003}
{Habart}, E., {Boulanger}, F., {Verstraete}, L., {et~al.} 2003, \aap, 397, 623

\bibitem[{{Habing}(1968)}]{habing1968}
{Habing}, H.~J. 1968, \bain, 19, 421

\bibitem[{{Harju} {et~al.}(2017){Harju}, {Daniel}, {Sipil{\"a}}, {Caselli},
  {Pineda}, {Friesen}, {Punanova}, {G{\"u}sten}, {Wiesenfeld}, {Myers},
  {Faure}, {Hily-Blant}, {Rist}, {Rosolowsky}, {Schlemmer}, \&
  {Shirley}}]{harju_2017}
{Harju}, J., {Daniel}, F., {Sipil{\"a}}, O., {et~al.} 2017, \aap, 600, A61

\bibitem[{{Hollenbach} \& {Tielens}(1999)}]{hollen_tielen_1999}
{Hollenbach}, D.~J. \& {Tielens}, A.~G.~G.~M. 1999, Reviews of Modern Physics,
  71, 173

\bibitem[{{Houk} \& {Smith-Moore}(1988)}]{hd_spec}
{Houk}, N. \& {Smith-Moore}, M. 1988, {Michigan Catalogue of Two-dimensional
  Spectral Types for the HD Stars. Volume 4, Declinations -26{\textdegree}.0 to
  -12{\textdegree}.0.}, Vol.~4

\bibitem[{{Ivlev} {et~al.}(2019){Ivlev}, {Silsbee}, {Sipil{\"a}}, \&
  {Caselli}}]{alexei_2019_cr_ir}
{Ivlev}, A.~V., {Silsbee}, K., {Sipil{\"a}}, O., \& {Caselli}, P. 2019, \apj,
  884, 176

\bibitem[{{Johnstone} {et~al.}(2004){Johnstone}, {Di Francesco}, \&
  {Kirk}}]{johnstone_2004_hmm1}
{Johnstone}, D., {Di Francesco}, J., \& {Kirk}, H. 2004, \apjl, 611, L45

\bibitem[{{Kamazaki} {et~al.}(2003){Kamazaki}, {Saito}, {Hirano}, {Umemoto}, \&
  {Kawabe}}]{kamakazi2003}
{Kamazaki}, T., {Saito}, M., {Hirano}, N., {Umemoto}, T., \& {Kawabe}, R. 2003,
  \apj, 584, 357

\bibitem[{{Kauffmann} {et~al.}(2008){Kauffmann}, {Bertoldi}, {Bourke}, {Evans},
  \& {Lee}}]{jens_2008_mu}
{Kauffmann}, J., {Bertoldi}, F., {Bourke}, T.~L., {Evans}, N.~J., I., \& {Lee},
  C.~W. 2008, \aap, 487, 993

\bibitem[{{Koch} {et~al.}(2018){Koch}, {Rosolowsky}, \&
  {Leroy}}]{koch_2018_chan-width}
{Koch}, E., {Rosolowsky}, E., \& {Leroy}, A.~K. 2018, Research Notes of the
  American Astronomical Society, 2, 220

\bibitem[{{Koumpia} {et~al.}(2015){Koumpia}, {Harvey}, {Ossenkopf}, {van der
  Tak}, {Mookerjea}, {Fuente}, \& {Kramer}}]{tk-td_pdr_koumpia2015}
{Koumpia}, E., {Harvey}, P.~M., {Ossenkopf}, V., {et~al.} 2015, \aap, 580, A68

\bibitem[{{Ladjelate} {et~al.}(2016){Ladjelate}, {Andr{\'e}}, {K{\"o}nyves}, \&
  {Men'shchikov}}]{hgbs_ladj2016}
{Ladjelate}, B., {Andr{\'e}}, P., {K{\"o}nyves}, V., \& {Men'shchikov}, A.
  2016, in IAU Symposium, Vol. 315, From Interstellar Clouds to Star-Forming
  Galaxies: Universal Processes?, ed. P.~{Jablonka}, P.~{Andr{\'e}}, \& F.~{van
  der Tak}, E46

\bibitem[{{Ladjelate} {et~al.}(2020){Ladjelate}, {Andr{\'e}}, {K{\"o}nyves},
  {Ward-Thompson}, {Men'shchikov}, {Bracco}, {Palmeirim}, {Roy}, {Shimajiri},
  {Kirk}, {Arzoumanian}, {Benedettini}, {Di Francesco}, {Fiorellino},
  {Schneider}, \& {Pezzuto}}]{ladje_2020_hgbs}
{Ladjelate}, B., {Andr{\'e}}, P., {K{\"o}nyves}, V., {et~al.} 2020, arXiv
  e-prints, arXiv:2001.11036

\bibitem[{{Launhardt} {et~al.}(2013){Launhardt}, {Stutz}, {Schmiedeke},
  {Henning}, {Krause}, {Balog}, {Beuther}, {Birkmann}, {Hennemann},
  {Kainulainen}, {Khanzadyan}, {Linz}, {Lippok}, {Nielbock}, {Pitann}, {Ragan},
  {Risacher}, {Schmalzl}, {Shirley}, {Stecklum}, {Steinacker}, \&
  {Tackenberg}}]{lht_2013_core_temp}
{Launhardt}, R., {Stutz}, A.~M., {Schmiedeke}, A., {et~al.} 2013, \aap, 551,
  A98

\bibitem[{{Leroy} {et~al.}(2016){Leroy}, {Hughes}, {Schruba}, {Rosolowsky},
  {Blanc}, {Bolatto}, {Colombo}, {Escala}, {Kramer}, {Kruijssen}, {Meidt},
  {Pety}, {Querejeta}, {Sandstrom}, {Schinnerer}, {Sliwa}, \&
  {Usero}}]{leroy_2016_chan-width}
{Leroy}, A.~K., {Hughes}, A., {Schruba}, A., {et~al.} 2016, \apj, 831, 16

\bibitem[{{Loren} {et~al.}(1990){Loren}, {Wootten}, \&
  {Wilking}}]{loren_1990_b3}
{Loren}, R.~B., {Wootten}, A., \& {Wilking}, B.~A. 1990, \apj, 365, 269

\bibitem[{{Lovas} {et~al.}(2009){Lovas}, {Bass}, {Dragoset}, \&
  {Olsen}}]{amm_db}
{Lovas}, F.~J., {Bass}, J.~E., {Dragoset}, R.~A., \& {Olsen}, K.~J. 2009, NIST
  Recommended Rest Frequencies for Observed Interstellar Molecular Microwave
  Transitions - 2009 Revision, (version 3.0),
  {http://physics.nist.gov/restfreq}, [Online; accessed 15-May-2020]

\bibitem[{{Markwardt}(2009)}]{markwardt2009}
{Markwardt}, C.~B. 2009, in Astronomical Society of the Pacific Conference
  Series, Vol. 411, Astronomical Data Analysis Software and Systems XVIII, ed.
  D.~A. {Bohlender}, D.~{Durand}, \& P.~{Dowler}, 251

\bibitem[{{Motte} {et~al.}(1998){Motte}, {Andre}, \& {Neri}}]{motte1998}
{Motte}, F., {Andre}, P., \& {Neri}, R. 1998, \aap, 336, 150

\bibitem[{{Ortiz-Le{\'o}n} {et~al.}(2018){Ortiz-Le{\'o}n}, {Loinard}, {Dzib},
  {Kounkel}, {Galli}, {Tobin}, {Evans}, {Hartmann}, {Rodr{\'\i}guez},
  {Brice{\~n}o}, {Torres}, \& {Mioduszewski}}]{l1688_dist_ortiz-leon}
{Ortiz-Le{\'o}n}, G.~N., {Loinard}, L., {Dzib}, S.~A., {et~al.} 2018, \apjl,
  869, L33

\bibitem[{{Pagani} {et~al.}(2007){Pagani}, {Bacmann}, {Cabrit}, \&
  {Vastel}}]{pagani_2007_core_temp}
{Pagani}, L., {Bacmann}, A., {Cabrit}, S., \& {Vastel}, C. 2007, \aap, 467, 179

\bibitem[{{Pineda} {et~al.}(2010){Pineda}, {Goodman}, {Arce}, {Caselli},
  {Foster}, {Myers}, \& {Rosolowsky}}]{pineda2010}
{Pineda}, J.~E., {Goodman}, A.~A., {Arce}, H.~G., {et~al.} 2010, \apj, 712,
  L116

\bibitem[{{Pon} {et~al.}(2012){Pon}, {Johnstone}, \&
  {Kaufman}}]{pon_2012_shock}
{Pon}, A., {Johnstone}, D., \& {Kaufman}, M.~J. 2012, \apj, 748, 25

\bibitem[{Scott(1992)}]{scott_1992_kde}
Scott, D.~W. 1992, Multivariate Density Estimation: Theory, Practice, and
  Visualization (New York: John Wiley \& Sons)

\bibitem[{{Tafalla} {et~al.}(2002){Tafalla}, {Myers}, {Caselli}, {Walmsley}, \&
  {Comito}}]{tafalla_2002}
{Tafalla}, M., {Myers}, P.~C., {Caselli}, P., {Walmsley}, C.~M., \& {Comito},
  C. 2002, \apj, 569, 815

\bibitem[{{White} {et~al.}(2015){White}, {Drabek-Maunder}, {Rosolowsky},
  {Ward-Thompson}, {Davis}, {Gregson}, {Hatchell}, {Etxaluze}, {Stickler},
  {Buckle}, {Johnstone}, {Friesen}, {Sadavoy}, {Natt}, {Currie}, {Richer},
  {Pattle}, {Spaans}, {di Francesco}, \& {Hogerheijde}}]{white2015}
{White}, G.~J., {Drabek-Maunder}, E., {Rosolowsky}, E., {et~al.} 2015, \mnras,
  447, 1996

\bibitem[{{Young} {et~al.}(2004){Young}, {Lee}, {Evans}, {Goldsmith}, \&
  {Doty}}]{young_2004_tmp_drp_l1544}
{Young}, K.~E., {Lee}, J.-E., {Evans}, Neal~J., I., {Goldsmith}, P.~F., \&
  {Doty}, S.~D. 2004, \apj, 614, 252

\bibitem[{Zucconi {et~al.}(2001)Zucconi, Walmsley, \&
  Galli}]{zucconi_2001_core-temp}
Zucconi, A., Walmsley, C.~M., \& Galli, D. 2001, Astron. Astrophys., 376, 650

\end{thebibliography}

\end{document}